\documentclass[12pt,twoside]{article} 
\usepackage[super,sort,comma]{natbib}

\usepackage{amsmath,graphicx}

\usepackage[english]{babel}
\usepackage[T1]{fontenc}  
\usepackage[latin1]{inputenc} 
\usepackage{url}
\usepackage{graphicx}
\usepackage{subfigure}
\usepackage{xcolor}
\usepackage{amsmath,amsfonts,amssymb}
\usepackage{booktabs}
\usepackage{epsfig}
\usepackage{pstricks}
\usepackage{pst-node}
\usepackage{pst-grad}
\usepackage{ifthen}
\usepackage{algorithm}
\usepackage{algorithmic}
\usepackage{setspace}      

\usepackage{epstopdf}
\usepackage{enumerate}
\usepackage{hyperref}
\usepackage{multicol}
\usepackage{enumitem}
\usepackage{appendix}

\allowdisplaybreaks

\newcommand{\mypar}[1]{\bigskip\noindent {\bf #1.}}

\definecolor{red}{RGB}{153,0,0}

\newcommand{\tabincell}[2]{\begin{tabular}{@{}#1@{}}#2\end{tabular}}

\newcommand{\bx}{\mathbf{x}}
\newcommand{\by}{\mathbf{y}}
\newcommand{\bX}{\mathbf{X}}
\newcommand{\bY}{\mathbf{Y}}
\newcommand{\bZ}{\mathbf{Z}}
\newcommand{\bU}{\mathbf{U}}
\newcommand{\bV}{\mathbf{V}}
\newcommand{\bD}{\mathbf{D}}

\newcommand{\bS}{\mathbf{S}}
\newcommand{\bW}{\mathbf{W}}

\usepackage{fancyhdr}		

\usepackage[section]{placeins} 

\usepackage{graphicx}

\makeatletter \renewcommand\@biblabel[1]{$^{#1}$} \makeatother
\newcommand{\note}[1]{\mbox{}\\ \noindent \rule{16cm}{0.5mm} \\
{\em #1} \\ \noindent \rule{16cm}{0.5mm}
\typeout{    }
\typeout{***********note active on this page *************************}
\typeout{Note: #1  }
\typeout{****************************************end Note}
}

\newcommand{\cen}[1]{\begin{center} #1 \end{center}}

\usepackage[mathlines]{lineno}

\usepackage{hyperref}
\hypersetup{ colorlinks,
    citecolor=blue,
    filecolor=blue,
    linkcolor=blue,
    urlcolor=blue
}

\usepackage{xcolor}
\definecolor{gray}{rgb}{0.6,0.6,0.6}
\definecolor{red}{rgb}{0.85,0,0}
\definecolor{green}{rgb}{0,0.85,0}
\definecolor{blue}{rgb}{0,0,0.85}
\definecolor{beige}{rgb}{0.92,0.87,0.78}
\usepackage[all]{hypcap}    

\begin{document}

\cen{\sf {\large {\bfseries HYDRA: Hybrid Deep Magnetic Resonance Fingerprinting} \\  
\vspace*{10mm}
\normalsize 
Pingfan Song$^{1}$ \quad
Yonina C. Eldar$^{2}$ \quad
Gal Mazor$^{3}$ \quad
Miguel R.\ D.\ Rodrigues$^{4}$  } \\
}

\pagenumbering{roman}
\setcounter{page}{1}
\pagestyle{plain}

\begin{abstract}
\noindent {\bf Purpose:} 
Magnetic resonance fingerprinting (MRF) methods typically rely on dictionary matching to map the temporal MRF signals to quantitative tissue parameters. Such approaches suffer from inherent discretization errors, as well as high computational complexity as the dictionary size grows. To alleviate these issues, we propose a HYbrid Deep magnetic ResonAnce fingerprinting approach, referred to as HYDRA.
\\
{\bf Methods:} 
HYDRA involves two stages: a model-based signature restoration phase and a learning-based parameter restoration phase. Signal restoration is implemented using low-rank based de-aliasing techniques while parameter restoration is performed using a deep nonlocal residual convolutional neural network. The designed network is trained on synthesized MRF data simulated with the Bloch equations and fast imaging with steady state precession (FISP) sequences. In test mode, it takes a temporal MRF signal as input and produces the corresponding tissue parameters. 
\\
{\bf Results:}
We validated our approach on both synthetic data and anatomical data generated from a healthy subject. The results demonstrate that, in contrast to conventional dictionary-matching based MRF techniques, our approach significantly improves inference speed by eliminating the time-consuming dictionary matching operation, and alleviates discretization errors by outputting continuous-valued parameters. We further avoid the need to store a large dictionary, thus reducing memory requirements.
\\
{\bf Conclusions:}
Our approach demonstrates advantages in terms of inference speed, accuracy and storage requirements over competing MRF methods.

~

\end{abstract}

\note{
\scriptsize
	This work was supported by the Royal Society International Exchange Scheme IE160348, by the European Union's Horizon 2020 grant ERC-BNYQ, by the Israel Science Foundation grant no. 335/14, by ICore: the Israeli Excellence Center 'Circle of Light', by the Ministry of Science and Technology, Israel, by UCL Overseas Research Scholarship (UCL-ORS) and by China Scholarship Council (CSC) and by EPSRC grant EP/K033166/1. \\
	$^{1}$ Department of Electronic and Electrical Engineering, Imperial College London, UK \\
	$^{2}$ Faculty of Mathematics and Computer Science, Weizmann Institute of Science, Israel \\
	$^{3}$ Department of Electrical Engineering, Technion -- Israel Institute of Technology, Israel \\
	$^{4}$ Department of Electronic and Electrical Engineering, University College London, UK \\
	p.song@imperial.ac.uk, yonina.eldar@weizmann.ac.il, galmazor@technion.ac.il, m.rodrigues@ucl.ac.uk
}

\newpage     

\setlength{\baselineskip}{0.7cm}      

\pagenumbering{arabic}
\setcounter{page}{1}
\pagestyle{fancy}

\section{Introduction}
\label{sec:Introduction}

\vspace{-0.5cm}

Magnetic Resonance Fingerprinting (MRF)~\cite{ma2013magnetic,jiang2015mr,davies2014compressed,wang2016magnetic,mazor2016low,mazor2018low,liao20173d,cao2017robust} has emerged as a promising Quantitative Magnetic Resonance Imaging (QMRI) approach, with the capability of providing multiple tissue's intrinsic spin parameters simultaneously, such as the spin-lattice magnetic relaxation time (T1) and the spin-spin magnetic relaxation time (T2). Based on the fact that the response from each tissue with respect to a given pseudo-random pulse sequence is unique, MRF exploits pseudo-randomized acquisition parameters to create unique temporal signal signatures, analogous to a "fingerprint", for different tissues. A dictionary matching operation is then performed to map an inquiry temporal signature to the best matching entry in a precomputed dictionary, leading to multiple tissue parameters directly. 

The temporal signatures are generated by varying the acquisition parameters of a pseudo-random excitation pulse sequence, such as repetition time (TR), time of echo (TE), and radio frequency flip angle (FA) over time. The dictionary is composed of a large number of entries that are usually simulated by the Bloch equations given pseudo-random pulse sequences. Each entry represents a unique temporal signature associated with a specific tissue and its quantitative parameters, such as the T1 and T2 relaxation times. Thus, once the best matching (i.e. most correlated) entry is found, it directly leads to multiple quantitative parameters simultaneously via a lookup-table operation.

MRI physics and physiological constraints make the MR scanning procedure time-consuming. To shorten acquisition time, subsampling is commonly performed in k-space (a.k.a conjugate Fourier transform domain) in order to reduce the number of samples and accelerate imaging speed. However, such k-space subsampling results in temporal signatures that are corrupted by aliasing, blurring and noise. This hampers the accuracy associated with estimation of the tissue parameters using a dictionary matching procedure. In order to alleviate the impact of such distortion and corruption, de-aliasing operations are often exploited to restore cleaner signatures before performing signature-to-parameter mapping. Therefore, MRF reconstruction usually involves two operations: signature restoration and parameter restoration. 

Inspired by the successful application of sparsity-driven image processing approaches in MRI reconstruction~\cite{lustig2008compressed,weizman2015compressed,weizman2016reference,eldar2015sampling}, several works~\cite{davies2014compressed,wang2016magnetic,mazor2016low,mazor2018low} suggest to incorporate prior knowledge such as sparsity and low-rank to attenuate distortion and corruption, improving the signature restoration performance, during the initial MRF reconstruction stage. This is then followed by a dictionary matching operation, performing mapping from purified temporal signatures to tissue's quantitative parameters. However, such dictionary matching based signature-to-parameter mapping exhibits several drawbacks~\cite{cohen2018mr,hoppe2017deep}. Since the simulated dictionary and lookup-table contain a finite number of elements, they can only cover a limited number of discrete values for each type of tissue parameter. We refer to the difference between a continuous-valued tissue parameter and its closest available discrete value on a lattice as the discretization error. For example, a pair of dictionary and lookup-table that contain 101 elements will lead to a discretization error of maximum 25 ms if they cover the range of 0 ms - 5000 ms with a fixed interval of 50 ms for a specific tissue parameter, e.g. T1. To reduce the discretization error, a huge dictionary that is composed of a large number of entries is needed to represent tissues with fine granularity over the entire value range of target tissue parameters. However, storing a large dictionary becomes prohibitively memory-consuming, as the dictionary size and density often increase exponentially with the number of tissue parameters. Specifically, the number of entries in a dictionary will be $P^s$ for $s$ parameters each containing $P$ values, since every combination of these $s$ parameters determines a specific tissue which is characterized by a specific signature. For example, given T1, T2 relaxation times, i.e. $s=2$, if each of them contains $1000$ values, the dictionary will have $1000^2$ entries. In addition, finding the best matching entry becomes computationally intense for a large dictionary, considerably limiting the inference speed.

In this paper, we propose an alternative approach to dictionary matching based on deep neural networks (a.k.a. deep learning)~\cite{lecun2015deep,goodfellow2016deep}, which we refer to as HYDRA: HYbrid Deep magnetic ResonAnce fingerprinting. The motivation derives from the fact that a well designed and tuned deep neural network is capable of approximating complex functions, leading to state-of-the-art results in a number of tasks such as image classification, super-resolution, speech recognition, and more~\cite{krizhevsky2012imagenet,he2016deep,he2016identity,dong2016image,kim2016deeply,gehring2017convolutional,hinton2012deep}. Recent work~\cite{cohen2018mr,hoppe2017deep} proposed to exploit neural networks to replace the dictionary and lookup-table used in conventional MRF reconstruction approaches. These proposed neural networks suffer from two limitations: First, these approaches are based on neural network models containing only 3-layers, thus suffer from limited capacity of capturing complex mapping functions. Second, these methods focused exclusively on parameter restoration stage (the second stage in MRF reconstruction), but not on signature restoration (the first stage in MRF reconstruction). Therefore, these techniques rely on fully-sampled data instead of typically available sub-sampled k-space data.

Different from Cohen et al.'s fully-connected feed-forward neural network~\cite{cohen2018mr}, and Hoppe et al.'s vanilla convolutional neural network (CNN)~\cite{hoppe2017deep}, the proposed HYDRA involves both a signature restoration and a parameter restoration phase. Signature restoration is implemented using a low-rank based de-aliasing method adapted from Mazor et al.~\cite{mazor2018low} while parameter restoration is implemented using a deep nonlocal residual convolutional neural network developed for this purpose. Our key contributions with respect to prior work are:

\vspace{-0.3cm}

\begin{itemize}
	\setlength\itemsep{0.0em}
	\item
	HYDRA is, to the best of our knowledge, the first deep network approach to combine model-based de-aliasing and learning-based parameter mapping. HYDRA eliminates the requirement for the memory and time-consuming dictionary matching operation, thus significantly improving inference speed without compromising on reconstruction performance.
	\item
	A 1D nonlocal residual convolutional neural network is designed to capture the mappings from temporal MRF signals to tissue parameters. Owing to residual learning and a self-attention mechanism, our network is deeper and more sophisticated than competing network models. This allows to capture complex parameter mappings more effectively, and output continuous parameters to alleviate discretization issues.
	\item
	The designed network is trained on synthesized MRF data simulated with the Bloch equations, but is still applicable to anatomical data. This contributes to eliminating the requirement for a large amount of real MRF data.
	\item
	A low-rank based de-aliasing technique is developed in order to take advantage of temporal similarity for signature restoration. 
	\item
	The low-rank based signature restoration is organically combined with the learning-based parameter restoration to achieve fast and accurate MRF reconstruction. Such strategy enables HYDRA to handle both fully-sampled k-space data and more importantly sub-sampled k-space data.
	\item
	A series of numerical experiments are conducted to evaluate the proposed approach on both synthetic and anatomical data. The results demonstrate improved inference speed, accuracy and discretization errors over competing methods~\cite{ma2013magnetic,jiang2015mr,davies2014compressed,wang2016magnetic,mazor2016low,mazor2018low,hoppe2017deep,cohen2018mr}.
\end{itemize}

\vspace{-0.3cm}

The rest of the paper is organized as follows. In Section~\ref{sec:Methods}, we formulate the MRF reconstruction problem, introduce related methods, and present our approach, involving the use of a low-rank based signature restoration procedure together with a deep network for parameter restoration. Section~\ref{sec:Experiments} is devoted to experimental results, followed by a discussion in Section~\ref{sec:Discussion} and a conclusion in Section~\ref{sec:Conclusion}.

\section{Materials and Methods}
\label{sec:Methods}

\subsection{The MRF Problem Formulation}
\label{ssec:ProblemForm}

\vspace{-0.2cm}

MRF data is composed of multiple frames sampled in k-space over time. A series of such frames are vectorized and then stacked together along the temporal dimension to construct a measurement matrix $\bY \in \mathcal{C}^{Q \times L}$, where $Q$ is the number of k-space samples in each frame, and $L$ is the number of frames. Due to k-space subsampling, every column vector $\bY_{:,i}$ represents a subsampled Fourier transform of a vectorized image frame $\bX_{:,i}$:
\begin{equation}
\label{Eq:Y_X}
\bY = [\bY_{:,1}, \cdots, \bY_{:,L} ] 
= [ F_u\{\bX_{:,1} \}, \cdots, F_u\{\bX_{:,L} \} ] \,,
\end{equation}
where $F_u\{\cdot \}$ denotes a subsampled 2D Fourier transform.

\begin{figure}[t]
	\centering
	\includegraphics[width = 10cm]{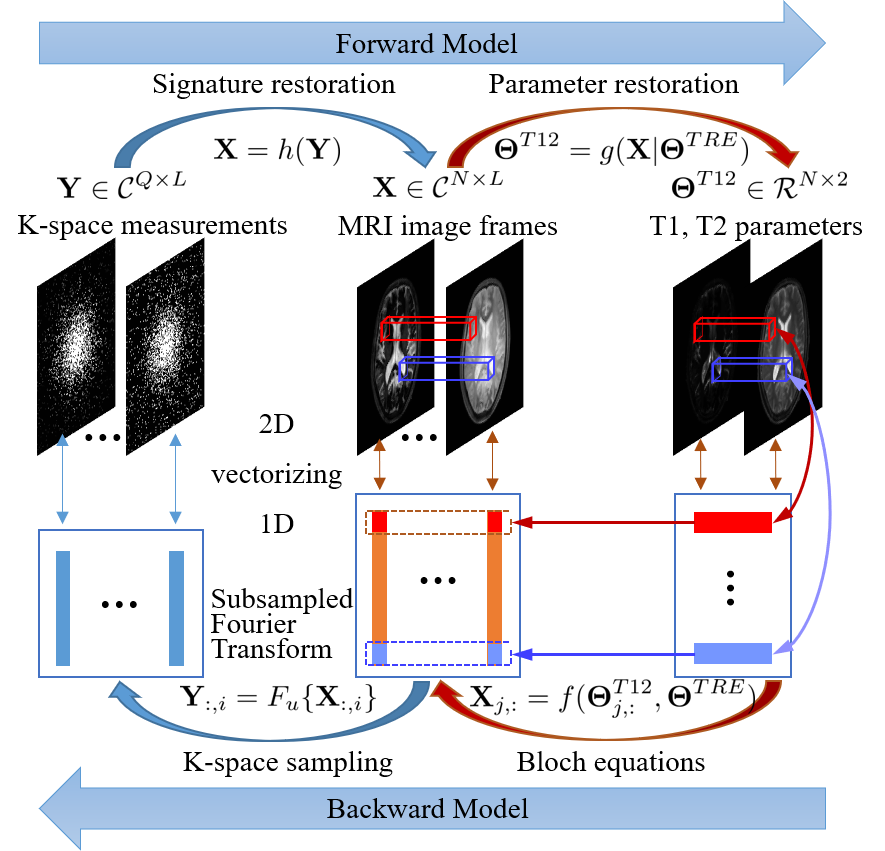}
	\caption{
		Relationship between key variables. The MR contrast matrix $\bX$ is associated with the k-space measurements $\bY$ per column by the subsampled Fourier transform. It is related to tissue parameters $\boldsymbol{\Theta}^{T12}$ per row by the Bloch equations. Given $\boldsymbol{\Theta}^{TRE}$ and $\bY$, the image stack $\bX$ is commonly first reconstructed from $\bY$, referred to as signature restoration, and then mapped to tissue parameters $\boldsymbol{\Theta}^{T12}$ via dictionary matching, referred to as parameter restoration.
		\label{Fig:XY}
	}
\end{figure}

Each column $\bX_{:,i}$ represents a MR contrast acquired with RF sequence parameters:
\begin{equation}
\label{Eq:TRE}
\boldsymbol{\Theta}^{TRE}_{:,i} = [TR^i, TE^i, FA^i]^T, \; i \in [1,L]
\end{equation}
where $TR^i$ and $TE^i$ denote the repetition time and echo time, respectively, and $FA^i$ denotes the flip angle of the RF pulse during sampling the $i$-th contrast. Every row $\bX_{j,:}$ represents a temporal signature, i.e. temporal signal evolution of a specific tissue at the $j$-th image pixel. The signature depends on the tissue's relaxation times, such as T1 and T2, grouped as a row vector:
\begin{equation}
\label{Eq:T12}
\boldsymbol{\Theta}^{T12}_{j,:} = [T1^j, T2^j], \; j \in [1,N]
\end{equation} 
where, $N$ denotes the number of pixels in each image frame. 
Note that, $j$ is the spatial index while $i$ is the temporal index throughout. Given RF sequence parameters $\boldsymbol{\Theta}^{TRE}$, and parameters $\boldsymbol{\Theta}^{T12}_{j,:}$ of a specific tissue, its temporal signature $\bX_{j,:}$ can be derived as:
\begin{equation}
\label{Eq:X_Bloch}
\bX_{j,:} = f(\boldsymbol{\Theta}^{T12}_{j,:}, \boldsymbol{\Theta}^{TRE})
\end{equation}
where $f(\cdot)$ denotes the Bloch equations. This MR contrast matrix $\bX$ is associated with the k-space measurements $\bY$ per column by the subsampled Fourier transform, and it is related to tissue parameters $\boldsymbol{\Theta}^{T12}$ per row by the Bloch equations, as illustrated in Fig.~\ref{Fig:XY}.

\begin{figure}[t]
	\centering
	\includegraphics[width = 11cm]{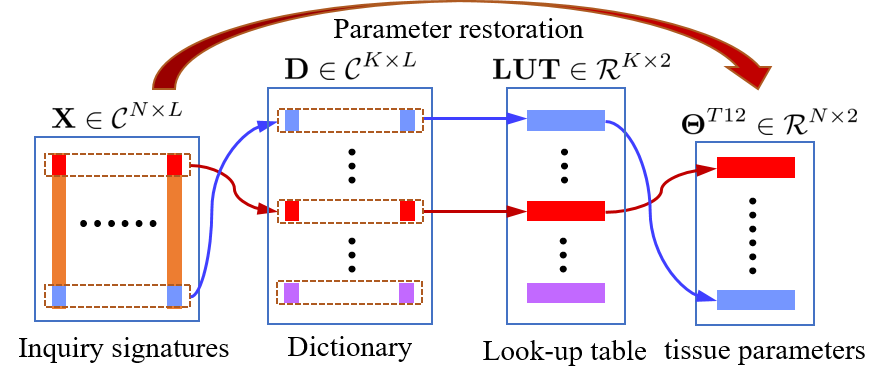}
	
	\vspace{-0.3cm}
	
	\caption{
		Parameter restoration using dictionary matching. Given an inquiry temporal signature, dictionary matching computes its inner product with each dictionary entry, and selects the most correlated one with the highest inner product as the best matching signature. Once the best matching entry is found, it directly leads to multiple tissue parameters, such as T1, T2, simultaneously, via searching a lookup-table.
		\label{Fig:DictMatching}
	}
	
\end{figure}

Given RF sequence parameters $\boldsymbol{\Theta}^{TRE}$ and k-space measurements $\bY$, the goal of MRF reconstruction is to estimate the tissue parameters $\boldsymbol{\Theta}^{T12}$. Typically, the image stack $\bX$ is first reconstructed from $\bY$, referred to as signature restoration, and then mapped to tissue parameters $\boldsymbol{\Theta}^{T12}$ via dictionary matching, referred to as parameter restoration~\cite{ma2013magnetic,jiang2015mr,davies2014compressed,wang2016magnetic,mazor2016low,mazor2018low}. This process is illustrated in Fig.~\ref{Fig:DictMatching}.

The dictionary is a collection of temporal signatures that are usually simulated by the Bloch equations for various typical tissues, given the pseudo-random RF pulse sequences and tissue parameters. Given an inquiry temporal signature, dictionary matching computes the inner product between the temporal signature with each dictionary entry, selecting the entry in the dictionary exhibiting the highest correlation with the inquiry one as the best matching signature. Once the best entry is found, it directly leads to multiple tissue parameters, such as T1, T2, simultaneously, via searching a lookup-table.

Let $\mathbf{LUT} \in \mathcal{R}^{K \times 2} $ denote a lookup-table composed of $K$ tissues, each containing 2 parameters, i.e., T1 and T2 relaxation times%
\footnote{Note that the off resonance parameter, which appeared in the original MRF paper~\cite{ma2013magnetic}, has been omitted here, since the sequence used in our experiments is derived from the FISP sequence, which is insensitive to off resonance effects~\cite{jiang2015mr,mazor2018low}.}. Let $\mathbf{D} \in \mathcal{C}^{K \times L} $ denote the corresponding dictionary simulated using Bloch equations given the RF sequence parameters $\boldsymbol{\Theta}^{TRE}$, formulated as $\mathbf{D}_{k,:} = f(\mathbf{LUT}_{k,:}, \boldsymbol{\Theta}^{TRE})$. Since each temporal signature $\mathbf{D}_{k,:}$ is linked with the $k$-th tissue's parameters $\mathbf{LUT}_{k,:}$, the choice of a large dictionary size $K$ can in principle provide enough granularity to capture a range of possible tissue values.

In conclusion, existing MRF reconstruction approaches involve two stages: signature restoration and parameter restoration, that can be succinctly written as
\begin{equation}
\label{Eq:OverallFormulation}
\boldsymbol{\Theta}^{T12}_{j,:} = g(h(\bY)_{j,:} | \boldsymbol{\Theta}^{TRE} ), \; j \in [1, N], 
\end{equation}
where the function $\bX = h(\bY)$ represents the signature restoration operation such as sparsity or low-rank based de-aliasing and denoising methods, whereas $ g(\bX_{j,:} | \boldsymbol{\Theta}^{TRE} )$ denotes the parameter restoration operation, such as dictionary matching based methods~\cite{ma2013magnetic,jiang2015mr,davies2014compressed,wang2016magnetic,mazor2016low,mazor2018low}.

Our approach aims to perform signature restoration via low-rank based de-aliasing and parameter restoration via a neural network in order to achieve improved MRF reconstruction performance. We highlight that our method only requires a simulated dictionary during network training. Once the network is trained, the dictionary is not needed anymore. In addition, our approach also eliminates a simulated dictionary for signature restoration, which is a key difference from FLOR~\cite{mazor2016low,mazor2018low} during signature restoration.

\begin{algorithm}[t]
	\small
	\caption{Original MRF method~\cite{ma2013magnetic}}
	\label{Alg:MRF_DictMatching}
	\textbf{Input:} 
	
	A set of subsampled k-space images: $\bY$
	
	A pre-simulated dictionary: $\bD$
	
	An appropriate lookup-table: $\mathbf{LUT}$
	
	\textbf{Output:} 
	
	Magnetic parameter maps: $\widehat{T}_1$, $\widehat{T}_2$ %
	
	\textbf{Step 1. Restore signatures:}	

	$$\widehat{\bX}_{:,i} = F_u^H\{ \bY_{:,i} \}, \, \forall i$$
	
	\textbf{Step 2. Restore parameters for every $j$ via dictionary matching:}
	
	$$
	\widehat{k_j} = \arg \underset{k}{\max} 
	\frac{\textbf{Re} \left< \bD_{k,:}, \widehat{\bX}_{j,:} \right> }{\| \bD_{k,:} \|_2^2} 
	\; , \; \quad
	\widehat{T}_1^j,  \widehat{T}_2^j = \mathbf{LUT}(\widehat{k_j})
	$$
\end{algorithm}

\subsection{Previous Methods}
\label{ssec:RelatedWork}

\vspace{-0.5cm}

\mypar{Dictionary Matching based MRF approaches}
The original MRF reconstruction algorithm~\cite{ma2013magnetic} is based on dictionary matching, as presented in Algorithm~\ref{Alg:MRF_DictMatching}. It finds the best matching dictionary entry for the acquired temporal signature according to their inner product and then searches the lookup-table to obtain corresponding tissue parameters. Here, $F_u^H\{\cdot\}$ denotes the inverse Fourier transform operating on the zero filled k-space data where zeros are filled at the unknown frequencies and symbol $\textbf{Re} \left< \mathbf{a}, \mathbf{b} \right>$ represents the real part of the inner product of two vectors $\mathbf{a}$ and $\mathbf{b}$.

Exploiting the nature of signals, by using appropriate prior knowledge, can often contribute to improved signal processing performance. In this spirit, later works suggested to incorporate sparsity in MRF reconstruction to further improve performance, inspired by successful applications of sparsity in MRI reconstruction~\cite{lustig2008compressed,weizman2015compressed,weizman2016reference}. Davies et al.~\cite{davies2014compressed} proposed BLoch response recovery via Iterative Projection (BLIP) which exploits sparsity in the dictionary domain. BLIP consists of iterating between two main steps: (a) a gradient step which enforces consistency with the measurements, based on the Projected Landweber Algorithm (PLA) generalized from the iterative hard thresholding method; (b) a projection which matches each row of $\bX$ to a single dictionary atom. Instead of exploiting sparsity in the dictionary domain, Wang et al.~\cite{wang2016magnetic} suggested to leverage sparsity in the wavelet domain of each imaging frame, $\bX_{:,i}$. They further replaced the Euclidean norm with the Mahalanobis distance for dictionary matching. Considering that adjacent MR image frames along the temporal dimension should exhibit high resemblance, Mazor et al.~\cite{mazor2016low,mazor2018low} proposed a magnetic resonance Fingerprint with LOw-Rank prior for reconstructing the image stack and quantitative parameters, referred to as FLOR, which achieved state-of-the-art performance. The algorithm, described in Algorithm~\ref{Alg:FLOR}, relies on two priors: a low rank prior on the the matrix $\bX$, and the fact that the rows of $\bX$ lie in the column space of the dictionary $\bD$.

\begin{algorithm}[thb] 
	\small
	\caption{FLOR~\cite{mazor2018low}}
	\label{Alg:FLOR}
	\textbf{Input:} 
	
	A set of subsampled k-space images: $\bY$;
	A pre-simulated dictionary: $\bD$;
	An appropriate lookup-table: $\mathbf{LUT}$;
	Parameters $\mu$ for gradient step and $\lambda$ for regularization
	
	\textbf{Output:}
	
	Magnetic parameter maps: $\widehat{T}_1$, $\widehat{T}_2$ 
	
	\textbf{Initialization:}
	
	$\widehat{\bX}^0=0$, $\mathbf{P}=\bD^\dagger \bD$, where $\bD^\dagger$ is the pseudo-inverse of $\bD$.
	
	\textbf{Step 1. Restore signatures via iterating until convergence:}
	
	\begin{itemize}
		\item
		Gradient step for every $i$:
		\begin{equation} \label{Eq:GradientStep}
		\widehat{\bZ}_{:,i}^{t+1} = \widehat{\bX}_{:,i}^t 
		- \mu F_u^H\{F_u\{\widehat{\bX}_{:,i}^t\} - \bY_{:,i}\}
		\end{equation}
		where the superscript $^t$ represents the index of iterations.
		\item
		Project onto the dictionary subspace:
		\begin{equation} \label{Eq:ProjectOnDict}
		[\bU, \bS, \bV] = \mathrm{svd}(\widehat{\bZ}^{t+1} \mathbf{P})
		\end{equation}
		where $\mathrm{svd}$ denotes the singular-value decomposition operation, and $\bS = diag(\{\sigma_j\})$ is a rectangular diagonal matrix with singular values $\{\sigma_j\}$ on its diagonal.  
		\item
		Soft-threshold the non-zero singular values with  $\lambda \mu$ and reconstruct signatures $\widehat{\bX}^{t+1}$:
		\begin{equation} \label{Eq:Soft-threshold}
		\begin{split}
		\sigma_j' = \max \{ \sigma_j - \lambda \mu, 0 \}
		\; , \; \quad
		\widehat{\bX}^{t+1} = \bU \bS' \bV^H
		\end{split}
		\end{equation}
		where $\bS' = diag(\{\sigma_j'\})$.
	\end{itemize}
	
	\textbf{Step 2. Restore parameters for every $j$ via dictionary matching:}
	
	$$
	\widehat{k_j} = \arg \underset{k}{\max} 
	\frac{\textbf{Re} \left< \bD_{k,:}, \widehat{\bX}_{j,:} \right> }{\| \bD_{k,:} \|_2^2} 
	\; , \; \quad
	 \widehat{T}_1^j,  \widehat{T}_2^j = \mathbf{LUT}(\widehat{k_j})
	$$
\end{algorithm}

\mypar{Learning-based MRF approaches}
The above techniques all use dictionary matching to perform mapping from temporal signatures to tissue parameters. Therefore, these methods suffer from drawbacks such as discretization error, slow inference speed and memory-consuming storage. In order to alleviate these issues, recent works~\cite{cohen2018mr,hoppe2017deep} propose to exploit neural networks to replace dictionaries and lookup-tables used in conventional MRF reconstruction approaches. Cohen et al. suggest a fully-connected feed-forward neural network (FNN)~\cite{cohen2018mr}. Since the input layer of the FNN is fully connected with the input temporal signature, the number of neurons in the input layer corresponds to the length of the input temporal signature. This makes the network structure less flexible, as a FNN network trained on temporal signatures with a certain length is not applicable to temporal signatures with a different length. In addition, the fully-connected structure results in rapid increase in the number of parameters along with the growth of depth, making the network more susceptible to overfitting. Hoppe et al.~\cite{hoppe2017deep} propose a 3-layer vanilla CNN for parameter restoration. Both \cite{cohen2018mr} and \cite{hoppe2017deep} focus exclusively on learning the signature-to-parameter mapping from a pair of dictionary and lookup-table simulated using the Bloch equations. During the validation, they assume that clean temporal signatures are available as input into the trained networks. However, since temporal signatures obtained from k-space subsampled MRF data are always contaminated by aliasing and noise, their approaches, when applied directly in such k-space subsampling situations, suffer from heavy artifacts introduced during the signature restoration phase, leading to poor performance.

\vspace{-0.2cm}

\subsection{Proposed Methods}
\label{ssec:ProposedMethods}

\vspace{-0.2cm}

\begin{figure*}[th]
	\centering
	\includegraphics[width = 17cm]{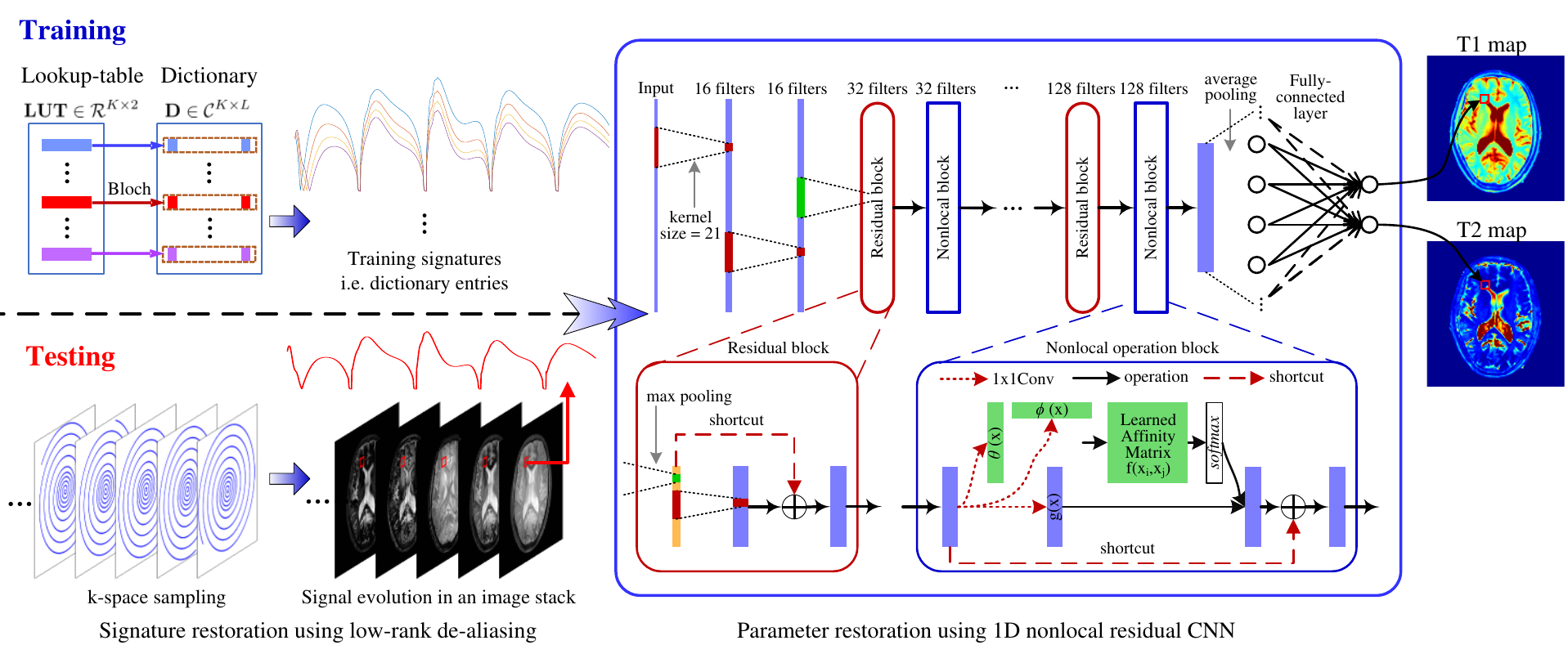}
	
	\vspace{-0.3cm}
	
	\caption{
		Diagram of the proposed MRF reconstruction approach. During the training stage, synthesized dictionary entries are used as training signatures to train the designed 1D nonlocal residual CNN until the outputs approximate parameter values in $\mathbf{LUT}$ well. In this way, the network captures the signature-to-parameter mapping. During the testing stage, a low-rank based algorithm is used to restore the image stack, a matrix containing signatures in rows, from k-space measurements. Then the restored signatures are input into the trained network to obtain corresponding tissue parameters directly. 
		\label{Fig:ResNet-CNN}
	}
\end{figure*}

The proposed hybrid deep magnetic resonance fingerprinting (HYDRA) approach, summarized in Algorithm \ref{Alg:LowRank_ResCNN}, consists of two stages: signature restoration and parameter restoration, (see also \eqref{Eq:OverallFormulation}). As illustrated in Fig.~\ref{Fig:ResNet-CNN}, a low-rank based de-aliasing method is used to restore signatures, and then a 1D nonlocal residual convolutional neural network is used to map each restored signature to corresponding tissue parameters.

In particular, given $\boldsymbol{\Theta}^{TRE}$ and k-space samples $\bY$, in our proposed approach, the function $\bX = h(\bY)$ in~\eqref{Eq:OverallFormulation} represents a signature restoration operation using low-rank based de-aliasing techniques without requiring a dictionary. The function $\boldsymbol{\Theta}^{T12}_{j,:} = g(\bX_{j,:} | \boldsymbol{\Theta}^{TRE} )$ in~\eqref{Eq:OverallFormulation} represents a parameter restoration operation that exploits a trained neural network to map each restored signature $\bX_{j,:}$ to corresponding tissue parameters $\boldsymbol{\Theta}^{T12}_{j,:}$ directly. In the subsequent sections, we provide a detailed description of both stages of our technique.

\vspace{-0.2cm}

\subsubsection{Low-rank based signature restoration}
\label{sssec:Low-rankSignature}

\vspace{-0.2cm}

Since MRF data consists of multiple frames exhibiting temporal similarity, the imaging contrasts matrix $\bX$ is typically a low-rank matrix~\cite{mazor2018low}. Therefore, $h(\cdot)$ leverages a low-rank prior for denoising and de-aliasing, formulated as 
\begin{equation}
\label{Eq:LowRank_Constraint}
\begin{array}{rl}
h(\bY) = 
\arg \underset{\bX}{\min}
&
\frac{1}{2} \sum_i \|\bY_{:,i} - F_u\{ \bX_{:,i} \} \|_2^2
\\
\text{s.t.}
& 
\textrm{rank}(\bX) < r
\end{array}
\end{equation}
where the parameter $r$ is the rank of the matrix, a fixed pre-chosen parameter. Since typically $r$ is not known in advance, we consider a relaxed regularized version:

\vspace{-0.5cm}

\begin{equation}
\label{Eq:LowRank_NuclearRegularization}
h(\bY) = \arg \underset{\bX}{\min}
\frac{1}{2} \sum\nolimits_i \|\bY_{:,i} - F_u\{ \bX_{:,i} \} \|_2^2
+ \lambda \, \|\bX \|_*
\end{equation}

\vspace{-0.2cm}

\noindent where $\|\bX \|_*$ denotes the nuclear norm~\cite{cai2010singular} of $\bX$, defined as the sum of the singular values of $\bX$, and $\lambda$ is the Lagrangian multiplier manually selected for balancing data fidelity and the rank. Problem~\eqref{Eq:LowRank_NuclearRegularization} can be solved using the incremental subgradient proximal method~\cite{sra2012optimization}, similar as to FLOR~\cite{mazor2018low}. The procedure for solving \eqref{Eq:OverallFormulation} is shown in Algorithm \ref{Alg:LowRank_ResCNN}.

One of differences from FLOR~\cite{mazor2018low} is the fact that we removed the operation of projecting the temporal signal onto a dictionary. This allows to eliminate the requirement for a simulated dictionary in the signature restoration stage, which also alleviates the memory consumption issue. In addition, the computational complexity is reduced by $ N \cdot L^2$ floating-point operations in each iteration, where $L$ is the dimension of a dictionary element, and $N$ is the number of pixels in each image frame. On the other hand, the gained benefits are at the price of requiring more iterations to converge. Another difference from FLOR~\cite{mazor2018low} is that we exploit a network, instead of dictionary matching, for signature-to-parameter mapping.

\begin{algorithm}[t]
	\small 
	\caption{Proposed MRF reconstruction approach: HYDRA}
	\label{Alg:LowRank_ResCNN}
	
	\textbf{Input:} 
	
	A set of subsampled k-space images: $\bY$
	
	The trained network: $g$
	
	Parameters $\mu$ for gradient step and $\lambda$ for regularization

	\textbf{Output:} 
	Magnetic parameter maps $\widehat{T}_1$, $\widehat{T}_2$ %
	
	\textbf{Initialization:}  $\widehat{\bX}^0 = 0$
	
	\textbf{Step 1. Restore signatures via iterating until convergence:} 
	
	\begin{itemize}
		\item
		Gradient step for every $i$, the same as \eqref{Eq:GradientStep}.
		\item
		Perform SVD:
		\begin{equation*}
		[\bU, \bS, \bV] = \mathrm{svd}(\widehat{\bZ}^{t+1})
		\end{equation*}
		\item
		Soft-threshold the non-zero singular values $\{\sigma_j\}$ of $\bS$ with parameter $\lambda \mu$ and reconstruct signatures $\widehat{\bX}^{t+1}$, the same as \eqref{Eq:Soft-threshold}.		
	\end{itemize}
	
	\textbf{Step 2. Restore parameters for every $j$ via the trained network:}
	
	$$ \widehat{T}_1^j, \widehat{T}_2^j = g(\widehat{\bX}_{j,:})$$
	
\end{algorithm}

\subsubsection{Learning-based parameter restoration}
\label{sssec:ParamRestore}

\vspace{-0.2cm}

Once the imaging contrasts matrix $\bX$ is recovered from the k-space samples $\bY$, each temporal signature $\bX_{j,:}$ is input into the trained network for parameter restoration, formulated as: 
\begin{equation}
\label{Eq:mapping}
\boldsymbol{\Theta}^{T12}_{j,:} = g(\bX_{j,:}  | \boldsymbol{\Theta}^{TRE}) , \; j \in [1, N]
\end{equation}
where $g(\cdot)$ denotes the trained network, $\boldsymbol{\Theta}^{TRE}$ denotes the fixed RF sequence parameters. We next describe the network structure, training and testing procedures.

\mypar{Network structure}

The proposed network has a 1D nonlocal residual CNN architecture with short-cuts for residual learning and nonlocal operations for achieving a self-attention mechanism. As illustrated in Fig.~\ref{Fig:ResNet-CNN}, it starts with two 1D convolutional layers before connecting with 4 residual / non-local operation blocks, and finally ends with a global-average-pooling layer followed by a fully-connected layer. Every residual block is followed by a non-local operation block. Four such blocks are interspersed with each other.

Each residual block contains a max-pooling layer with stride 2, two convolution layers and a shortcut that enforces the network to learn the residual content. The filter size is set to be equal to 21 throughout convolutional layers. The number of channels, a.k.a feature maps, in the first two convolutional layers is set to 16 and then is doubled in subsequent four residual blocks until 128 in the final residual block. The size of feature maps in the next block halves in contrast with the previous one due to max-pooling. In this way, we gradually reduce temporal resolution while extract more features to increase content information.

Inspired by the self-attention scheme and nonlocal neural networks~\cite{vaswani2017attention,wang2018non}, non-local operations are incorporated into the designed network to achieve the attention mechanism, in order to capture long-range dependencies with fewer layers. In contrast to the progressive behavior of convolutional operations that process one local neighborhood at a time, the non-local operations compute the response at a position as a weighted sum of the features at all positions in the feature maps. Formally, the nonlocal operation is formulated as~\cite{wang2018non}:
\begin{equation}
	\by_i = \frac{1}{C(\bx)} \sum_{\forall j} f(\bx_i, \bx_j) g(\bx_j) \,.
	\label{Eq:NonlocalOperation}
\end{equation} 
Here, $i$ is the index of an output position and $j$ is the index that enumerates all possible temporal positions, $\bx$ is the input temporal signal or its features and $\by$ is the output signal of the same size as $\bx$. A pairwise function $f$ computes a scalar between $i$ and all $j$ to represent the affinity relationship of these two positions. The unary function $g$ computes a representation of the input signal at position $j$. The response is normalized by a factor $C(\bx)$. There exists a few instantiations for function $f$ and $g$. For simplicity, the unary function $g$ is chosen as a linear embedding: $g(\bx_j) = \bW_g (\bx_j)$, where $\bW_g$ is a weight matrix to be learned. Regarding the affinity matrix $f$, we adopt the embedded Gaussian to compute similarity in an embedding space, which is formulated as:
$f(\bx_i, \bx_j) = e^{{\theta(\bx_i)}^\top \phi(\bx_j)}.$ Here, $\theta(\bx_i) = \bW_\theta \bx_i$ and $\phi(\bx_j) = \bW_\phi \bx_j$ are two learned embeddings. The normalization factor is set as $C(\bx) = \sum_{\forall j} f(\bx_i, \bx_j)$. For a given $i$, $\frac{1}{C(\bx)} f(\bx_i, \bx_j)$ becomes the softmax computation along the dimension $j$, which leads to $\by = \textit{softmax}(\bx^\top \bW_\theta^\top \bW_\phi \bx) g(\bx)$, which is the self-attention form~\cite{vaswani2017attention}.

The non-local behavior in \eqref{Eq:NonlocalOperation} is due to the fact that all positions $(j)$ are considered in the operation. As a comparison, a convolutional operation sums up the weighted input in a local neighborhood~\cite{wang2018non}. It implies that the non-local operation directly captures long-range dependencies in the temporal dimension via computing interactions between any two points, regardless of their positional distance. In this way, the network is able to extract global features and take advantage of the full receptive field in each layer. 

The global-average-pooling layer is used to average each feature map in order to integrate information in every channel for improved robustness to corrupted input data. This global-average-pooling layer also reduces the number of parameters significantly, thus lessening the computation cost as well as preventing over-fitting. The last fully-connected layer outputs estimated parameters -- T1 and T2 relaxation times. The designed network contains around 0.27 million parameters. The weights are initialized using He-normal-distribution~\cite{he2015delving}. The max-norm kernel constraint~\cite{srivastava2014dropout} is exploited to regularize the weight matrix directly in order to prevent over-fitting. The designed network can also be adapted for various MRF sequences, such as the original MRF sequence -- inversion-recovery balanced steady state free-precession (IR-bSSFP) sequence, that depends also on the intrinsic df parameter. It is possible to adjust the number of outputs to adapt to more parameters, such as proton density, B0. 

To summarize, our network is motivated and inspired by recent successful applications of convolutional neural networks and variants. Convolutional neural networks have been proved to be a powerful model to capture useful features from signals and images. By introducing convolution, local receptive field and weight sharing design, a CNN is capable of taking advantage of local spatial coherence and translation invariance characteristics in the input signal, thus become especially well suited to extract relevant information at a low computational cost~\cite{krizhevsky2012imagenet,he2016deep,he2016identity,dong2016image,kim2016deeply,gehring2017convolutional}. On the other hand, the residual network architecture~\cite{he2016deep,he2016identity} provides an effective way to design and train a deeper model, since it alleviates the gradient vanishing or exploding problems by propagating gradients throughout the model via short-cuts, a.k.a skip connections. By leveraging non-local operation based attention mechanism, neural networks are endowed with capability of extracting global features and capturing long-range dependencies.

\vspace{-0.2cm}

\mypar{Network training}

The designed network is trained on a synthesized dictionary $\mathbf{D}$ and corresponding lookup-table $\mathbf{LUT}$ to learn the signature-to-parameter mappings $\mathbf{LUT}_{k,:} = g(\mathbf{D}_{k,:} | \boldsymbol{\Theta}^{TRE})$. 

The training dataset is synthesized as follows. First, we determine the range of tissue parameters. For example, one may set T1 relaxation times to cover a range of [1, 5000] ms and T2 relaxation times to cover a range of [1, 2000] ms with an increment of 10 ms for both. Thus, the T1 and T2 values constitute a grid with dimension 500 $\times$ 200, in which each point represents a specific combination of T1 and T2 values, and hence characterizes a specific tissue. Points corresponding to T1 < T2 have been excluded as such combinations have no physical meaning. All the valid points are stacked together to generate a lookup-table. For instance, the above setting for T1 and T2 leads to a lookup-table of dimension $80100 \times 2$. The RF pulse sequences used in our work are fast imaging with steady state precession (FISP) pulse sequences with parameters that have been used in previous publications in the field of MRF~\cite{mazor2018low,jiang2015mr,cao2017robust}. Given the lookup-table and RF pulse sequences, dictionary entries can be synthesized by solving the Bloch equations using the extended phase graph formalism~\cite{hennig1991echoes,weigel2015extended}.

When the training dataset is ready, the dictionary entries are used as input signals and corresponding lookup-table entries serve as the groundtruth. All the dictionary entries are input into the designed network batch by batch which outputs estimated parameters. The root mean square errors (RMSE) of the outputs are calculated with respect to corresponding groundtruth. The resulting RMSE loss is then backpropagated from the output layer to the first layer to update the weights and bias by using Adam~\cite{kingma2014adam} as the optimization algorithm. More training details are provided in the subsequent experiment section. Once the training procedure is completed, given an inquiry signal evolution $\bX_{j,:}$, it is able to map such a time sequence directly to corresponding tissue parameters, as formulated in \eqref{Eq:mapping}, implying that no dictionary or lookup-table are required during the inference. Since we only need to store the trained network which is a compact model, it consumes less memory than storing the dictionary and lookup-table.

\begin{figure*}[tb]
	\begin{multicols}{2}  
		\centering
		\begin{minipage}[b]{0.48\linewidth}
			\centering
			\includegraphics[width = 3.5cm, height=2.7cm]{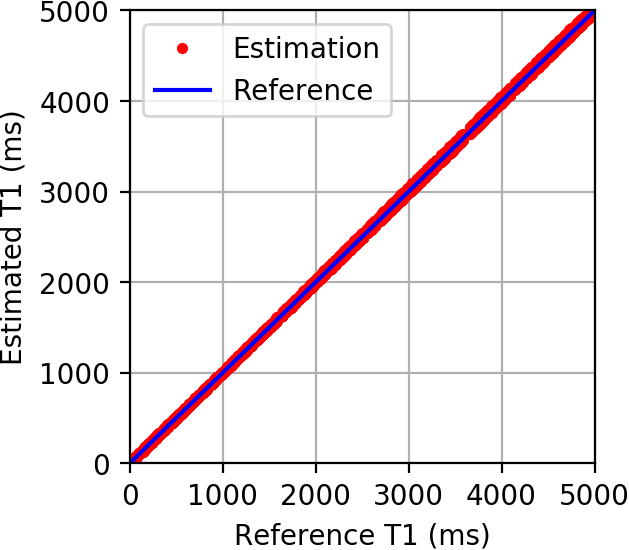}
		\end{minipage} 
		\begin{minipage}[b]{0.48\linewidth}
			\centering
			\includegraphics[width = 3.5cm, height=2.7cm]{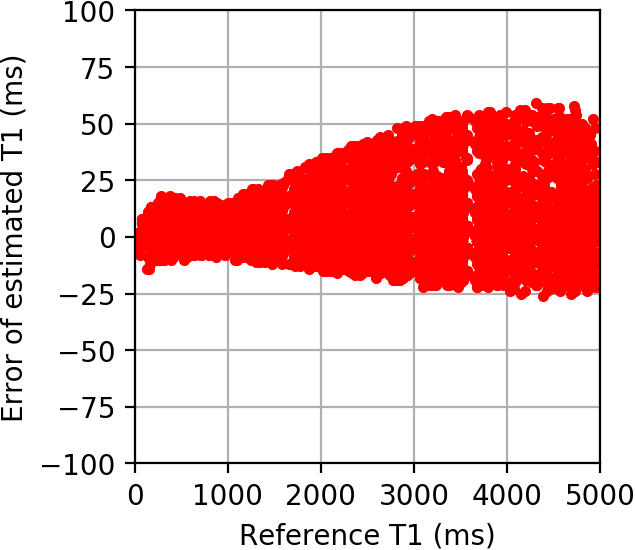}
		\end{minipage} 
		\\
		\vspace{+0.2cm}
		\begin{minipage}[b]{0.48\linewidth}
			\centering
			\includegraphics[width = 3.5cm, height=2.7cm]{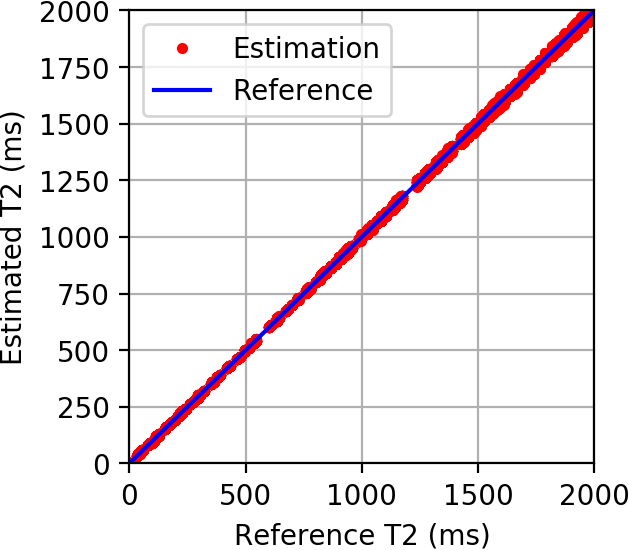}
		\end{minipage} 
		\begin{minipage}[b]{0.48\linewidth}
			\centering
			\includegraphics[width = 3.5cm, height=2.7cm]{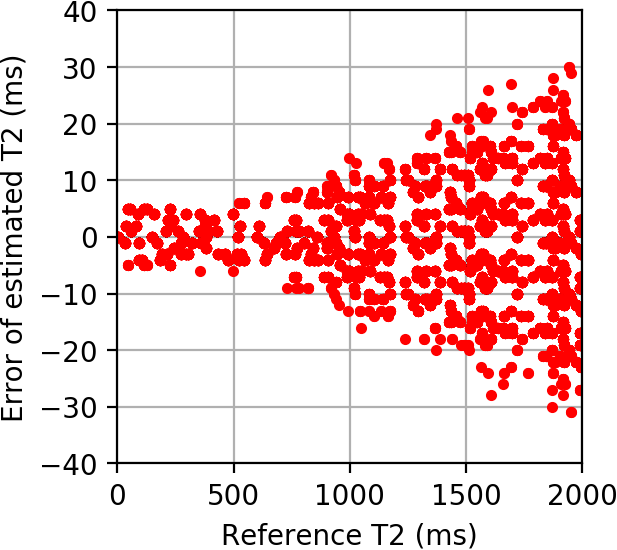}
		\end{minipage} 
		\\
		\begin{minipage}[b]{1\linewidth}
			\centering
			\scriptsize (a) T1, T2 estimations using dictionary matching. 
		\end{minipage}
		\\
		\vspace{+0.3cm}
		\begin{minipage}[b]{0.48\linewidth}
			\centering
			\includegraphics[width = 3.5cm, height=2.7cm]{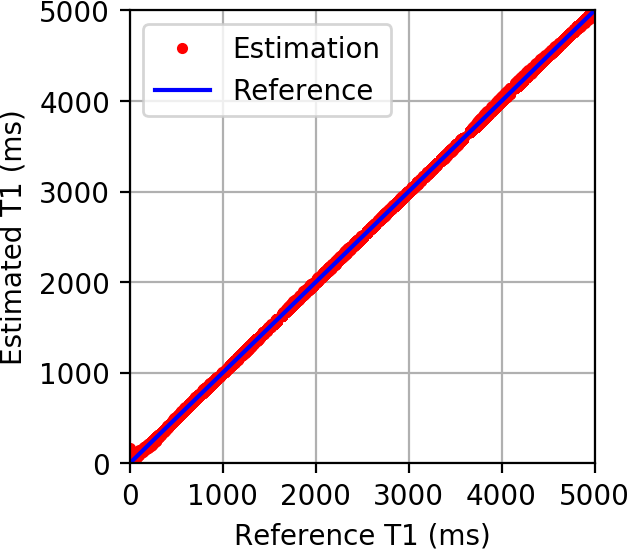}
		\end{minipage} 
		\begin{minipage}[b]{0.48\linewidth}
			\centering
			\includegraphics[width = 3.5cm, height=2.7cm]{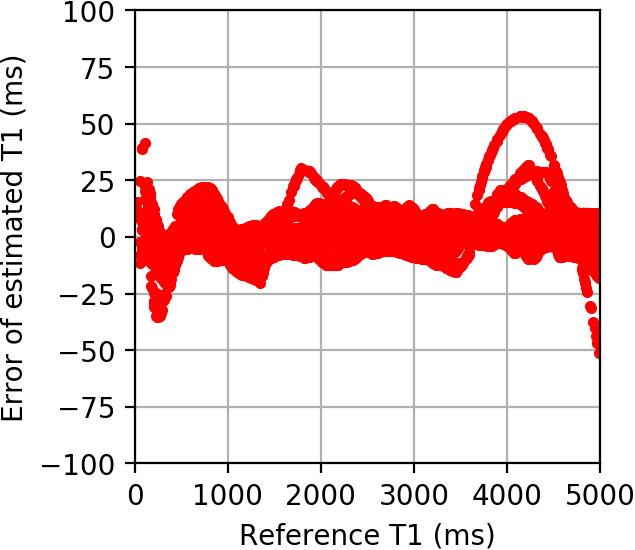}
		\end{minipage} 
		\\
		\vspace{+0.2cm}
		\begin{minipage}[b]{0.48\linewidth}
			\centering
			\includegraphics[width = 3.5cm, height=2.7cm]{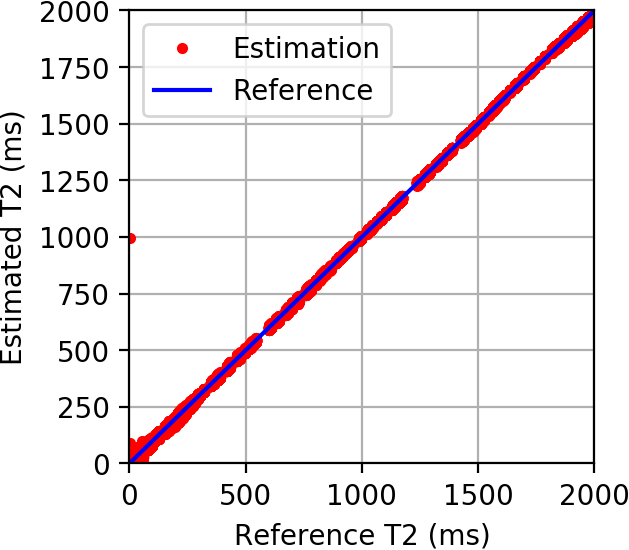}
		\end{minipage} 
		\begin{minipage}[b]{0.48\linewidth}
			\centering
			\includegraphics[width = 3.5cm, height=2.7cm]{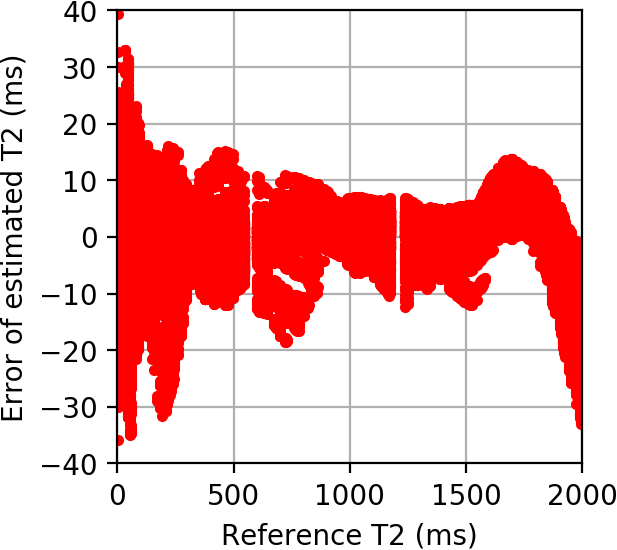}
		\end{minipage} 
		\\
		\begin{minipage}[b]{1\linewidth}
			\centering
			\footnotesize (b) T1, T2 estimations using CNN~\cite{hoppe2017deep}. 
		\end{minipage}
		\\
		\vspace{+0.3cm}
		\begin{minipage}[b]{0.48\linewidth}
			\centering
			\includegraphics[width = 3.5cm, height=2.7cm]{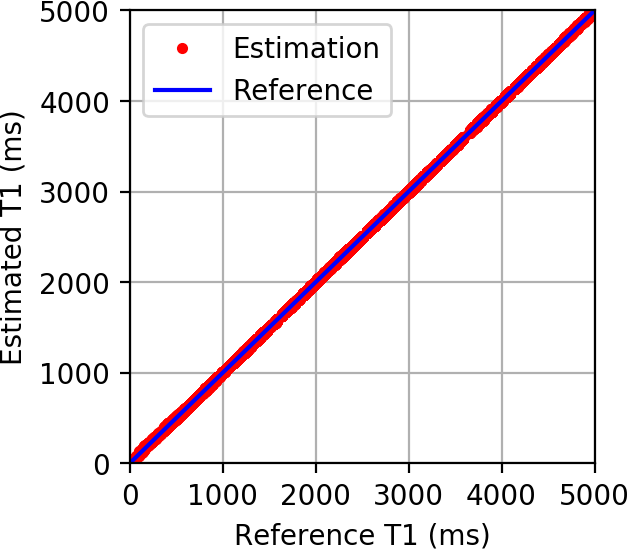}
		\end{minipage} 
		\begin{minipage}[b]{0.48\linewidth}
			\centering
			\includegraphics[width = 3.5cm, height=2.7cm]{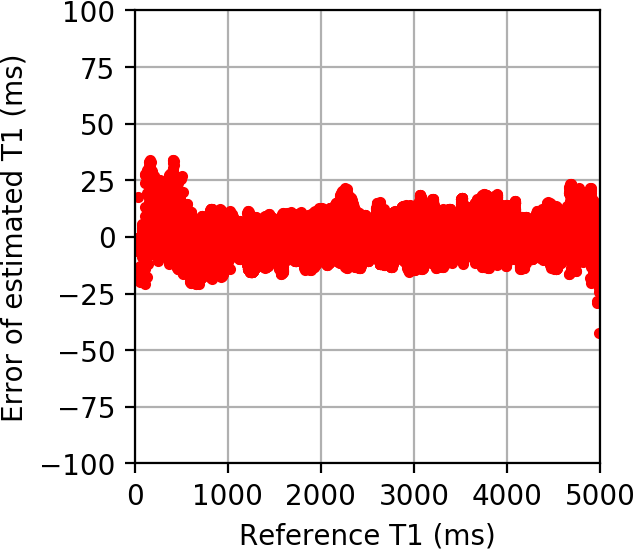}
		\end{minipage} 
		\\
		\vspace{+0.2cm}
		\begin{minipage}[b]{0.48\linewidth}
			\centering
			\includegraphics[width = 3.5cm, height=2.7cm]{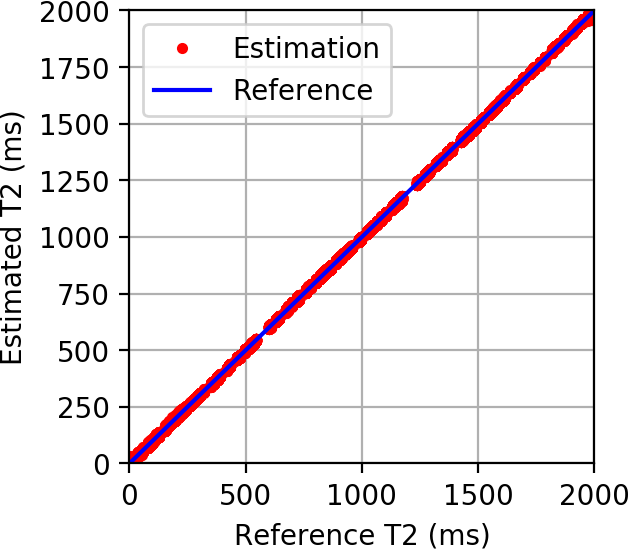}
		\end{minipage} 
		\begin{minipage}[b]{0.48\linewidth}
			\centering
			\includegraphics[width = 3.5cm, height=2.7cm]{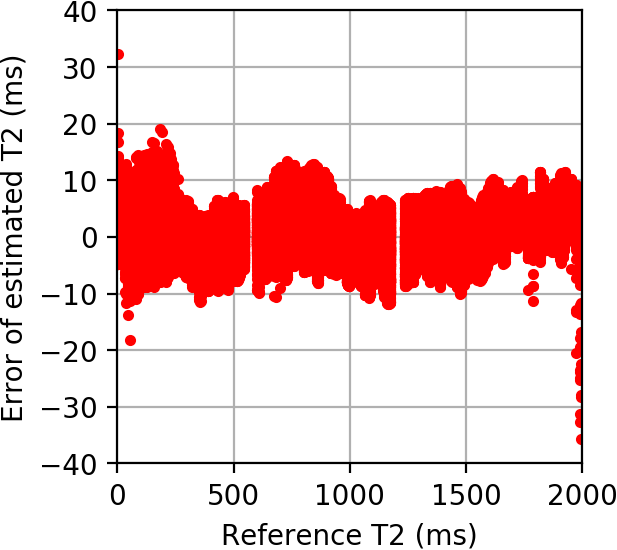}
		\end{minipage}  
		\\
		\begin{minipage}[b]{1\linewidth}
			\centering
			\footnotesize (c) T1, T2 estimations using FNN~\cite{cohen2018mr}. 
		\end{minipage}
		\\
		\vspace{+0.3cm}
		\begin{minipage}[b]{0.48\linewidth}
			\centering
			\includegraphics[width = 3.5cm, height=2.7cm]{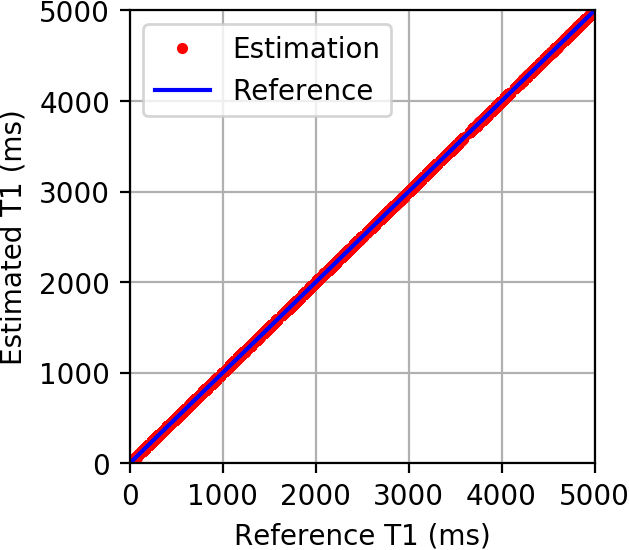}
		\end{minipage} 
		\begin{minipage}[b]{0.48\linewidth}
			\centering
			\includegraphics[width = 3.5cm, height=2.7cm]{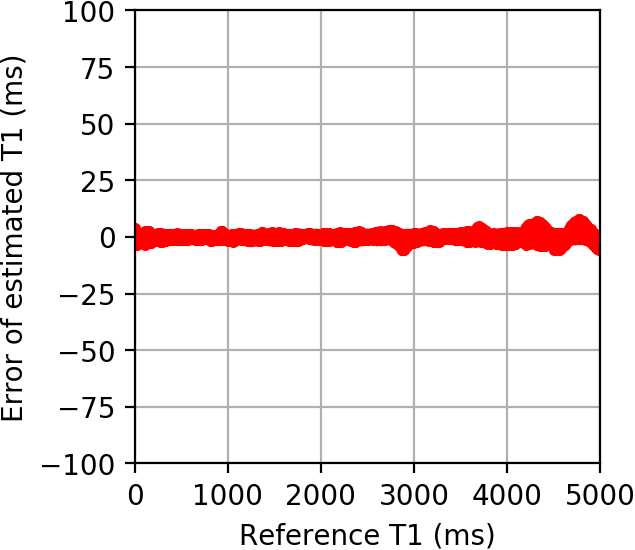}
		\end{minipage} 
		\\
		\vspace{+0.2cm}
		\begin{minipage}[b]{0.48\linewidth}
			\centering
			\includegraphics[width = 3.5cm, height=2.7cm]{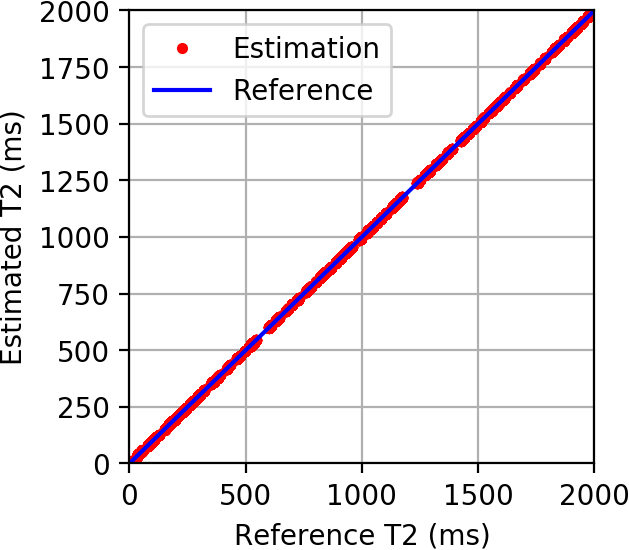}
		\end{minipage} 
		\begin{minipage}[b]{0.48\linewidth}
			\centering
			\includegraphics[width = 3.5cm, height=2.7cm]{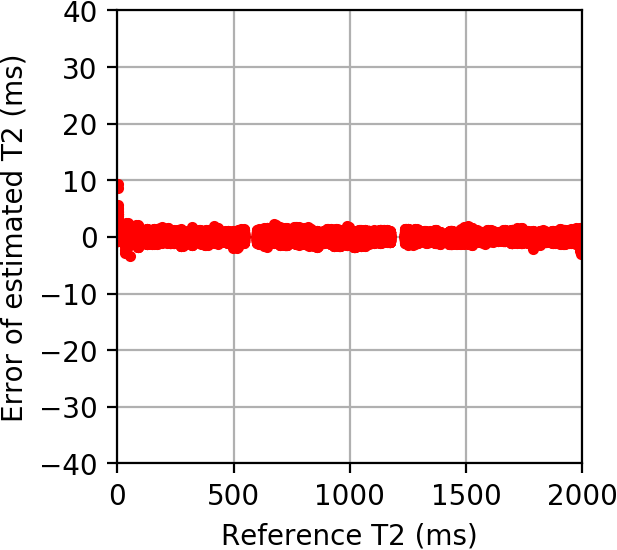}
		\end{minipage} 
		\\
		\begin{minipage}[b]{1\linewidth}
			\centering
			\footnotesize (d) T1, T2 estimations using HYDRA. 
		\end{minipage}
		\\
		\vspace{+0.3cm}
		
	\end{multicols}
	
	\vspace{-0.6cm}
	
	\caption{
		\footnotesize
		Testing on the synthetic dataset for comparing parameter restoration performance. Subfig. (a) - (d) show the results using dictionary matching~\cite{ma2013magnetic,jiang2015mr,davies2014compressed,wang2016magnetic,mazor2016low,mazor2018low}, FNN~\cite{cohen2018mr}, CNN~\cite{hoppe2017deep} and HYDRA. In each subfigure, the left figure compares the estimated T1 or T2 values (marked with red dot) with groundtruth values (marked with blue line), and the right figure shows the deviations of the estimation from the groundtruth. Parameter mapping performance of HYDRA is much better than competing methods,  in the entire value range of T1 and T2 parameters, resulting in smaller deviations. 
	}
	\label{Fig:ResNet-CNN_1Dtest}
\end{figure*}

We emphasize that even though the network is trained on a grid of tissue values, it is expected to capture the mapping function from temporal signatures to tissue parameters. Thus the trained network is capable of outputting tissue values not existing in the grid of training values. Detailed results can be found in Fig.~\ref{Fig:ResNet-CNN_1Dtest} and Table~\ref{Tab:TestingSynthetic}. This feature is favorable, as it implies that well designed and trained networks have an ability to overcome discretization issues. The overall procedures for solving \eqref{Eq:LowRank_NuclearRegularization} and \eqref{Eq:mapping} are shown in Algorithm \ref{Alg:LowRank_ResCNN}. 

\section{Experimental Results}
\label{sec:Experiments}

\vspace{-0.3cm}

In this section, we conduct a series of experiments to evaluate our approach, comparing it with other state-of-the-art MRF methods~\cite{ma2013magnetic,davies2014compressed,mazor2018low,hoppe2017deep,cohen2018mr}.

\begin{figure}[t]
	\centering
	\begin{minipage}[b]{0.48\linewidth}
		\centering
		\includegraphics[width = 8.5cm, trim=1cm 0cm 0cm 0cm,clip ]{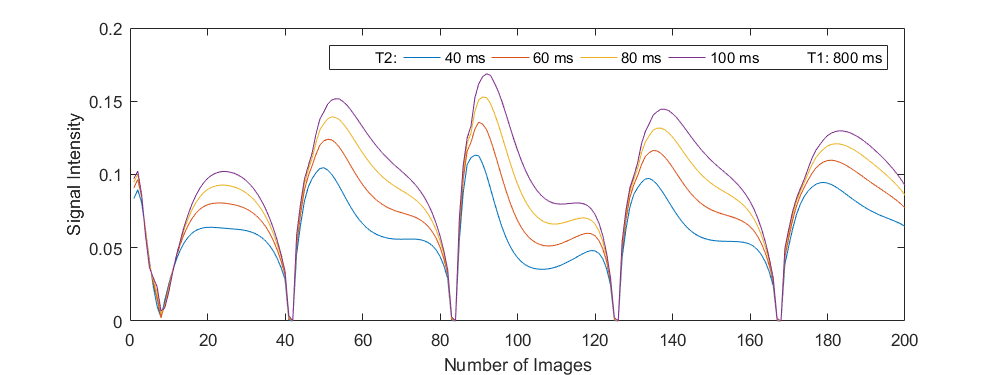} 
		\\
		\scriptsize (a) 
	\end{minipage} 
	\begin{minipage}[b]{0.48\linewidth}
		\centering
		\includegraphics[width = 8.5cm, trim=1cm 0cm 0cm 0cm,clip ]{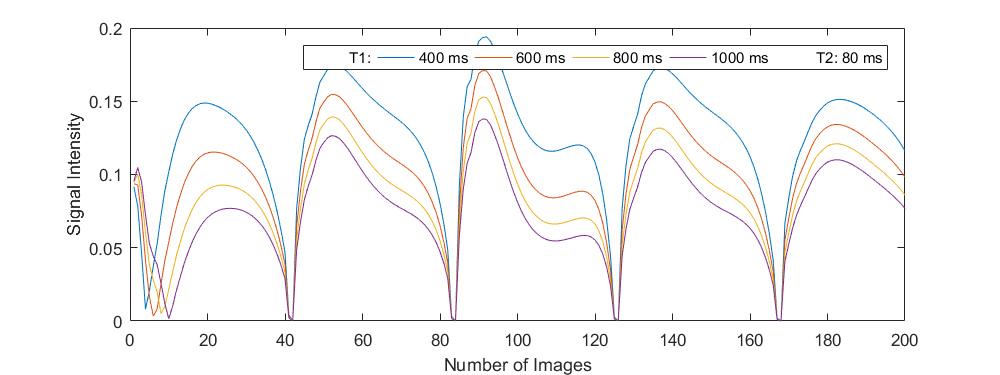} 
		\\
		\scriptsize (b) 
	\end{minipage} 
	
	\vspace{-0.3cm}
	
	\caption{
		\footnotesize
		Synthetic MRF temporal signatures with 200 time frames. (a) Temporal signatures corresponding to parameter values \{(T1, T2)\} ms = \{(800,40),(800,60),(800,80),(800,100)\} ms. (2) Temporal signatures corresponding to parameter values \{(T1, T2)\} ms = \{(400,80),(600,80),(800,80),(1000,80)\} ms.
	}
	\label{Fig:SynSignatures}
\end{figure}

The experiments are categorized into a few types: training, testing on synthetic data, testing on anatomical data using variable density Gaussian sampling patterns and spiral sampling patterns at different sampling ratios and number of time frames, as described in Table~\ref{Tab:ExperimentsTypes}. For the network training, synthesized temporal signatures, i.e. simulated dictionary entries of $\mathbf{D}$ shown as Fig.~\ref{Fig:SynSignatures}, are used as input signals and corresponding parameter values in the lookup-table $\mathbf{LUT}$ serve as the groundtruth. The proposed network is trained to capture the signature-to-parameter mappings. For testing on synthetic data, synthesized temporal signatures in $\bX$ are used as input signals and corresponding parameter values in $\boldsymbol{\Theta}^{T12}$ serve as the groundtruth. The aim is to test the parameter restoration performance only. For testing on anatomical data, the k-space measurements $\bY$ which are derived from the Fourier transform (for Gaussian patterns) or non-uniform FFT (for spiral trajectories)~\cite{fessler2003nonuniform} of $\bX$, are used as input and corresponding parameter values in $\boldsymbol{\Theta}^{T12}$ serve as reference. When there is no k-space subsampling, the aim is to test the parameter restoration performance only. When there exists k-space subsampling, the aim is to test the overall performance, including both signature restoration and parameter restoration. More detailed descriptions are provided in each subsection.

\vspace{-0.2cm}

\subsection{Training}
\label{ssec:Exp_Training}

\vspace{-0.2cm}

As mentioned in Section~\ref{sssec:ParamRestore}, the designed network is trained on a pair of synthesized dictionary $\mathbf{D}$ and lookup-table $\mathbf{LUT}$, simulated using Bloch equations and FISP pulse sequences~\cite{jiang2015mr,mazor2018low}.

\begin{figure}[t]
	\centering
	\begin{minipage}[b]{0.48\linewidth}
		\centering
		\includegraphics[width = 7cm
		]{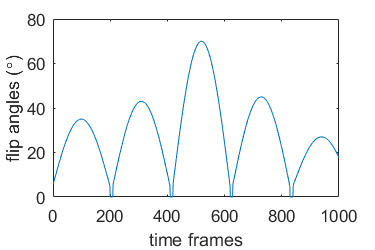} 
	\end{minipage} 
	\begin{minipage}[b]{0.48\linewidth}
		\centering
		\includegraphics[width = 7cm
		]{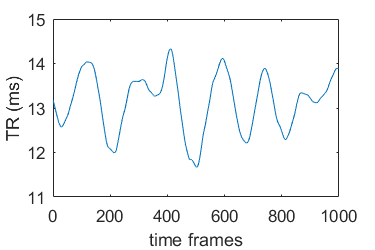} 
	\end{minipage} 
	
	\caption{
		FISP pulse sequence parameters. All the flip angles (FA) constituted a sinusoidal variation in the range of 0 - 70 degrees to ensure smoothly varying transient state of the magnetization. The repetition time (TR) was randomly varied in the range of 11.5 - 14.5 ms with a Perlin noise pattern.
	}
	\label{Fig:FA_TR}
\end{figure}

The FISP pulse sequence used in our experiments was designed with parameters $\boldsymbol{\Theta}^{TRE}_{:,i} = [TR^i, TE^i, FA^i]^T, i \in [1,L]$ that have been used in previous publications in the field of MRF~\cite{mazor2018low,jiang2015mr,cao2017robust}. The echo time $TE^i$ was constant of 2ms. The repetition time $TR^i$ was randomly varied in the range of 11.5 - 14.5 ms with a Perlin noise pattern. All the flip angles $FA^i, i \in [1,L]$ constituted a sinusoidal variation in the range of 0 - 70 degrees to ensure smoothly varying transient state of the magnetization, as shown in Figure~\ref{Fig:FA_TR}. 

For the range of tissue parameters, T1 relaxation times are set to cover a range of [1, 5000] ms and T2 relaxation times to cover a range of [1, 2000] ms with an increment of 10 ms for both. Such parameter ranges cover the relaxation time values that can be commonly found in a brain scan~\cite{vymazal1999t1}. All the valid combinations of T1 and T2 values are stacked together, generating a lookup-table $\mathbf{LUT}$ of dimension $K \times 2$ where $K=80100$. Given the lookup-table and RF pulse sequences, dictionary entries are synthesized by solving the Bloch equations using the extended phase graph formalism, leading to a dictionary of dimension $K \times L$ where $L = 200$ or $1000$ is the number of time frames.

When the training dataset is ready, the dictionary entries are used as input signals and corresponding lookup-table entries serve as the groundtruth to train the designed network, as mentioned in Section~\ref{sssec:ParamRestore}. The model was trained for 50 epochs. It takes around 30 seconds for running one epoch on average, thus around 25 minutes for completing 50 epochs, on a NVIDIA GeForce GTX 1080 Ti GPU. In each training epoch, 20\% of the training samples are separated aside for validation dataset. The learning rate decays from 1e-2 to 1e-6 every 10 epochs. Each batch was experimentally set to contain 256 time-sequences in order to balance the convergence rate and weights updating rate well. For comparison purposes, we also implemented Hoppe et al.'s CNN referring to~\cite{hoppe2017deep}, and Cohen et al.'s FNN referring to~\cite{cohen2018mr} with the same structure and parameters as specified in their papers. Then we use the same GPU and training dataset to train their networks with specified learning rate and number of epochs until convergence.

We adopt a few widely used metrics, such as root mean square error (RMSE), signal-to-noise ratio (SNR) and peak signal-to-noise ratio (PSNR) to evaluate the image quality quantitatively. The definitions of RMSE, SNR and PSNR are given as follows:
\begin{align} 
\textrm{RMSE} &= \sqrt{ 
	\frac{
		\| \mathbf{X} - \widehat{\mathbf{X}} \|_F^2
	}
	{ N } } \,, 
\label{Eq:Def_RMSE}
\\
\textrm{SNR} &= 20 \log_{10} \frac{\| \mathbf{X} \|_F^2}{\textrm{RMSE}} \,,
\label{Eq:Def_SNR}
\\
\textrm{PSNR} &= 20 \log_{10} \frac{\textrm{PeakVal}}{\textrm{RMSE}} \,,
\label{Eq:Def_PSNR}
\end{align}
where matrices $\mathbf{X}$ and $\widehat{\mathbf{X}}$ denote the ground truth signal and its reconstructed version, respectively, $N$ denotes the total number of elements in the signal and $\| \cdot \|_F$ denotes the Frobenius norm. PeakVal stands for the pixel peak value in an image, e.g., 1 for a normalized signal.

\subsection{Testing on synthetic dataset}
\label{ssec:Testingonsynthetic}

\vspace{-0.2cm}

In this subsection, we evaluate the performance of HYDRA on a synthetic testing dataset. The procedures of constructing a synthetic testing dataset is similar to the construction of the training dataset: 500 different T1 values are randomly selected from 1 - 5000 ms, while 200 different T2 values are randomly selected from 1 - 2000 ms, using random permutation based on uniformly distributed pseudorandom numbers. All the valid combinations from the selected T1 and T2 values are stacked together, generating a parameter matrix $\boldsymbol{\Theta}^{T12}$ of dimension $80000 \times 2$ with $N=80000$. The RF pulse sequences are the same as in the training stage. Given the parameter matrix and RF pulse sequences, input signal signatures are synthesized by solving the Bloch equations using the extended phase graph formalism, leading to a signature matrix $\bX$ of dimension $N \times L = 80000 \times 200$, with each row representing a temporal signature corresponding to a specific combination of T1 and T2 values. The signature matrix $\bX$ and parameter matrix $\boldsymbol{\Theta}^{T12}$ constitute the synthetic testing dataset, with $\bX$ as input and $\boldsymbol{\Theta}^{T12}$ as the groundtruth.

\begin{figure*}[t]
	\centering
	\begin{minipage}[b]{0.24\linewidth}
		\centering
		\includegraphics[width = 4cm]{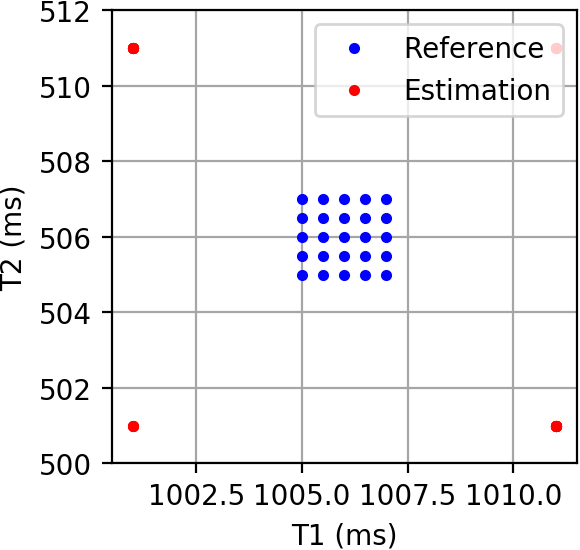}
		\scriptsize Dictionary Matching
	\end{minipage} 
	\begin{minipage}[b]{0.24\linewidth}
		\centering
		\includegraphics[width = 4cm]{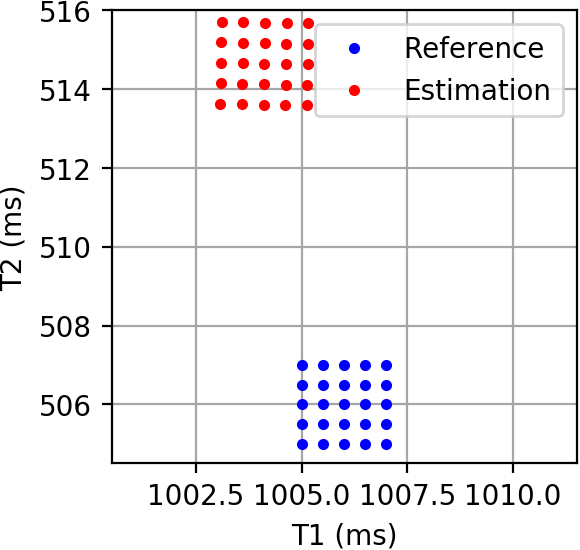}
		\scriptsize  CNN~\cite{hoppe2017deep}
	\end{minipage} 
	\begin{minipage}[b]{0.24\linewidth}
		\centering
		\includegraphics[width = 4cm]{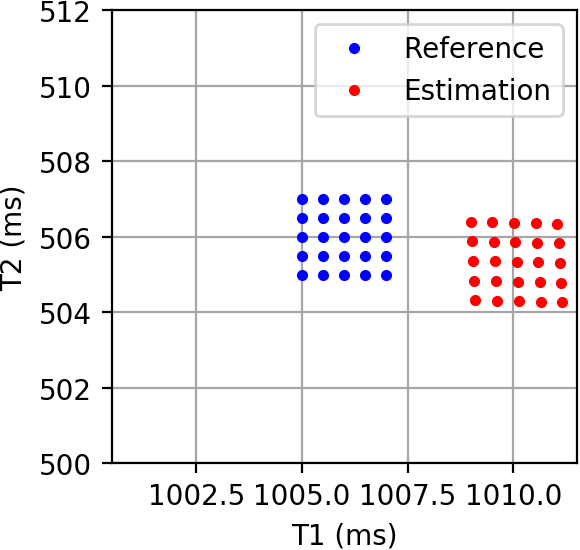}
		\scriptsize FNN~\cite{cohen2018mr}
	\end{minipage} 
	\begin{minipage}[b]{0.24\linewidth}
		\centering
		\includegraphics[width = 4cm]{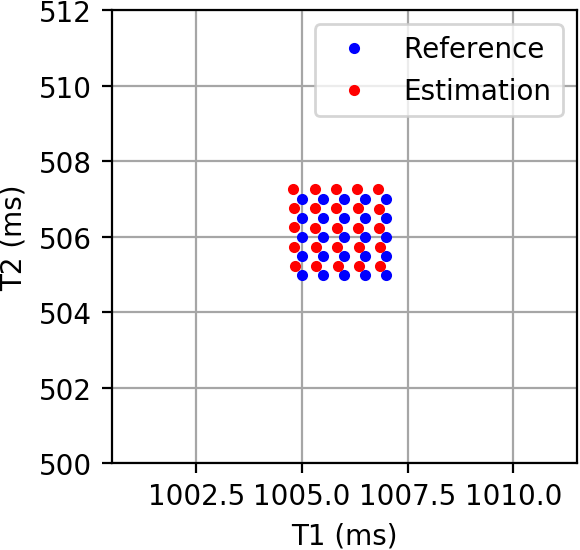}
		\scriptsize HYDRA
	\end{minipage} 
	
	\vspace{-0.3cm}
	
	\caption{
		Testing on synthetic dataset involving detailed T1 / T2 examples that are not on the training grid and their intervals are much smaller than the training grid intervals. Dictionary matching finds best adjacent values from the dictionary, i.e. 1001, 1011 for T1, and 501, 511 for T2. In contrast, owing to the captured mapping functions, neural networks output continuous values. Proposed HYDRA leads to the smallest deviations and bias.
	}
	\label{Fig:1DtestReal_Continuous}
\end{figure*}

We input the synthetic testing signatures $\bX$ into Hoppe et al.'s CNN~\cite{hoppe2017deep}, Cohen et al.'s FNN~\cite{cohen2018mr}, and the network of HYDRA to compare the outputs with groundtruth T1 and T2 values in $\boldsymbol{\Theta}^{T12}$. We also compare with dictionary matching methods~\cite{ma2013magnetic,jiang2015mr,davies2014compressed,wang2016magnetic,mazor2016low,mazor2018low} which exploit the same dictionary $\mathbf{D}$ and lookup-table $\mathbf{LUT}$ to find the best matching entry for each signature in $\bX$ and then estimate parameter values by searching the lookup-table. As shown in Table~\ref{Tab:TestingSynthetic}, Table~\ref{Tab:TestingSynthetic_Contin}, Fig.~\ref{Fig:1DtestReal_Continuous} and Fig.~\ref{Fig:ResNet-CNN_1Dtest}, the estimated parameter values using the proposed network obtained outstanding agreement with the groundtruth, yielding higher PSNR, SNR and smaller RMSE than the dictionary matching method~\cite{ma2013magnetic,jiang2015mr,davies2014compressed,wang2016magnetic,mazor2016low,mazor2018low}, as well as competing networks~\cite{hoppe2017deep,cohen2018mr}. 

In particular, to illustrate in detail how well neural networks tackle the discretization issue inherent to dictionary matching, we show the testing performance on continuous-valued T1, T2 parameters which have small intervals, e.g. 0.5ms, that is 20 times smaller than the training grid intervals 10ms, between neighboring values in Table~\ref{Tab:TestingSynthetic_Contin}. Since these values and their corresponding MRF signatures do not exist in the training dictionary and lookup-table%
\footnote{
	As mentioned in the experiment setting in Section~\ref{ssec:Exp_Training}, in the training dataset, T1 relaxation times are set to cover a range of [1, 5000] ms and T2 relaxation times to cover a range of [1, 2000] ms with an increment of 10 ms for both, that is, T1 values = \{1, 11, 21, $\cdots$, 4991\}, and T2 values = \{1, 11, 21, $\cdots$, 1991\}.
}, 
the dictionary matching methods report a T1 and T2 value -- the closest discretized value present in the dictionary -- that can be quite distinct from the groundtruth. In contrast, the various neural network approaches can potentially learn an underlying mapping from the temporal signatures to the respective T1 and T2 values, leading to estimates that are much closer to the groundtruth. Interestingly, our approach outperforms previous networks~\cite{hoppe2017deep,cohen2018mr} as shown in Table~\ref{Tab:TestingSynthetic_Contin} and Fig.~\ref{Fig:1DtestReal_Continuous}. Evidently, neural networks demonstrate much better robustness to discretization issues, leading to improved parameter restoration in comparison to dictionary based methods.

Another impressive advantage of HYDRA is the fast inference speed. HYDRA takes only 8.2 s to complete the mapping operation for eighty thousand temporal signatures, that is, 53$\times$ faster than dictionary matching. Furthermore, the inference speed of HYDRA is subject to the network topology. That is, once the network structure is fixed, the complexity is fixed. In contrast, the complexity of dictionary matching is limited by the dictionary density. This implies that our advantage will be more prominent in comparison with competing techniques using a dictionary with higher density.

\subsection{Testing on anatomical dataset}

\vspace{-0.3cm}

In this subsection, we evaluate our approach on an anatomical testing dataset. We construct the dataset from brain scans that were acquired with GE Signa 3T HDXT scanner from a healthy subject.\footnote{The experiment procedures involving human subjects described in this paper were approved by the Institutional Review Board of Tel-Aviv Sourasky Medical Center, Israel.} Since there are no groundtruth parameter values for the T1 and T2 parameter maps, we obtain gold standard data by acquiring Fast Imaging Employing Steady-state Acquisition (FIESTA) and Spoiled Gradient Recalled Acquisition in Steady State (SPGR) images, at 4 different flip angles (3$^\circ$,5$^\circ$,12$^\circ$ and 20$^\circ$), and implementing corrections~\cite{liberman2014t1} followed by DESPOT1 and DESPOT2~\cite{deoni2005high} algorithms. The constructed gold standard T1, T2 parameter maps have a dimension of $128 \times 128$ for each map, accordingly leading to a parameter matrix $\boldsymbol{\Theta}^{T12}$ of size $16384 \times 2$ by stacking vectorized T1, T2 maps together. Based on the parameter matrix $\boldsymbol{\Theta}^{T12}$ and pre-defined RF pulse sequences, we generate temporal signatures using Bloch equations, the same mechanism as generating the synthetic testing dataset, leading to a signature matrix $\bX$ of dimension $N \times L = 16384 \times 200$. The signature matrix $\bX$ and parameter matrix $\boldsymbol{\Theta}^{T12}$ constitute the anatomical testing dataset, with $\bX$ as input and $\boldsymbol{\Theta}^{T12}$ as the gold standard reference.

Note that, since the gold standard T1, T2 maps exhibit spatial structures in the image domain, the resulting signature matrix $\bX$ can be regarded as a stack of $L=200$ vectorized image frames, where each frame exhibits specific spatial structures. Therefore, it makes sense to perform Fourier transform and k-space subsampling for each column of $\bX$ to get k-space measurements $\bY$. This is the key difference between the anatomical dataset and the synthetic dataset.

We first explore the case with full k-space sampling in order to evaluate the parameter restoration performance of HYDRA. Then, we consider situations with k-space subsampling in order to evaluate both the signature restoration and the parameter restoration performance of HYDRA. 

\vspace{-0.2cm}

\subsubsection{Full k-space sampling}

\vspace{-0.2cm}

In the first case, the fully-sampled k-space measurements $\bY$, derived from the Fourier transform of $\bX$, are used as input to obtain the estimated $\boldsymbol{\Theta}^{T12}$. This is equivalent to inputting $\bX$ into the network of HYDRA, or performing dictionary matching based on $\bX$ directly, since the inverse Fourier transform of the fully-sampled measurements $\bY$ is exactly the same as $\bX$. The aim is to test the parameter restoration performance only. In the experiment, corresponding parameter values in $\boldsymbol{\Theta}^{T12}$ serve as the gold standard reference. For comparison, dictionary matching methods~\cite{ma2013magnetic,jiang2015mr,davies2014compressed,wang2016magnetic,mazor2016low,mazor2018low} exploit the same dictionary $\mathbf{D}$ and lookup-table $\mathbf{LUT}$ as in our training stage to find the best matching entry and estimate parameter values for each signature in $\bX$.

\begin{figure*}[tb]
	\begin{multicols}{2}  
		\centering
		\begin{minipage}[b]{0.48\linewidth}
			\centering
			\includegraphics[width = 4cm, height=3.2cm]{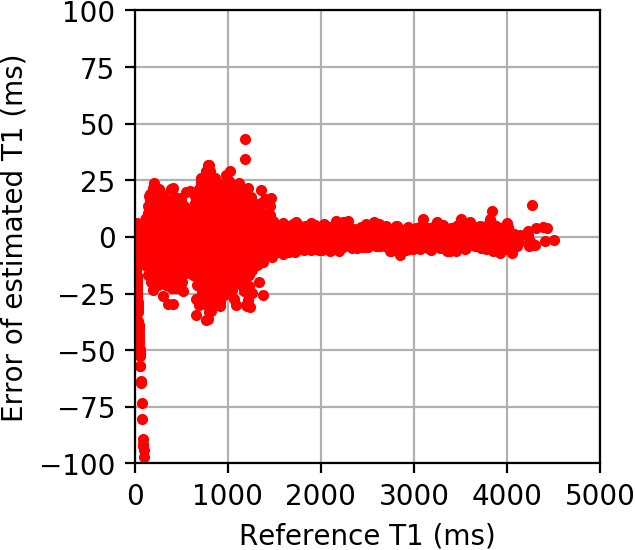}
		\end{minipage} 
		\begin{minipage}[b]{0.48\linewidth}
			\centering
			\includegraphics[width = 4cm, height=3.2cm]{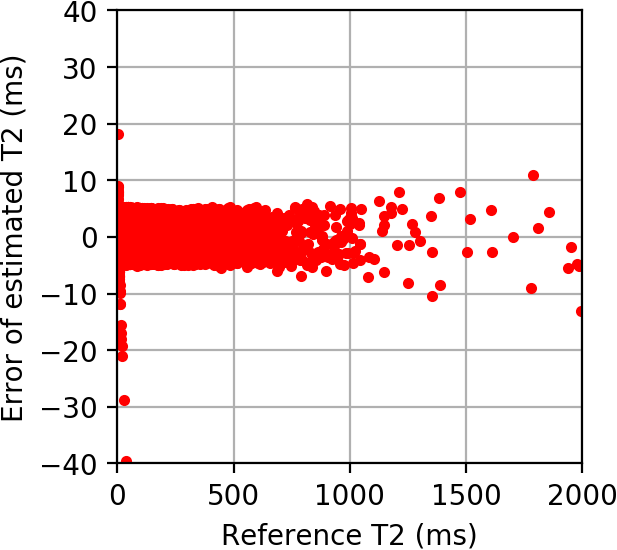}
		\end{minipage} 
		\\
		\begin{minipage}[b]{1\linewidth}
			\centering
			\scriptsize (a) T1, T2 estimations using dictionary matching. 
		\end{minipage}
		\\
		\vspace{+0.3cm}
		\begin{minipage}[b]{0.48\linewidth}
			\centering
			\includegraphics[width = 4cm, height=3.2cm]{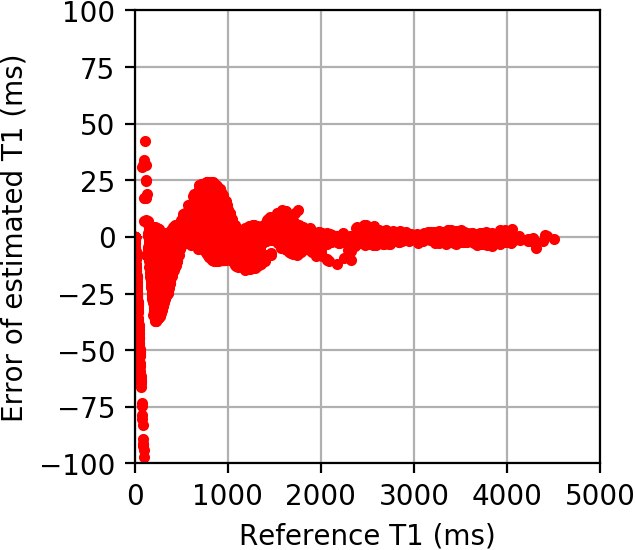}
		\end{minipage} 
		\begin{minipage}[b]{0.48\linewidth}
			\centering
			\includegraphics[width = 4cm, height=3.2cm]{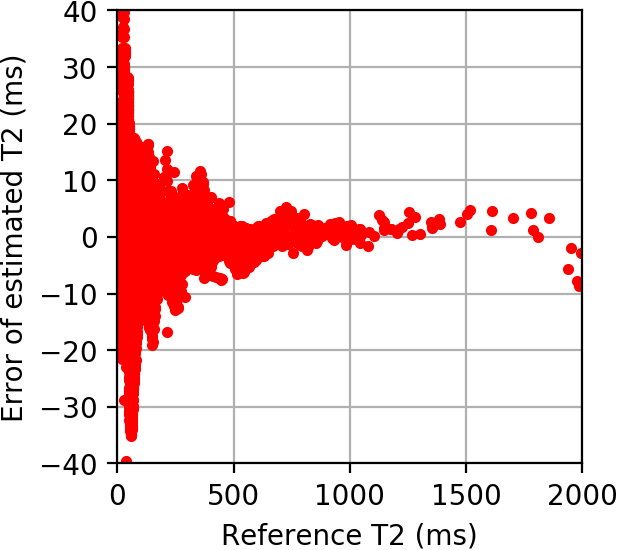}
		\end{minipage} 
		\\
		\begin{minipage}[b]{1\linewidth}
			\centering
			\footnotesize (b) T1, T2 estimations using CNN~\cite{hoppe2017deep}.
		\end{minipage}
		\\
		\vspace{+0.3cm}
		\begin{minipage}[b]{0.48\linewidth}
			\centering
			\includegraphics[width = 4cm, height=3.2cm]{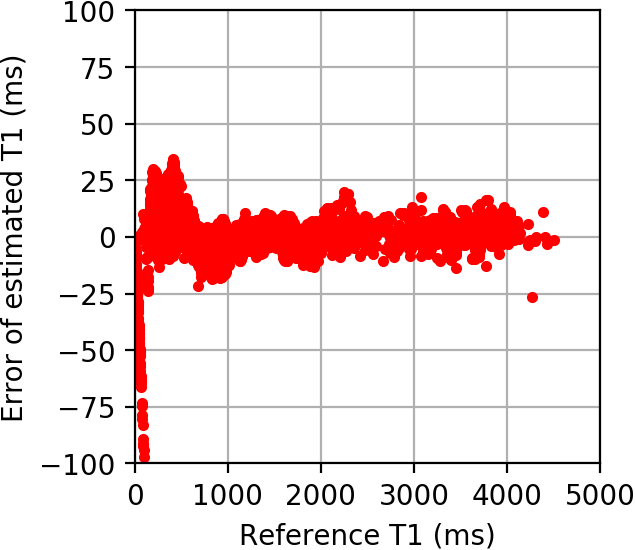}
		\end{minipage} 
		\begin{minipage}[b]{0.48\linewidth}
			\centering
			\includegraphics[width = 4cm, height=3.2cm]{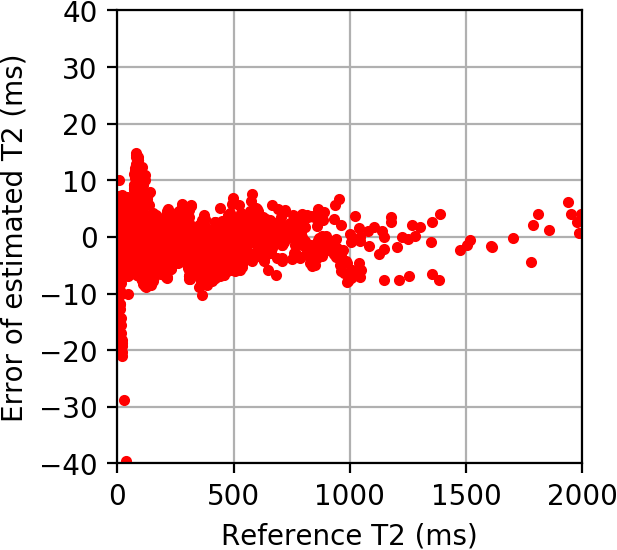}
		\end{minipage}  
		\\
		\begin{minipage}[b]{1\linewidth}
			\centering
			\footnotesize (c) T1, T2 estimations using FNN~\cite{cohen2018mr}. 
		\end{minipage}
		\\
		\vspace{+0.3cm}
		\begin{minipage}[b]{0.48\linewidth}
			\centering
			\includegraphics[width = 4cm, height=3.2cm]{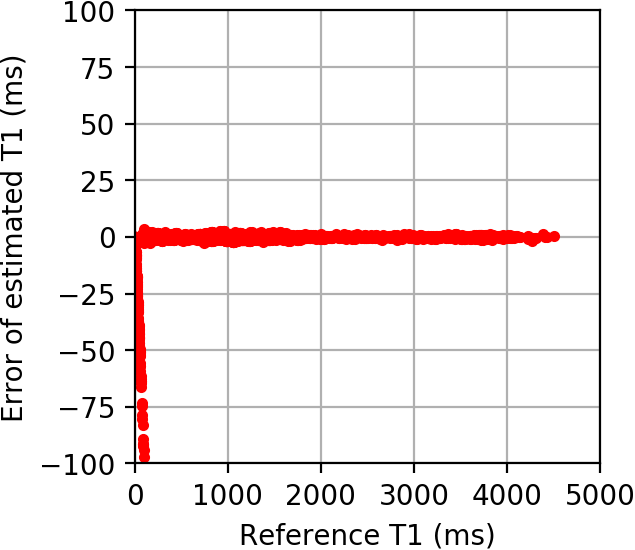}
		\end{minipage}  
		\begin{minipage}[b]{0.48\linewidth}
			\centering
			\includegraphics[width = 4cm, height=3.2cm]{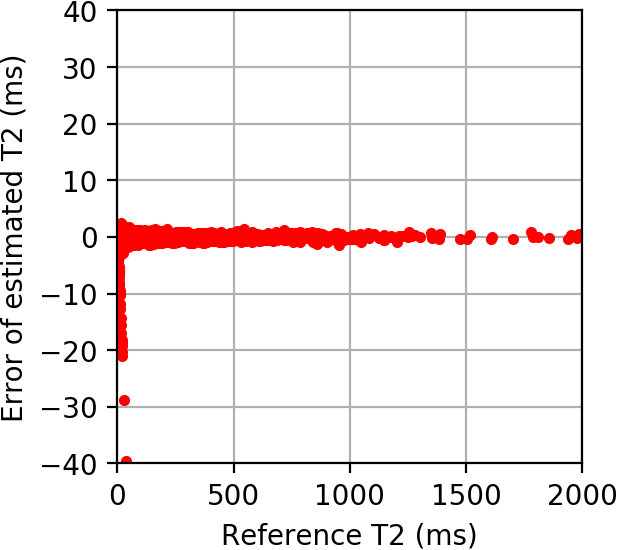}
		\end{minipage} 
		\\
		\begin{minipage}[b]{1\linewidth}
			\centering
			\footnotesize (d) T1, T2 estimations using HYDRA.
		\end{minipage}
		\\
		\vspace{+0.3cm}
		
	\end{multicols}
	
	\vspace{-0.6cm}
	
	\caption{
		\footnotesize
		Testing on the anatomical dataset with full k-space sampling for comparing parameter restoration performance. Subfig. (a) - (d) show the results using dictionary matching~\cite{ma2013magnetic,jiang2015mr,davies2014compressed,wang2016magnetic,mazor2016low,mazor2018low}, FNN~\cite{cohen2018mr}, CNN~\cite{hoppe2017deep} and HYDRA. Each subfigure shows the deviations of the estimation from the reference. Parameter mapping performance using HYDRA outperforms competing methods significantly, resulting in smaller deviations. The performance is also verified by quantitative metrics, as shown in Table~\ref{Tab:TestingPhantom}. 
	}
	\label{Fig:ResNet-CNN_1DtestReal}
\end{figure*}

\begin{figure*}[t]
	\centering
	\begin{minipage}[b]{0.19\linewidth}
		\raggedright 
		\includegraphics[width=3.6cm, height=3cm,trim=0cm 0cm 0cm 0.4cm,clip]{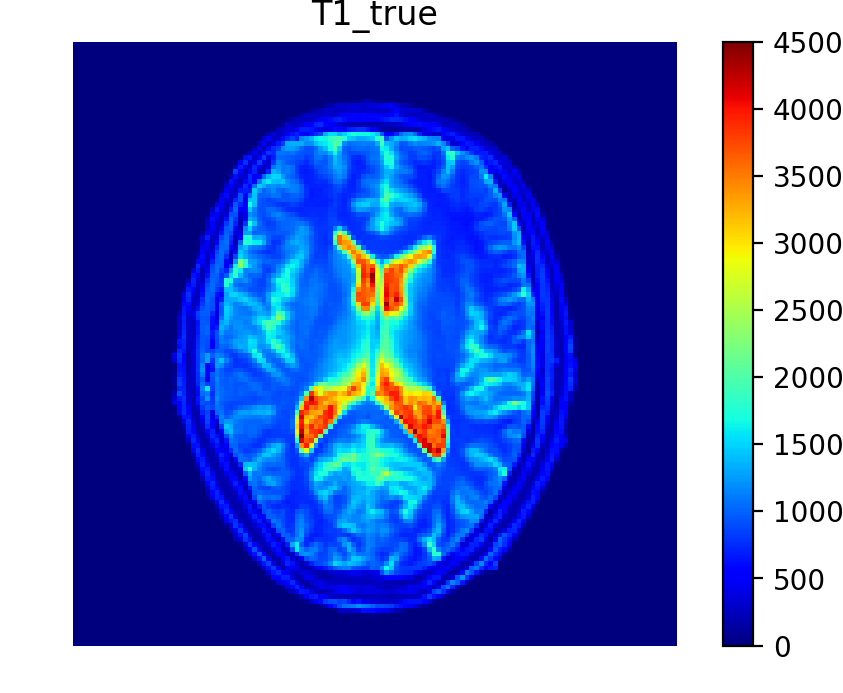}
	\end{minipage} 
	\begin{minipage}[b]{0.19\linewidth}
		\raggedright 
		\includegraphics[width=3.6cm, height=3cm,trim=0cm 0cm 0cm 0cm,clip]{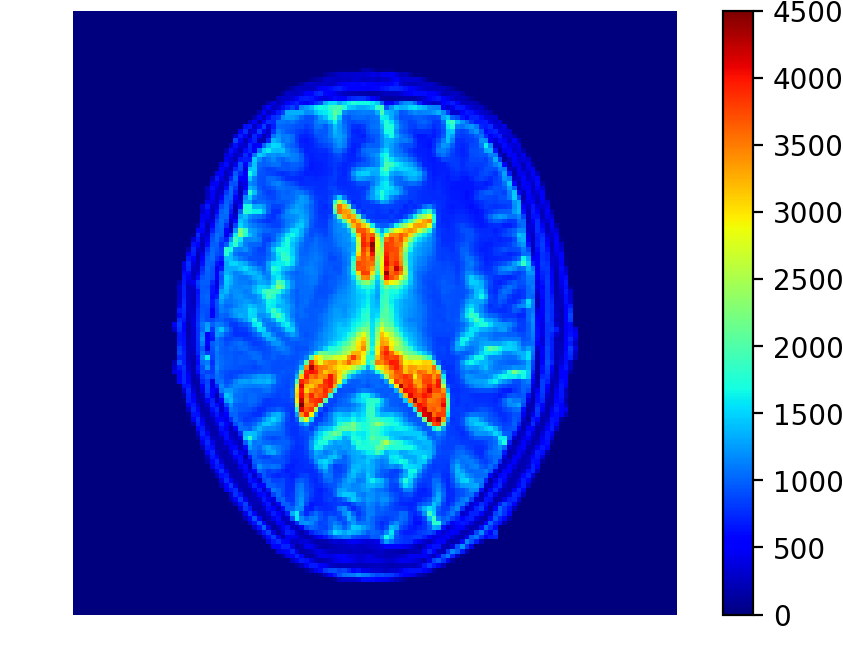}
	\end{minipage} 
	\begin{minipage}[b]{0.19\linewidth}
		\raggedright 
		\includegraphics[width=3.6cm, height=3cm,trim=0cm 0cm 0cm 0cm,clip]{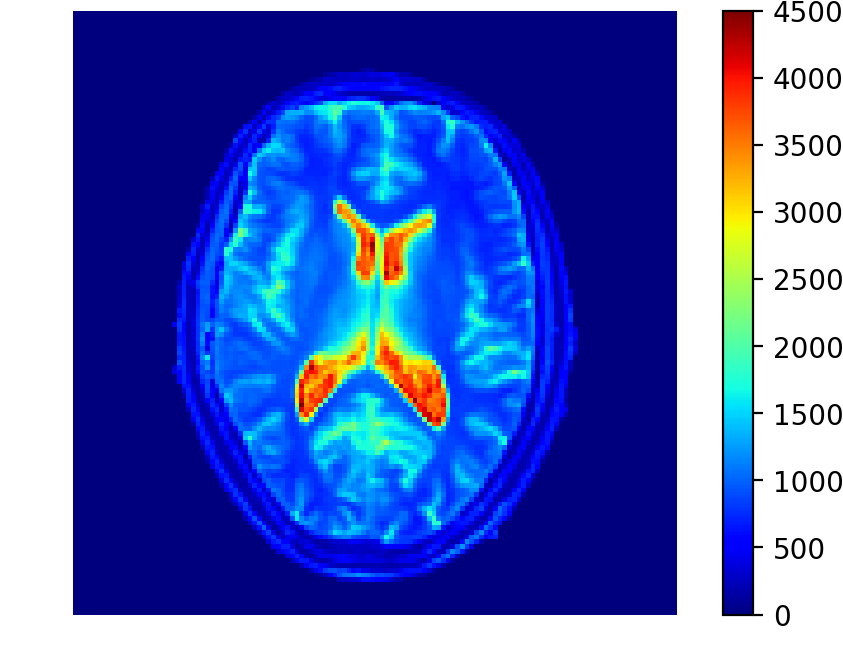}
	\end{minipage} 
	\begin{minipage}[b]{0.19\linewidth}
		\raggedright 
		\includegraphics[width=3.6cm, height=3cm,trim=0cm 0cm 0cm 0cm,clip]{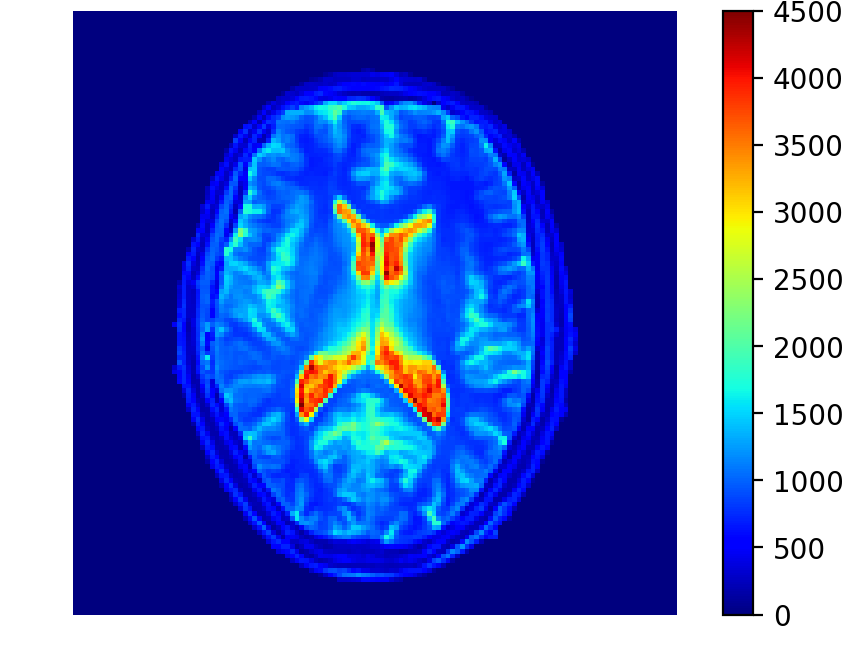}
	\end{minipage} 
	\begin{minipage}[b]{0.19\linewidth}
		\raggedright 
		\includegraphics[width=3.6cm, height=3cm,trim=0cm 0cm 0cm 0cm,clip]{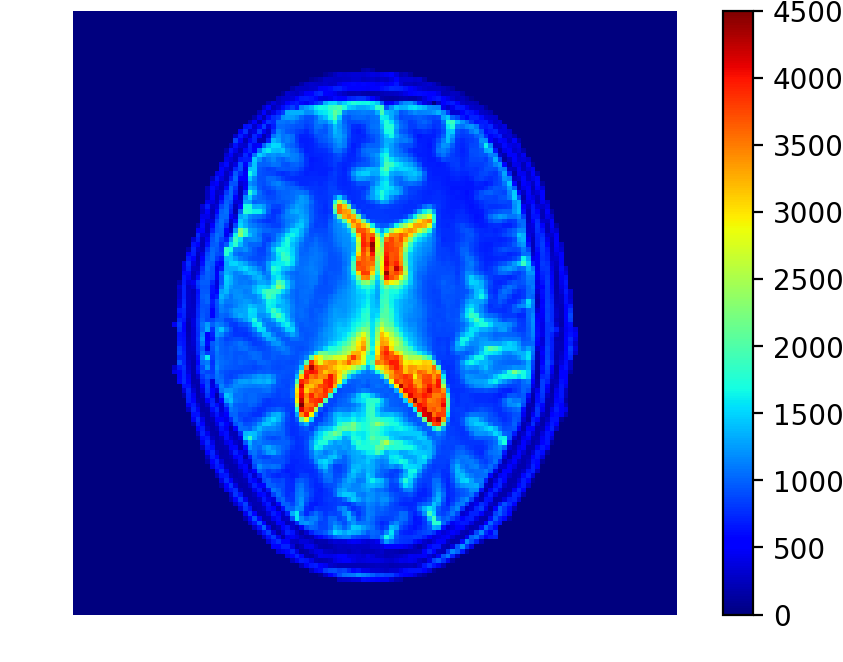}
	\end{minipage} 
	\\
	\begin{minipage}[b]{0.2\linewidth}
		\raggedright 
		\phantom{
			\includegraphics[width=3cm, height=3cm,trim=0cm 0cm 0cm 0.4cm,clip]{./Figures/DeepMRF_ResNet_CNN/T1_true.png}
		}
	\end{minipage} 
	\begin{minipage}[b]{0.19\linewidth}
		\raggedright 
		\includegraphics[width=3.5cm, height=3cm,trim=0cm 0cm 0cm 0cm,clip]{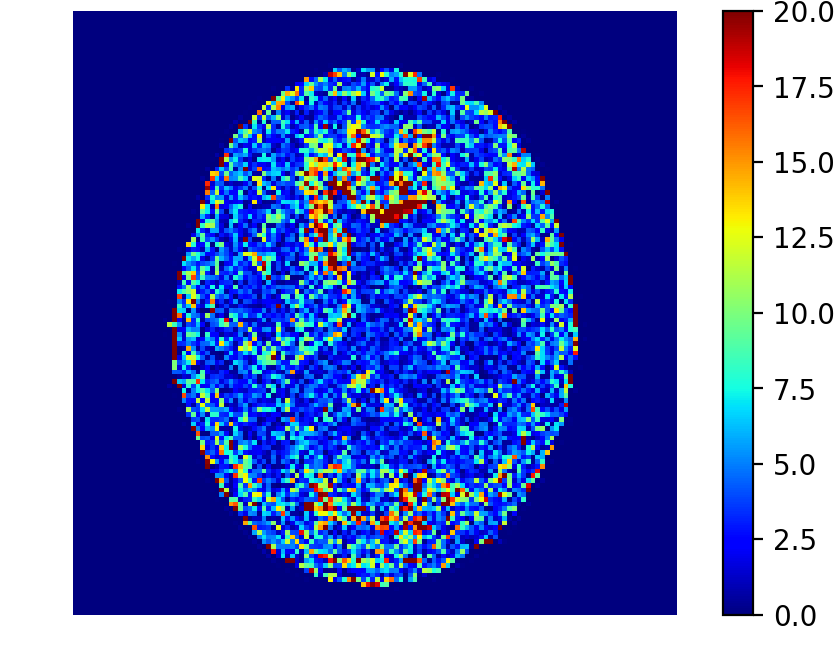}
	\end{minipage} 
	\begin{minipage}[b]{0.19\linewidth}
		\raggedright 
		\includegraphics[width=3.5cm, height=3cm,trim=0cm 0cm 0cm 0cm,clip]{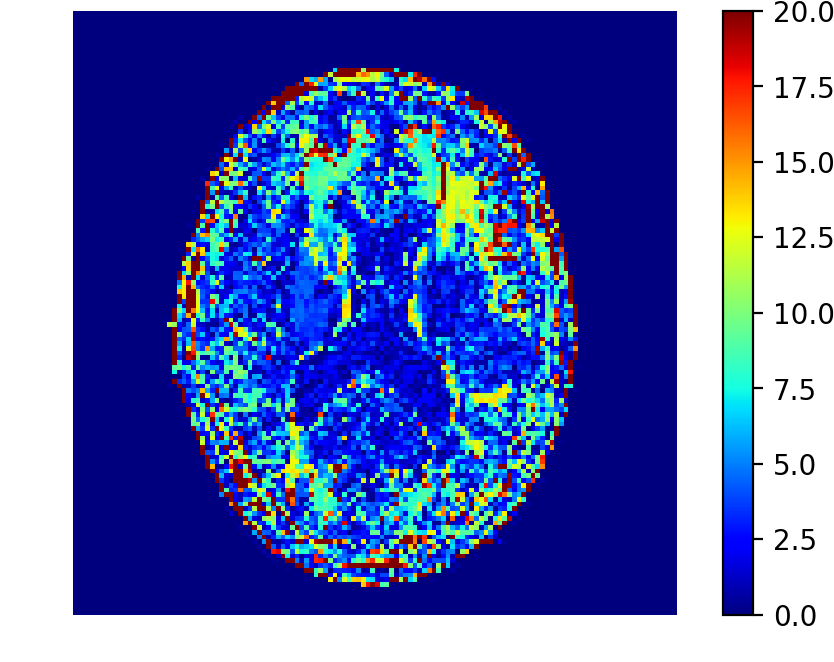}
	\end{minipage} 
	\begin{minipage}[b]{0.19\linewidth}
		\raggedright
		\includegraphics[width=3.5cm, height=3cm,trim=0cm 0cm 0cm 0cm,clip]{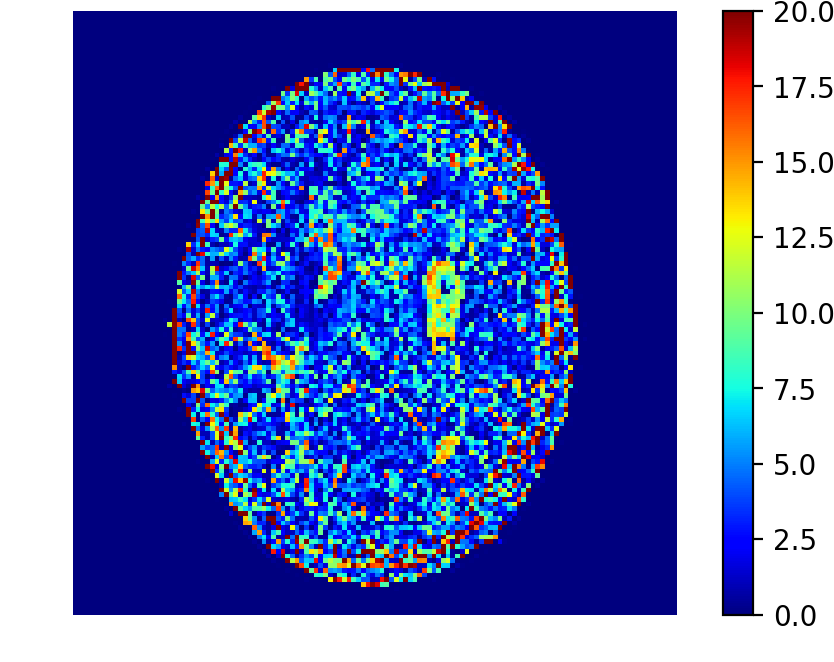}
	\end{minipage} 
	\begin{minipage}[b]{0.19\linewidth}
		\raggedright
		\includegraphics[width=3.5cm, height=3cm,trim=0cm 0cm 0cm 0cm,clip]{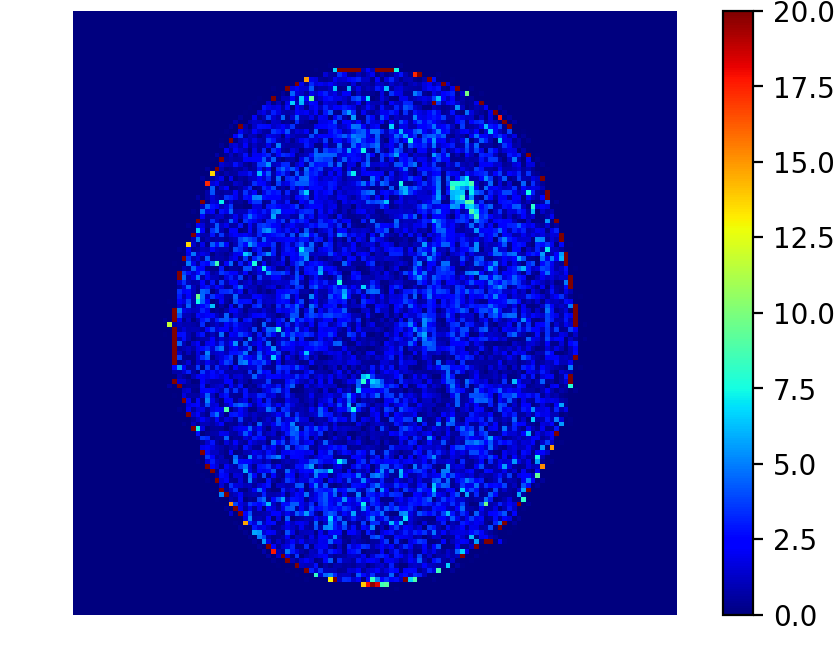}
	\end{minipage} 
	\\
	\begin{minipage}[b]{0.2\linewidth}
		\raggedright 
		\includegraphics[width=3.6cm, height=3cm,trim=0cm 0cm 0cm 0.4cm,clip]{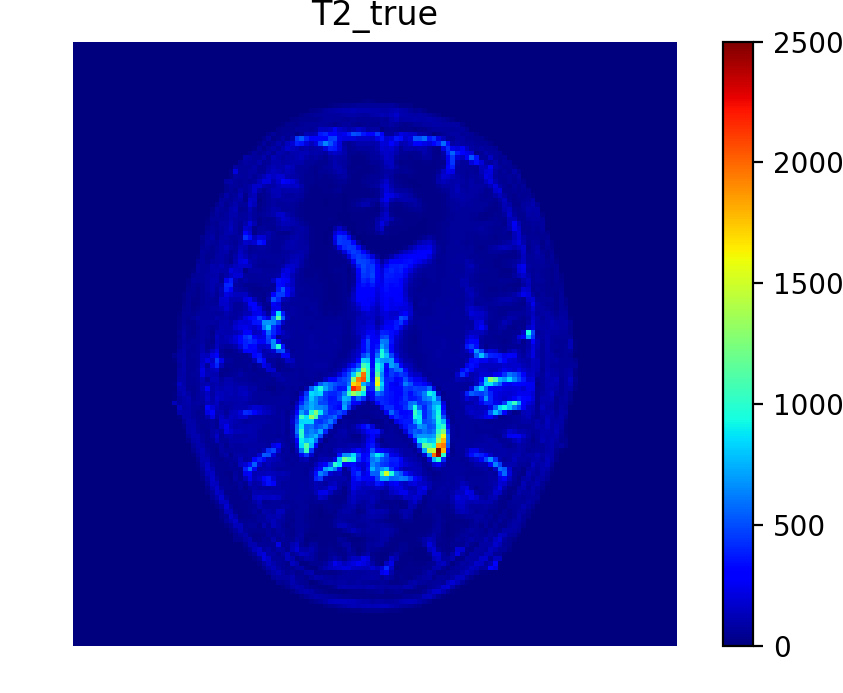}
	\end{minipage} 
	\begin{minipage}[b]{0.19\linewidth}
		\raggedright 
		\includegraphics[width=3.6cm, height=3cm,trim=0cm 0cm 0cm 0cm,clip]{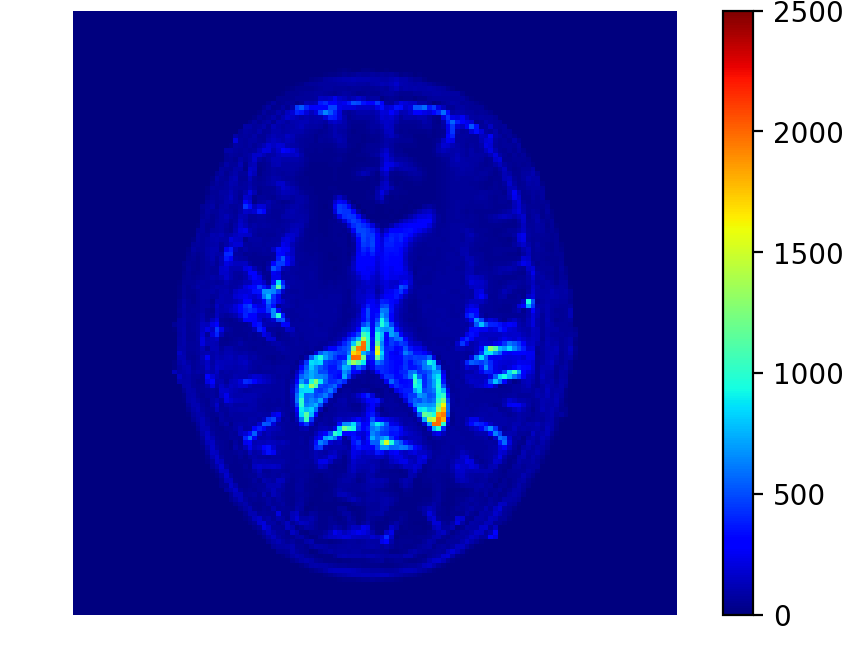}
	\end{minipage} 
	\begin{minipage}[b]{0.19\linewidth}
		\raggedright 
		\includegraphics[width=3.6cm, height=3cm,trim=0cm 0cm 0cm 0cm,clip]{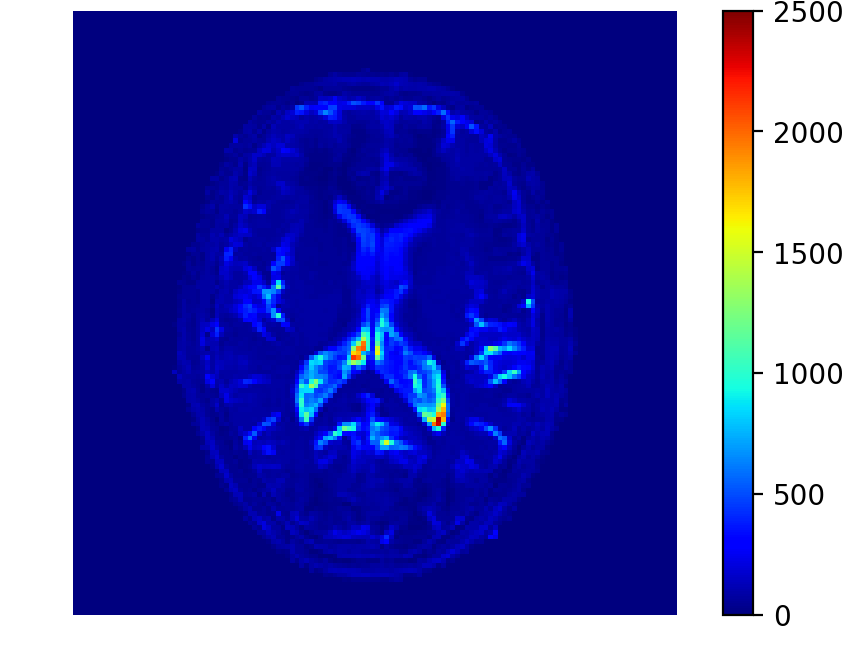}
	\end{minipage} 
	\begin{minipage}[b]{0.19\linewidth}
		\raggedright 
		\includegraphics[width=3.6cm, height=3cm,trim=0cm 0cm 0cm 0cm,clip]{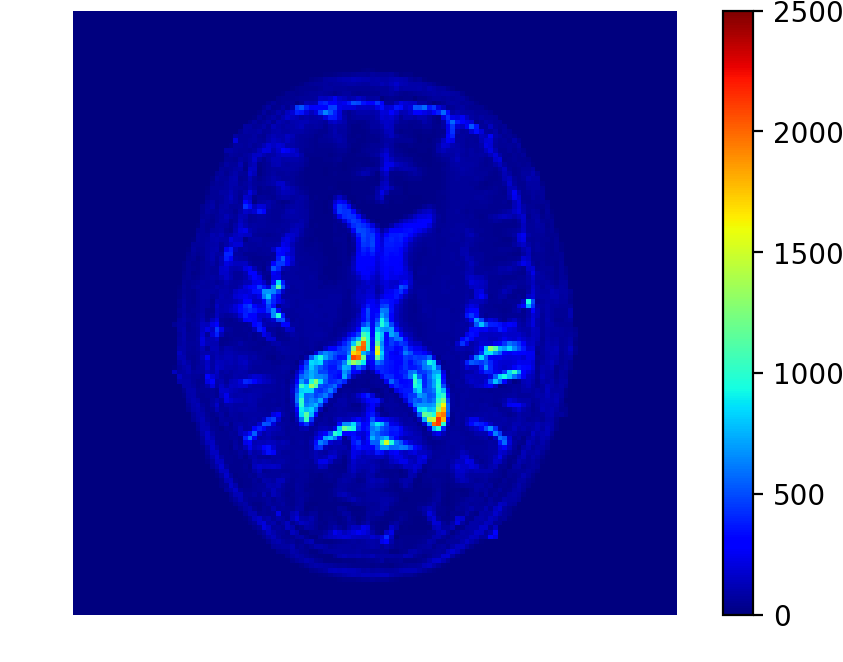}
	\end{minipage} 
	\begin{minipage}[b]{0.19\linewidth}
		\raggedright 
		\includegraphics[width=3.6cm, height=3cm,trim=0cm 0cm 0cm 0cm,clip]{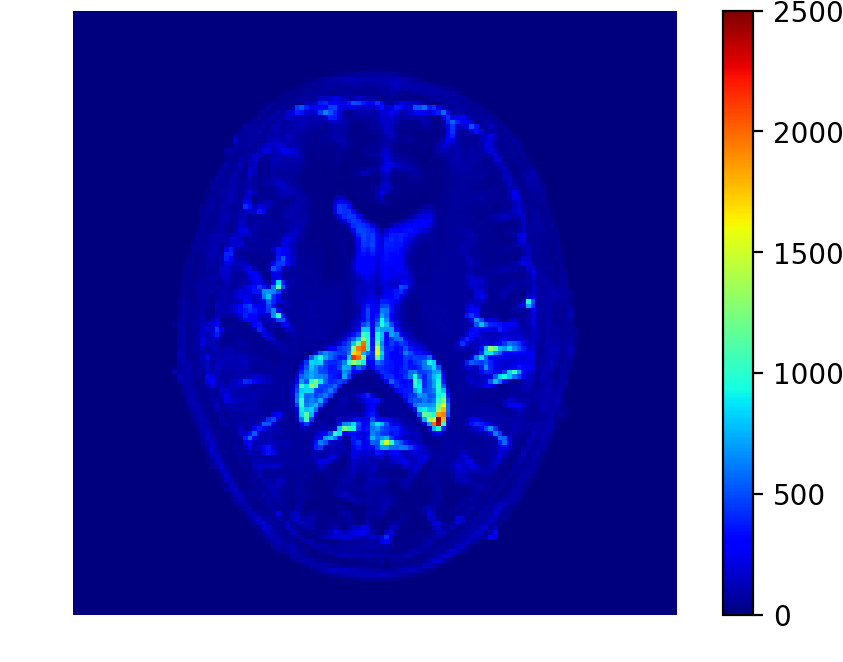}
	\end{minipage} 
	\\
	\begin{minipage}[b]{0.2\linewidth}
		\raggedright 
		\phantom{
			\includegraphics[width=3cm, height=3cm,trim=0cm 0cm 0cm 0.4cm,clip]{./Figures/DeepMRF_ResNet_CNN/T2_true.png}
		}
	\end{minipage} 
	\begin{minipage}[b]{0.19\linewidth}
		\raggedright 
		\includegraphics[width=3.5cm, height=3cm,trim=0cm 0cm 0cm 0cm,clip]{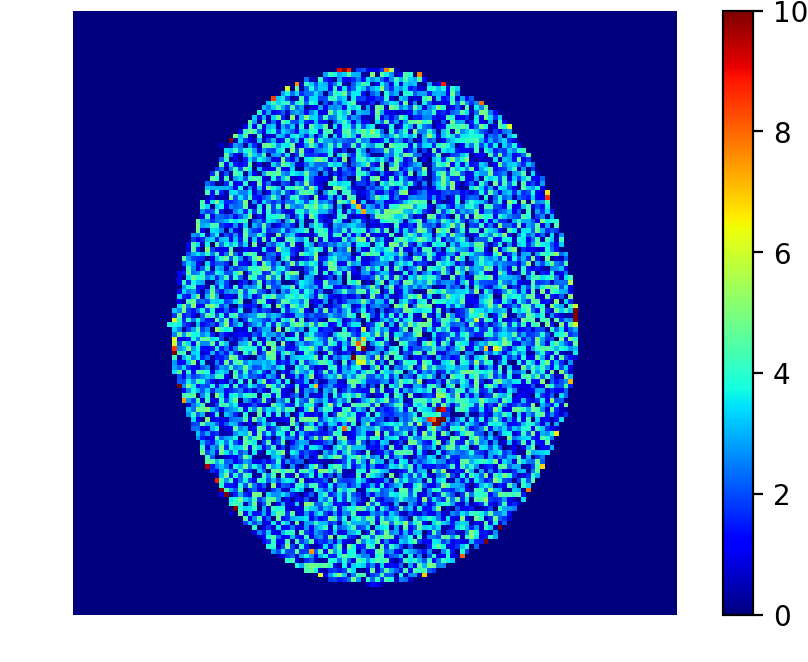}
	\end{minipage} 
	\begin{minipage}[b]{0.19\linewidth}
		\raggedright 
		\includegraphics[width=3.5cm, height=3cm,trim=0cm 0cm 0cm 0cm,clip]{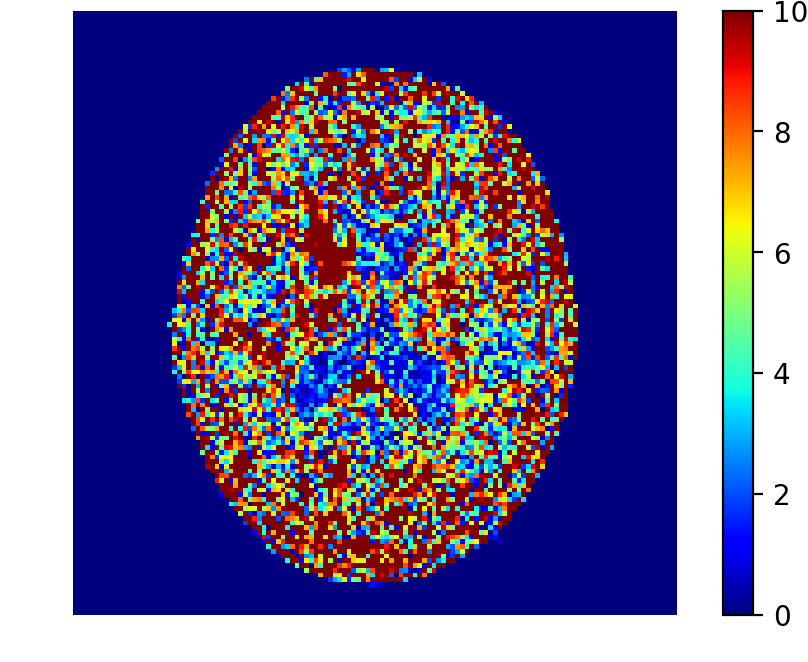}
	\end{minipage} 
	\begin{minipage}[b]{0.19\linewidth}
		\raggedright 
		\includegraphics[width=3.5cm, height=3cm,trim=0cm 0cm 0cm 0cm,clip]{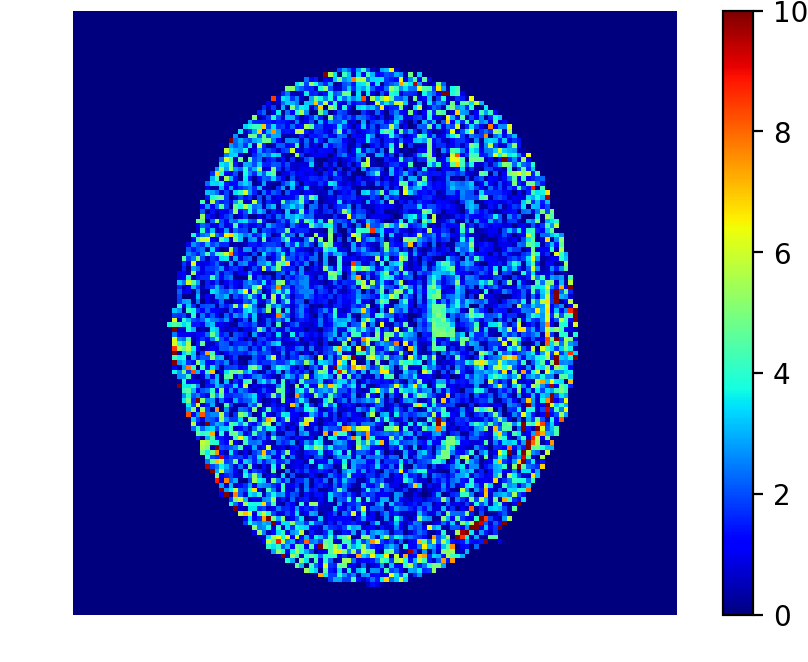}
	\end{minipage}
	\begin{minipage}[b]{0.19\linewidth}
		\raggedright 
		\includegraphics[width=3.5cm, height=3cm,trim=0cm 0cm 0cm 0cm,clip]{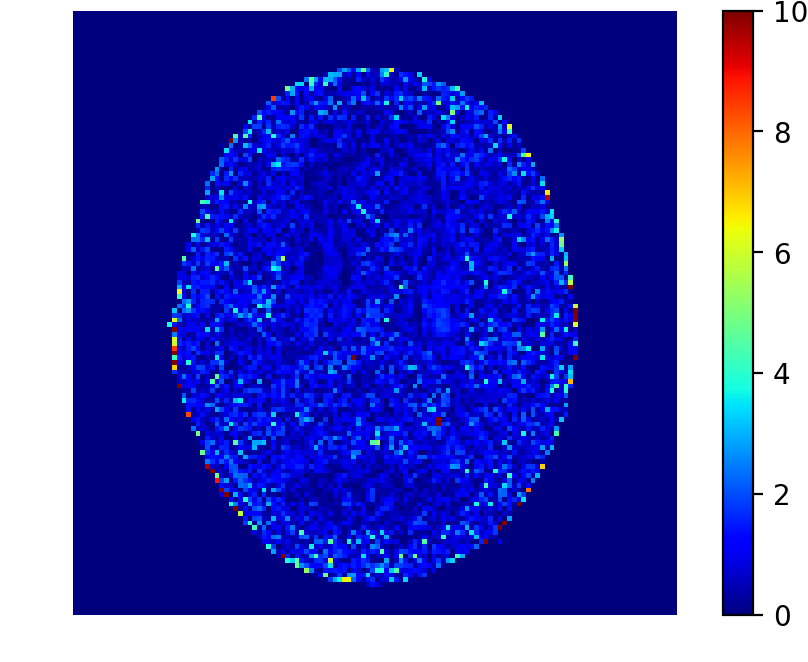}
	\end{minipage} 
	\\
	\begin{minipage}[b]{0.2\linewidth}
		\centering \scriptsize T1/T2 Reference
	\end{minipage} 
	\begin{minipage}[b]{0.19\linewidth}
		\centering \scriptsize Dictionary Matching  			
		\\SNR = 42.20/27.81 dB 		
	\end{minipage} 
	\begin{minipage}[b]{0.19\linewidth}
		\centering \scriptsize CNN~\cite{hoppe2017deep}
		\\SNR = 39.63/25.66 dB
	\end{minipage} 
	\begin{minipage}[b]{0.19\linewidth}
		\centering \scriptsize FNN~\cite{cohen2018mr}			
		\\SNR = 40.09/30.07dB
	\end{minipage} 
	\begin{minipage}[b]{0.19\linewidth}
		\centering \scriptsize HYDRA			
		\\SNR = 42.03/38.32 dB 
	\end{minipage} 
	
	\vspace{-0.3cm}
	
	\caption{
		\footnotesize
		Visual results of testing on anatomical dataset with full k-space sampling for comparing parameter restoration performance. Top two rows correspond to T1 maps and residual errors while bottom two rows correspond to T2 maps and residual errors. Proposed HYDRA results in comparable performance for T1 mapping and yields much better performance for T2 mapping, obtaining 10dB higher SNR gains than competing dictionary-matching based methods~\cite{ma2013magnetic,jiang2015mr,davies2014compressed,wang2016magnetic,mazor2016low,mazor2018low}. HYDRA also outperforms previous networks, such as CNN by Hoppe et al.~\cite{hoppe2017deep} and FNN by Cohen et al.~\cite{cohen2018mr}.
	}
	\label{Fig:ResNet-CNN_test}
\end{figure*}

Visual and quantitative results are shown in Fig.~\ref{Fig:ResNet-CNN_1DtestReal}, Fig.~\ref{Fig:ResNet-CNN_test} and Table~\ref{Tab:TestingPhantom}. It can be seen that our basic version of HYDRA outperforms dictionary matching~\cite{ma2013magnetic,jiang2015mr,davies2014compressed,wang2016magnetic,mazor2016low,mazor2018low}, yielding better visual and quantitative performance, e.g., 7.9 dB SNR gains for T2 mapping. The RMSE of T2 mapping is also reduced to 2.498 from 6.252, accordingly. Our nonlocal version of HYDRA achieves even better performance, leading to 10 dB SNR gains with RMSE as small as 1.86. This is owing to the advantage that the trained network is a powerful function approximator, which is able to output well-estimated parameter values based on learnt mapping functions, even though these values do not exist in the training dictionary and lookup-table. In contrast, dictionary matching only matches signatures to discrete parameters existing in the training dataset. In other words, if there are no exact matching dictionary element and parameter values for an inquiry MRF signature, it will find adjacent values as approximations, thus introducing discretization error. On the other hand, the advantage of HYDRA over dictionary matching on T1 mapping is not as significant as on T2 mapping quantitatively. But the visual improvements are evident. A similar trend is observed when comparing our network with competing networks such as Hoppe et al.'s CNN~\cite{hoppe2017deep} and Cohen et al.'s FNN~\cite{cohen2018mr}. In addition, HYDRA takes around 2 s to accomplish the mapping for 16384 signatures, 40$\times$ faster than dictionary matching~\cite{ma2013magnetic,jiang2015mr,davies2014compressed,wang2016magnetic,mazor2016low,mazor2018low}. 

\vspace{-0.2cm}

\subsubsection{k-space subsampling using Gaussian patterns}

\vspace{-0.2cm}

\begin{figure}[t]
	\centering
	\begin{minipage}[b]{0.24\linewidth}
		\centering
		\includegraphics[width = 2.5cm, trim=0cm 0cm 0cm 0cm,clip ]{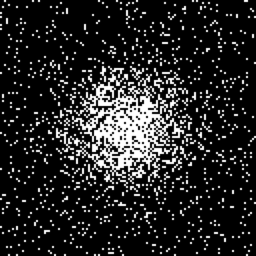}
	\end{minipage} 
	\begin{minipage}[b]{0.24\linewidth}
		\centering
		\includegraphics[width = 2.5cm, trim=0cm 0cm 0cm 0cm,clip ]{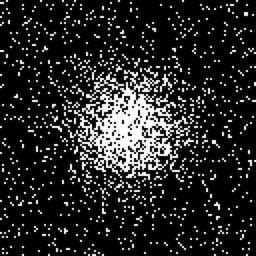}
	\end{minipage} 
	\begin{minipage}[b]{0.24\linewidth}
		\centering
		\includegraphics[width = 2.5cm, trim=0cm 0cm 0cm 0cm,clip ]{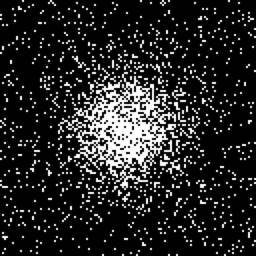}
	\end{minipage} 
	\begin{minipage}[b]{0.24\linewidth}
		\centering
		\includegraphics[width = 2.5cm, trim=0cm 0cm 0cm 0cm,clip ]{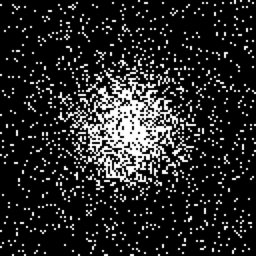}
	\end{minipage} 
	
	\caption{
		A series of Gaussian patterns used for k-space subsampling.
	}
	\label{Fig:SamplePatterns}
\end{figure}

In k-space subsampling situations, the developed low-rank based de-aliasing method is applied to restore the signature matrix $\bX$ from the measurements matrix $\bY$. Then, the reconstructed $\bX$ is used as input into the network for parameter mapping to obtain the corresponding tissue parameter values. In the experiments, the sub-sampling factor $\beta$ is set to be 70\% and 15\%. For $\beta$ = 15\%, 15\% k-space data is acquired by a series of 2D random Gaussian sampling patterns, shown in Fig.~\ref{Fig:SamplePatterns}, leading to a k-space measurement matrix $\bY$ of size $Q \times L = 2458 \times 200$. Similarly, $\beta$ = 70\% gives rise to a k-space measurement matrix $\bY$ of size $Q \times L = 11469 \times 200$. A larger $\lambda $ enforces lower rank for the restored signature matrix $\bX$ to strengthen the de-aliasing effect, while a smaller $\lambda $ encourages $\bX$ to have a subsampled Fourier transform that approximates the k-space measurements matrix $\bY$ better. Therefore, we tried a range of values from 1 to 20 for $\lambda $ and experimentally select the best one $\lambda = 5$. Since the low-rank based signature restoration involves gradient descent steps, a larger step size $\mu$ accelerates gradient descent speed, but tends to result in oscillation or even divergence, while a smaller $\mu$ leads to a slower convergence. We experimentally find that $\mu = 1$ gives a good balance. The same k-space measurements $\bY$ are also used by dictionary-matching based methods~\cite{ma2013magnetic,davies2014compressed,mazor2018low} for comparison, and the same signature restoration approach is used to convert $\bY$ onto $\bX$ for learning based methods~\cite{hoppe2017deep,cohen2018mr}. The aim is to evaluate the overall performance on both signature restoration and parameter restoration.

Quantitative performance is shown in Table~\ref{Tab:TestingPhantomSubsamplingNonlocal}. Note that the advantage of learning-based methods over dictionary matching degrades when the subsampling factor increases. This is due to the fact that the restored signatures from highly subsampled k-space data exhibit deviations and distortions, thus leading to poorer input for the trained networks. In spite of this, the proposed approach outperforms the dictionary matching based methods~\cite{ma2013magnetic,davies2014compressed} with significant gains, and also yields better or comparable performance as the state-of-the-art methods FLOR~\cite{mazor2018low}, CNN~\cite{hoppe2017deep} and FNN~\cite{cohen2018mr}. In addition, it takes around 23s for low-rank based signature restoration and less than 3s for network based parameter restoration. Thus, the total time cost is around 26s, almost 4.8$\times$ faster than FLOR~\cite{mazor2018low}. Furthermore, the speed of our method is 60$\times$ faster than FLOR~\cite{mazor2018low} for parameter restoration.

We compared the performance with/without nonlocal operations in our developed network. The results in Table~\ref{Tab:TestingPhantom} and \ref{Tab:TestingPhantomSubsamplingNonlocal} show that the proposed network with nonlocal operations based self-attention scheme outperforms the basic counterpart. In particular, the nonlocal version achieves 6 dB gains in terms of SNR over the basic version for T2 mapping. Such significant improvement demonstrates the benefits of capturing long-range dependencies and global features using the nonlocal operation based attention scheme. 

\begin{figure*}[tb]
	\begin{multicols}{2}  
		\centering
		\begin{minipage}[b]{0.48\linewidth}
			\centering
			\includegraphics[width = 4cm, height=3.2cm]{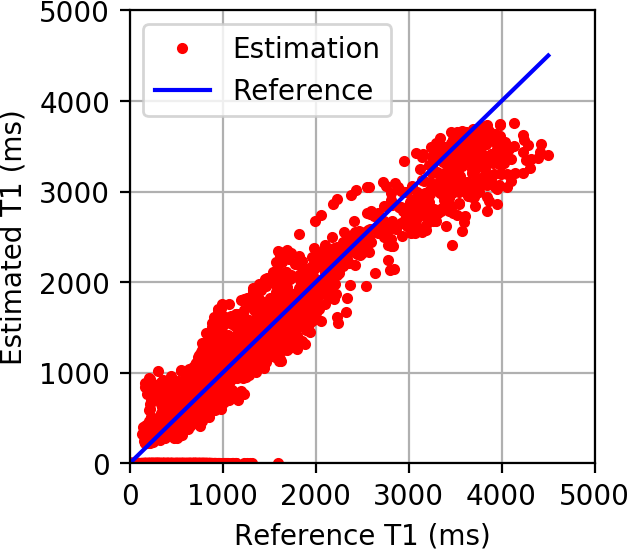}
		\end{minipage} 
		\begin{minipage}[b]{0.48\linewidth}
			\centering
			\includegraphics[width = 4cm, height=3.2cm]{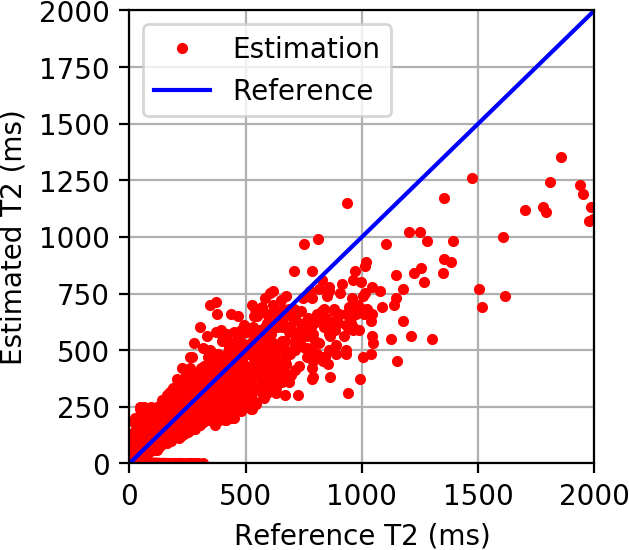}
		\end{minipage} 
		\\
		\begin{minipage}[b]{1\linewidth}
			\centering
			\footnotesize (a) T1, T2 estimations using Ma et al.~\cite{ma2013magnetic}. 
		\end{minipage}
		\\
		\vspace{+0.3cm}
		\begin{minipage}[b]{0.48\linewidth}
			\centering
			\includegraphics[width = 4cm, height=3.2cm]{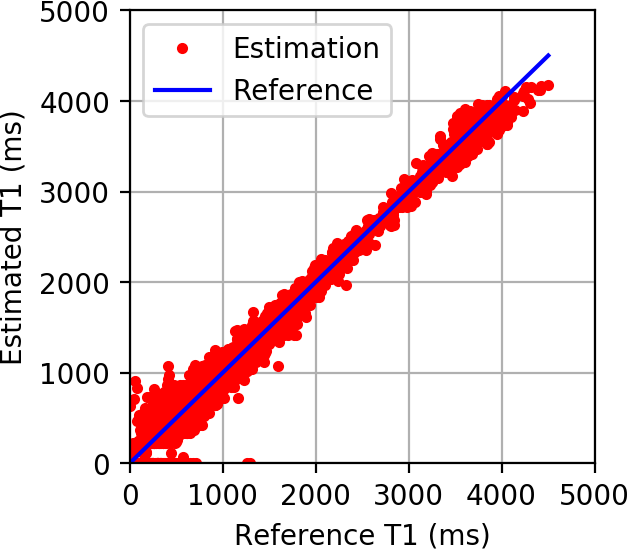}
		\end{minipage} 
		\begin{minipage}[b]{0.48\linewidth}
			\centering
			\includegraphics[width = 4cm, height=3.2cm]{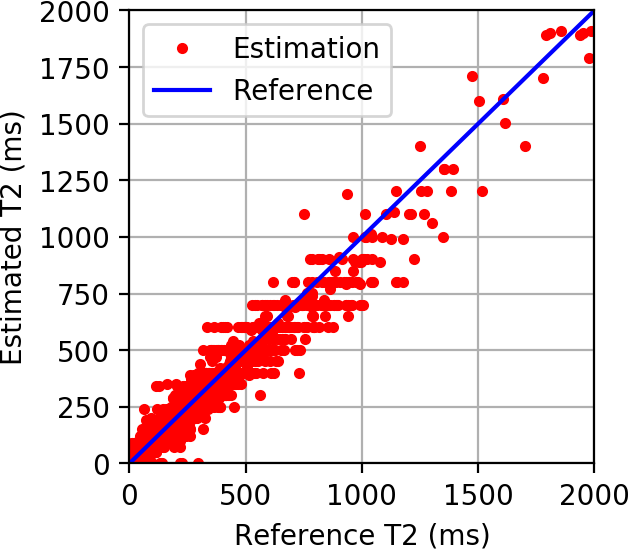}
		\end{minipage} 
		\\
		\begin{minipage}[b]{1\linewidth}
			\centering
			\footnotesize (b) T1, T2 estimations using BLIP~\cite{davies2014compressed} 
		\end{minipage}
		\\
		\vspace{+0.3cm}
		\begin{minipage}[b]{0.48\linewidth}
			\centering
			\includegraphics[width = 4cm, height=3.2cm]{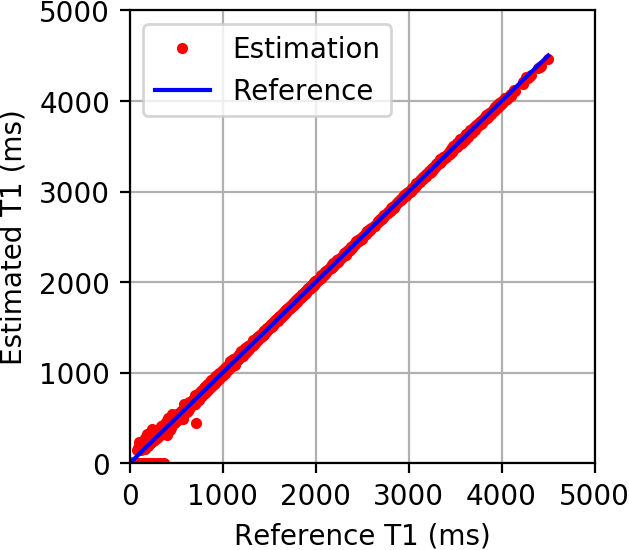}
		\end{minipage} 
		\begin{minipage}[b]{0.48\linewidth}
			\centering
			\includegraphics[width = 4cm, height=3.2cm]{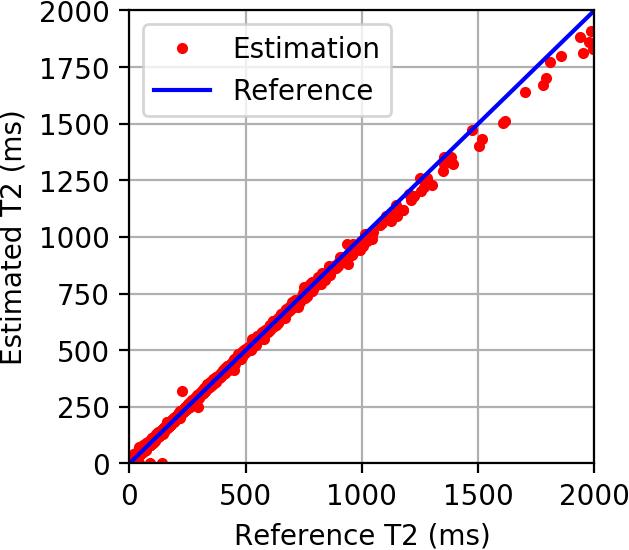}
		\end{minipage} 
		\\
		\begin{minipage}[b]{1\linewidth}
			\centering
			\footnotesize (c) T1, T2 estimations using FLOR~\cite{mazor2018low} 
		\end{minipage}
		\\
		\vspace{+0.3cm}
		\begin{minipage}[b]{0.48\linewidth}
			\centering
			\includegraphics[width = 4cm, height=3.2cm]{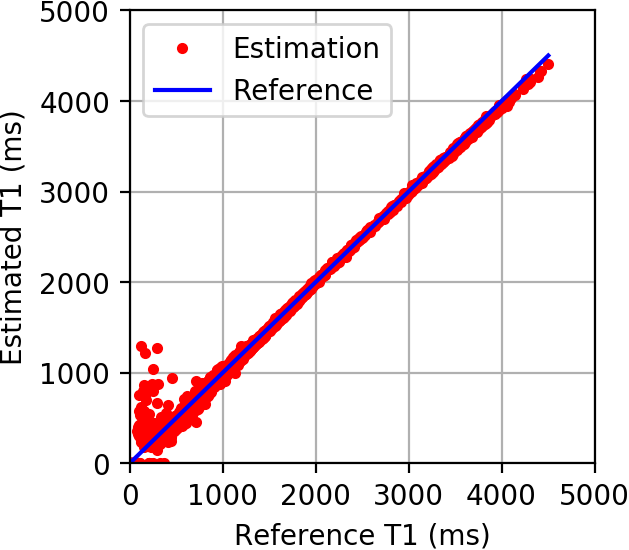}
		\end{minipage} 
		\begin{minipage}[b]{0.48\linewidth}
			\centering
			\includegraphics[width = 4cm, height=3.2cm]{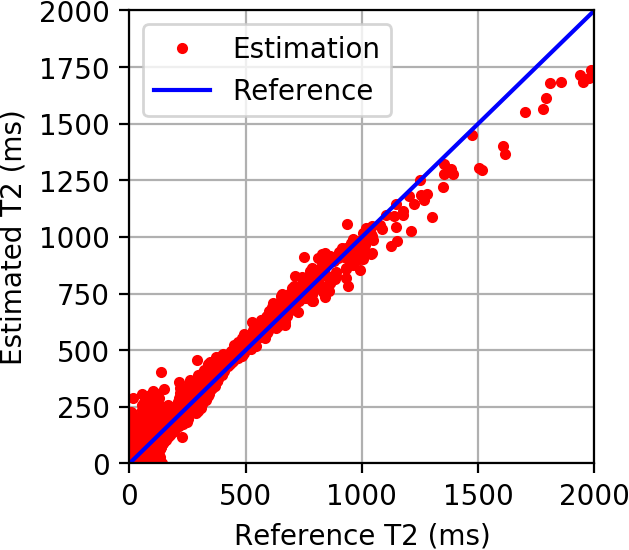}
		\end{minipage} 
		\\
		\begin{minipage}[b]{1\linewidth}
			\centering
			\footnotesize (d) T1, T2 estimations using CNN~\cite{hoppe2017deep}.
		\end{minipage}
		\\
		\vspace{+0.3cm}
		\begin{minipage}[b]{0.48\linewidth}
			\centering
			\includegraphics[width = 4cm, height=3.2cm]{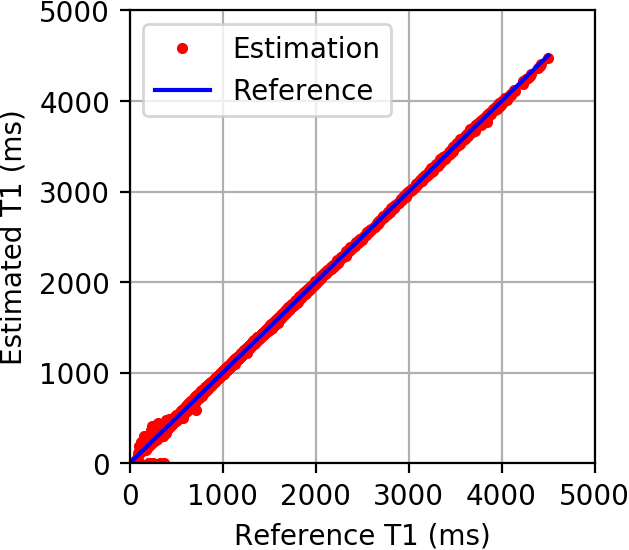}
		\end{minipage} 
		\begin{minipage}[b]{0.48\linewidth}
			\centering
			\includegraphics[width = 4cm, height=3.2cm]{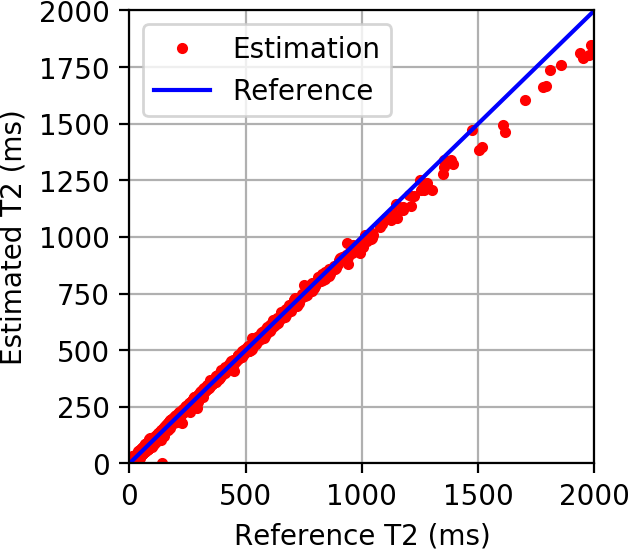}
		\end{minipage} 
		\\
		\begin{minipage}[b]{1\linewidth}
			\centering
			\footnotesize (e) T1, T2 estimations using FNN~\cite{cohen2018mr}.
		\end{minipage}
		\\
		\vspace{+0.3cm}
		\begin{minipage}[b]{0.48\linewidth}
			\centering
			\includegraphics[width = 4cm, height=3.2cm]{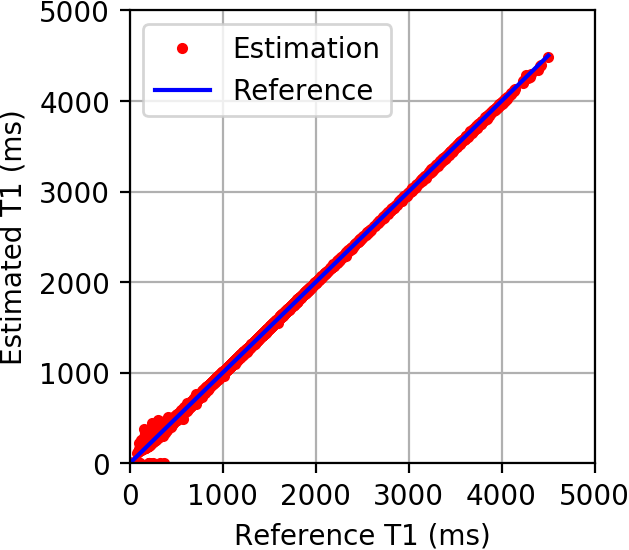}
		\end{minipage} 
		\begin{minipage}[b]{0.48\linewidth}
			\centering
			\includegraphics[width = 4cm, height=3.2cm]{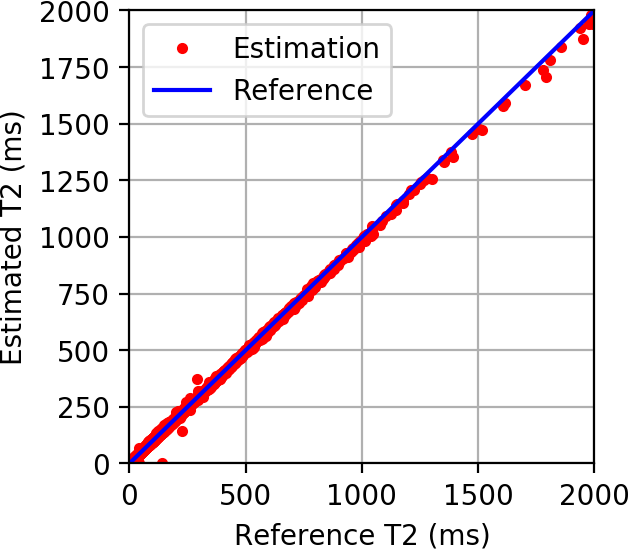}
		\end{minipage} 
		\\
		\begin{minipage}[b]{1\linewidth}
			\centering
			\footnotesize (f) T1, T2 estimations using HYDRA.
		\end{minipage}
		\\
		\vspace{+0.3cm}
		
	\end{multicols}
	
	\vspace{-0.5cm}
	
	\caption{
		\footnotesize
		Testing on the anatomical dataset with k-space subsampling factor 15\% using Gaussian patterns and 1000 time frames. Subfig. (a) - (f) show the results using Ma et al.~\cite{ma2013magnetic}, BLIP~\cite{davies2014compressed}, FLOR~\cite{mazor2018low}, CNN by Hoppe et al.~\cite{hoppe2017deep}, FNN by Cohen et al.~\cite{cohen2018mr} and HYDRA.
	}
	\label{Fig:KSpaceGaussRatio_15}
\end{figure*}

\begin{figure*}[tb]
	\centering 
	\begin{minipage}[b]{0.19\linewidth}
		\raggedright 
		\includegraphics[width=3.6cm, height=3cm,trim=0cm 0cm 0cm 0cm,clip]{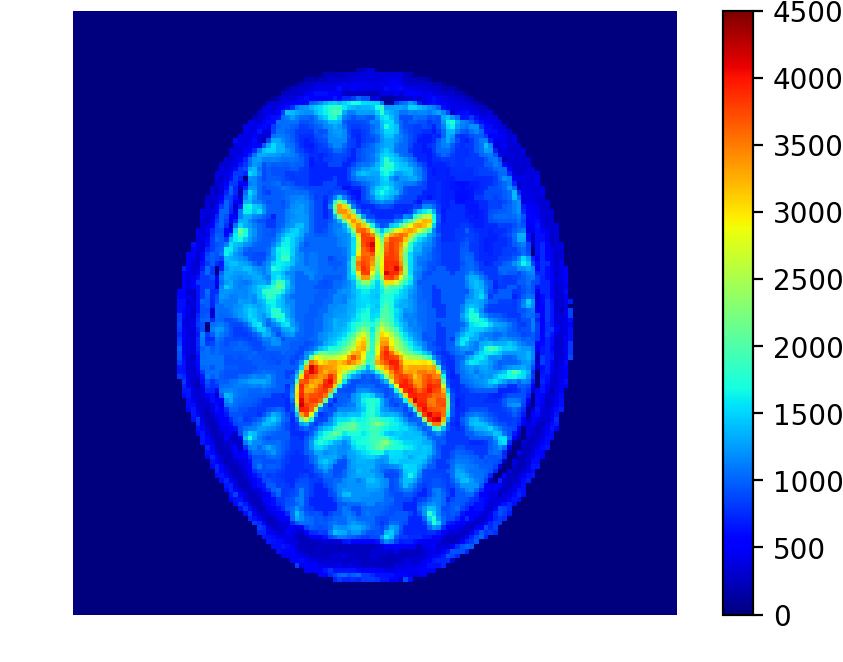}
	\end{minipage} 
	\begin{minipage}[b]{0.19\linewidth}
		\raggedright 
		\includegraphics[width=3.6cm, height=3cm,trim=0cm 0cm 0cm 0cm,clip]{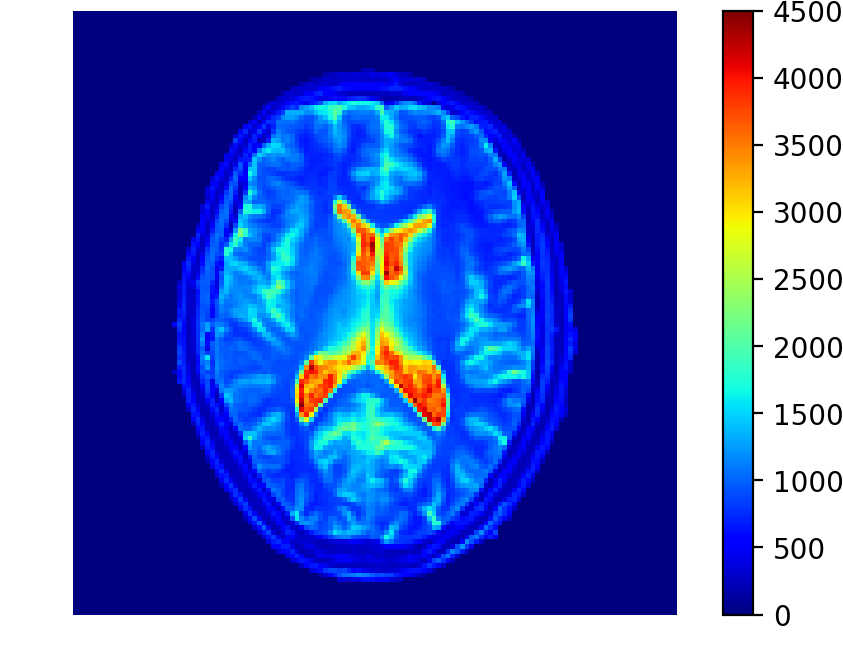}
	\end{minipage} 
	\begin{minipage}[b]{0.19\linewidth}
		\raggedright 
		\includegraphics[width=3.6cm, height=3cm,trim=0cm 0cm 0cm 0cm,clip]{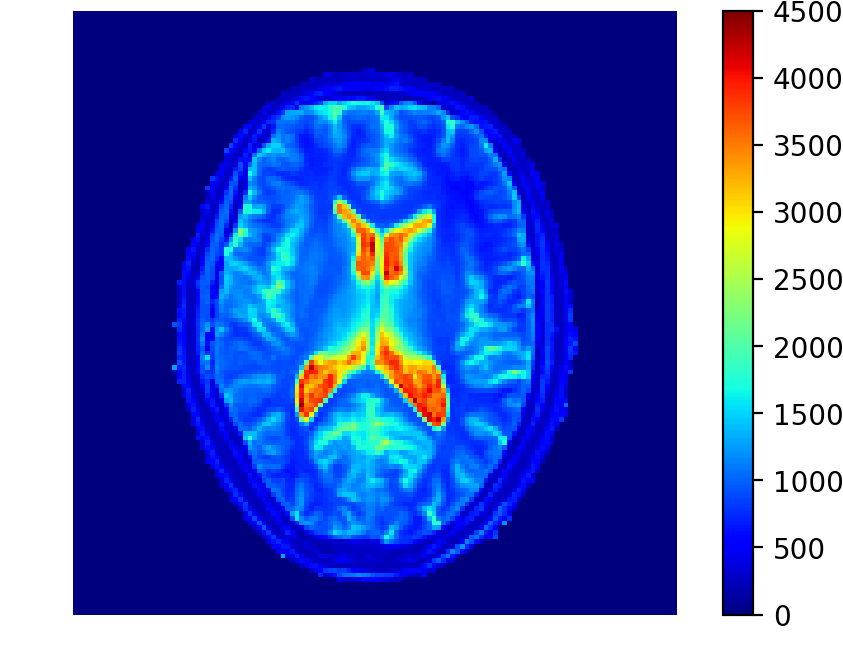}
	\end{minipage} 
	\begin{minipage}[b]{0.19\linewidth}
		\raggedright 
		\includegraphics[width=3.6cm, height=3cm,trim=0cm 0cm 0cm 0cm,clip]{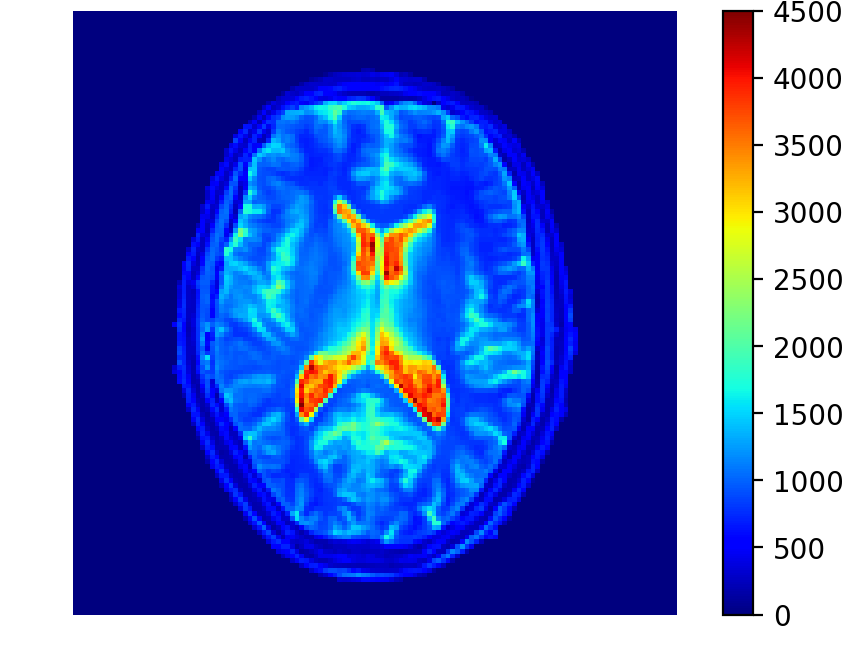}
	\end{minipage} 
	\begin{minipage}[b]{0.19\linewidth}
		\raggedright 
		\includegraphics[width=3.6cm, height=3cm,trim=0cm 0cm 0cm 0cm,clip]{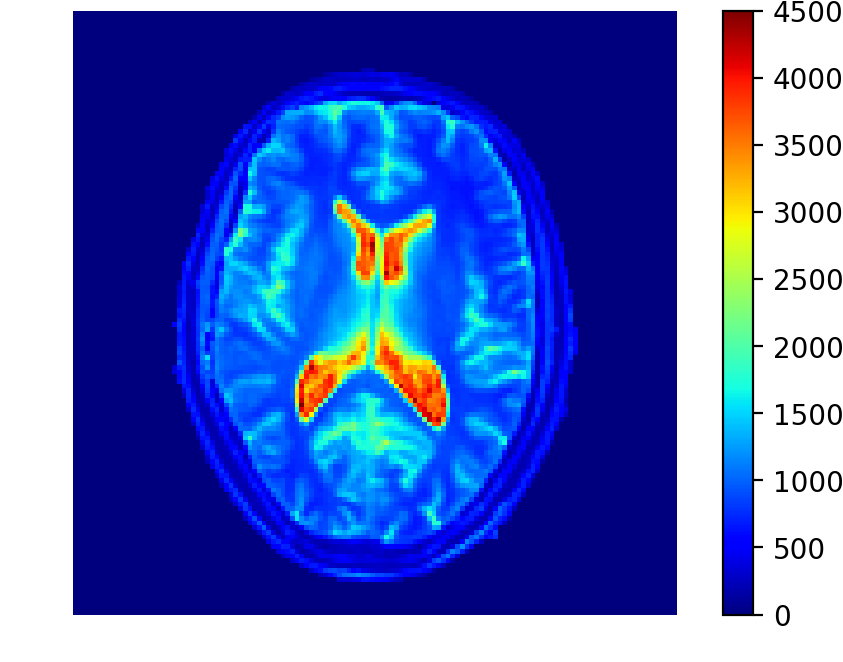}
	\end{minipage} 
	\\
	\begin{minipage}[b]{0.19\linewidth}
		\raggedright 
		\includegraphics[width=3.5cm, height=3cm,trim=0cm 0cm 0cm 0cm,clip]{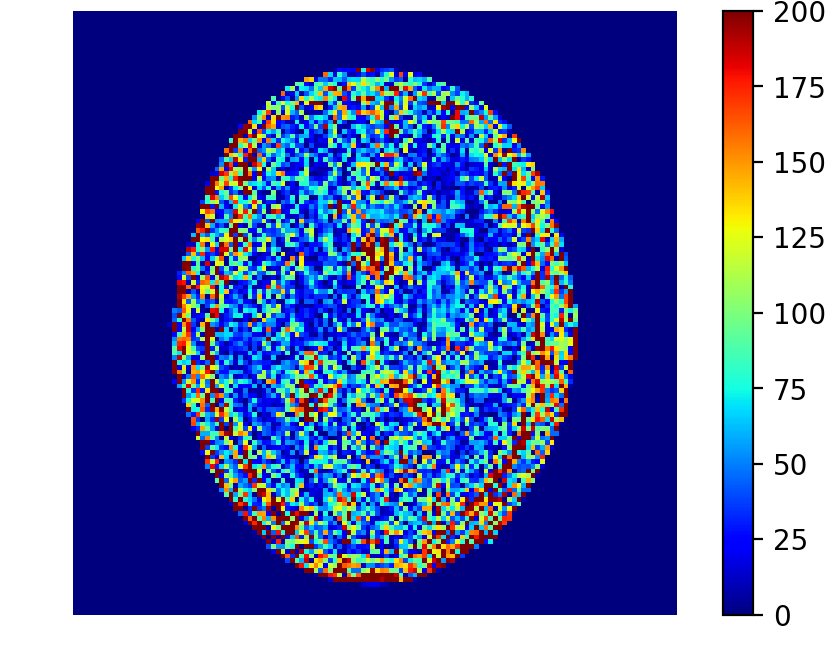}
	\end{minipage} 
	\begin{minipage}[b]{0.19\linewidth}
		\raggedright 
		\includegraphics[width=3.5cm, height=3cm,trim=0cm 0cm 0cm 0cm,clip]{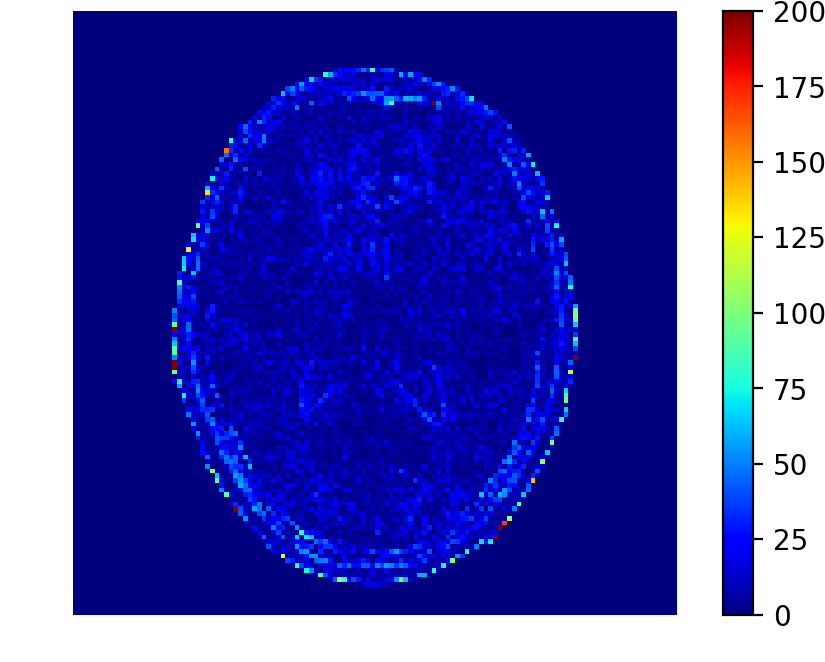}
	\end{minipage} 
	\begin{minipage}[b]{0.19\linewidth}
		\raggedright 
		\includegraphics[width=3.6cm, height=3cm,trim=0cm 0cm 0cm 0cm,clip]{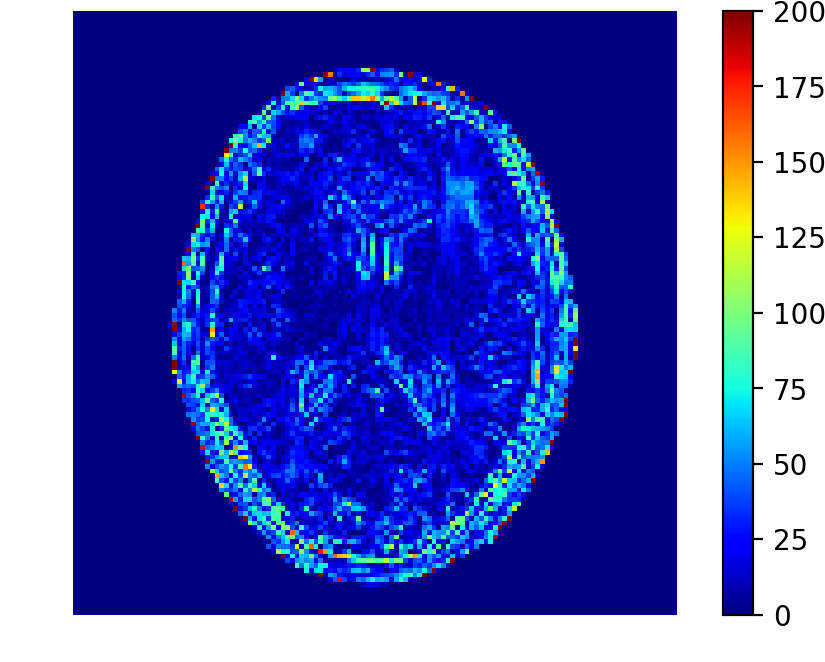}
	\end{minipage} 
	\begin{minipage}[b]{0.19\linewidth}
		\raggedright 
		\includegraphics[width=3.6cm, height=3cm,trim=0cm 0cm 0cm 0cm,clip]{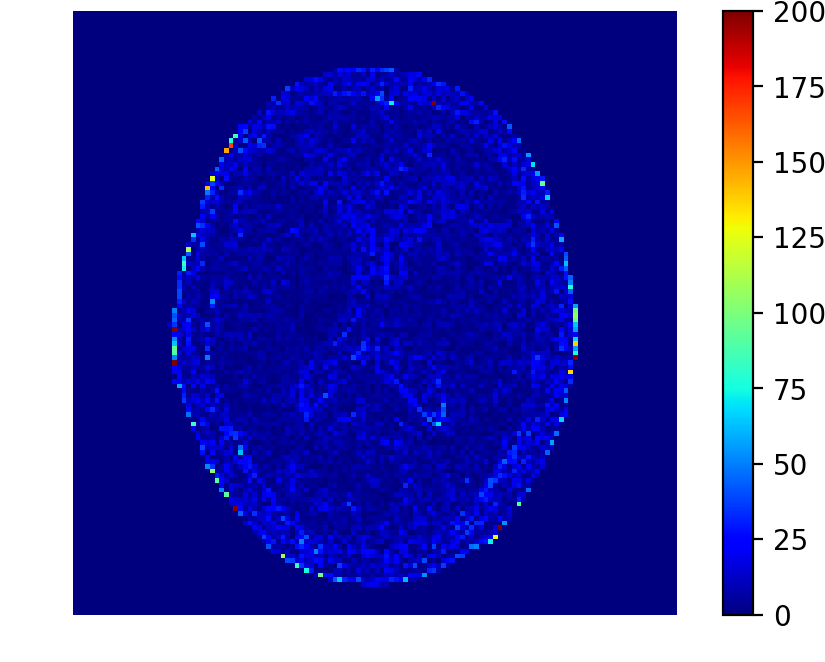}
	\end{minipage} 
	\begin{minipage}[b]{0.19\linewidth}
		\raggedright 
		\includegraphics[width=3.5cm, height=3cm,trim=0cm 0cm 0cm 0cm,clip]{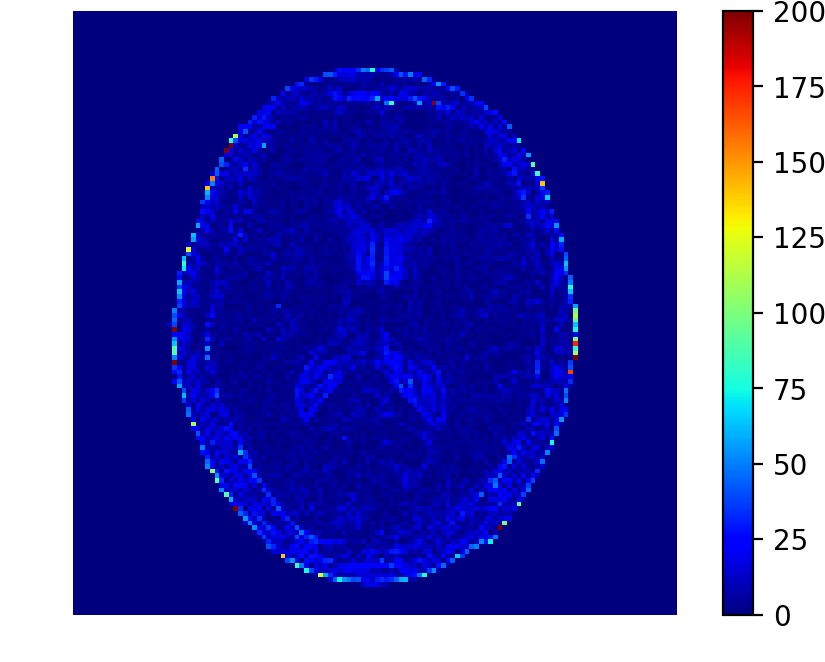}
	\end{minipage} 
	\\
	\begin{minipage}[b]{1\linewidth}
		\centering \footnotesize 
		(a) T1 estimation (top row) and residual errors (bottom row).
	\end{minipage}
	\\
	\vspace{+0.2cm}
	\begin{minipage}[b]{0.19\linewidth}
		\raggedright 
		\includegraphics[width=3.6cm, height=3cm,trim=0cm 0cm 0cm 0cm,clip]{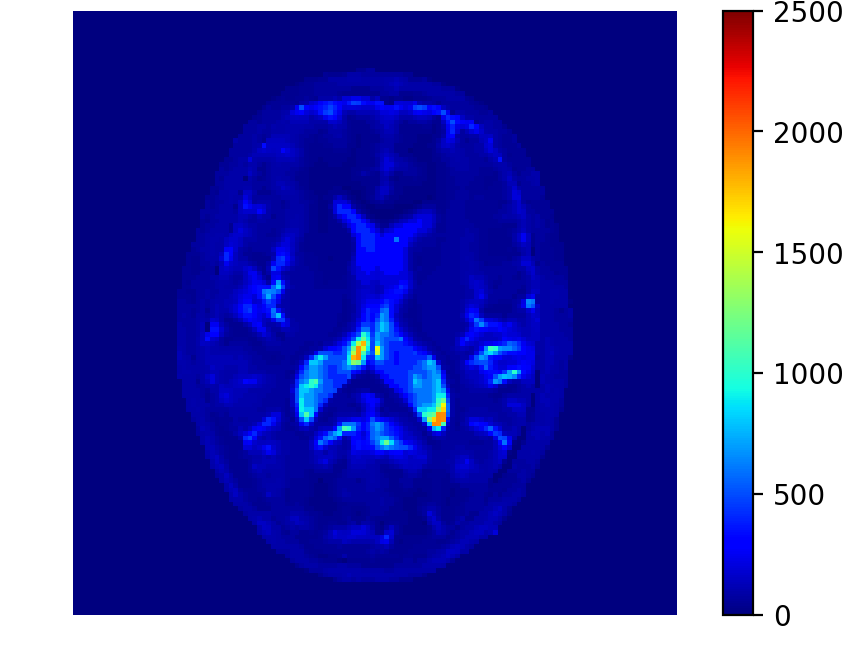}
	\end{minipage} 
	\begin{minipage}[b]{0.19\linewidth}
		\raggedright 
		\includegraphics[width=3.6cm, height=3cm,trim=0cm 0cm 0cm 0cm,clip]{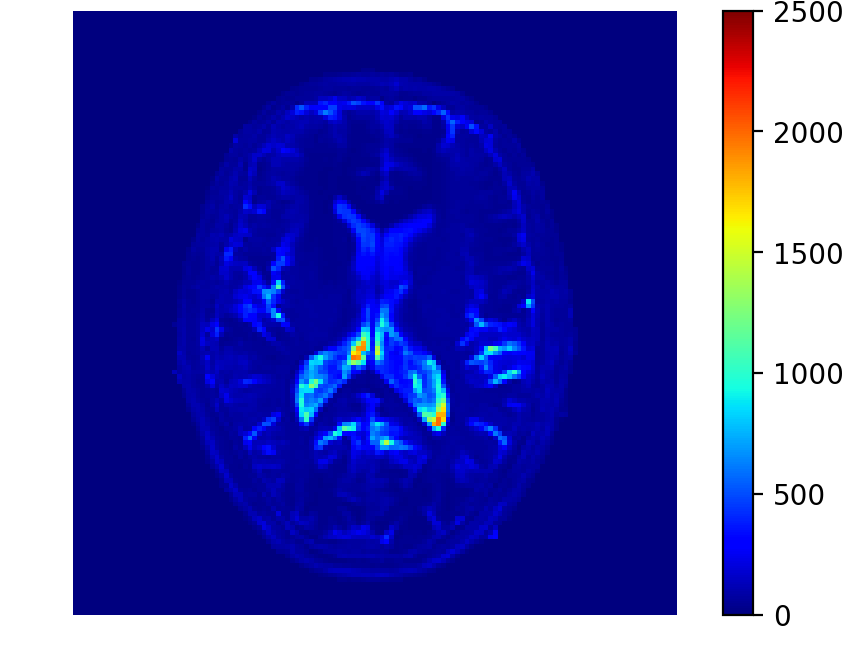}
	\end{minipage} 
	\begin{minipage}[b]{0.19\linewidth}
		\raggedright 
		\includegraphics[width=3.6cm, height=3cm,trim=0cm 0cm 0cm 0cm,clip]{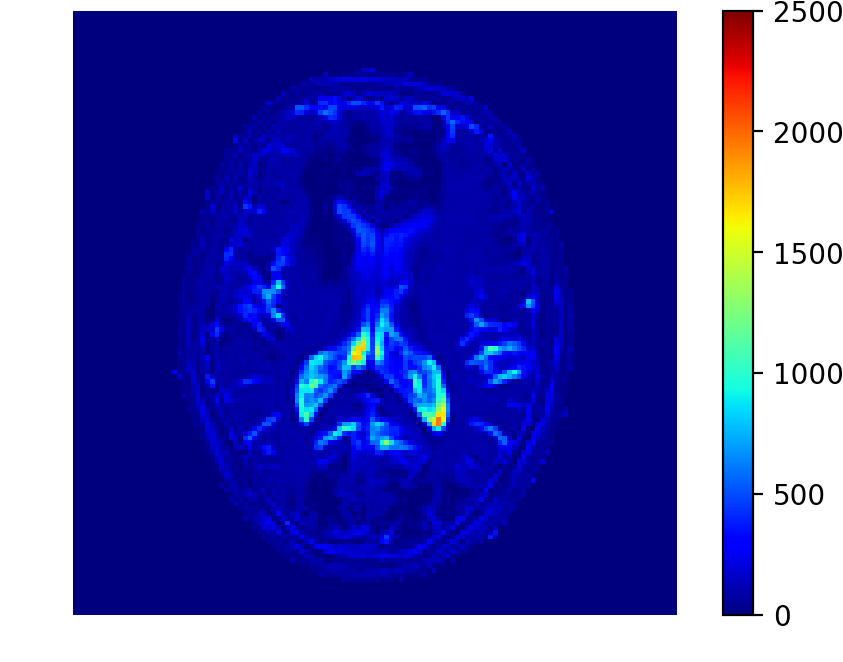}
	\end{minipage} 
	\begin{minipage}[b]{0.19\linewidth}
		\raggedright 
		\includegraphics[width=3.6cm, height=3cm,trim=0cm 0cm 0cm 0cm,clip]{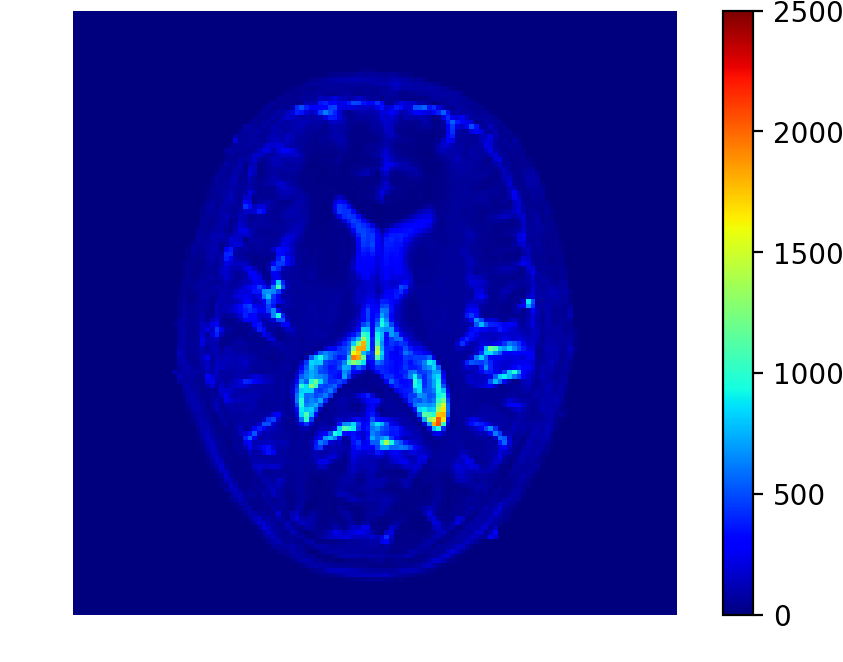}
	\end{minipage} 
	\begin{minipage}[b]{0.19\linewidth}
		\raggedright 
		\includegraphics[width=3.6cm, height=3cm,trim=0cm 0cm 0cm 0cm,clip]{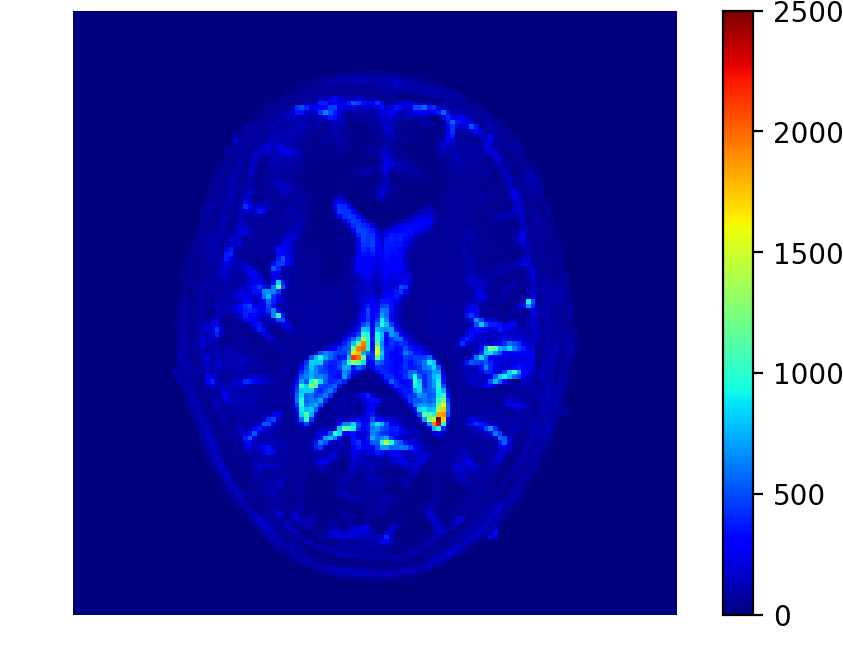}
	\end{minipage} 
	\\
	\begin{minipage}[b]{0.19\linewidth}
		\raggedright 
		\includegraphics[width=3.5cm, height=3cm,trim=0cm 0cm 0cm 0cm,clip]{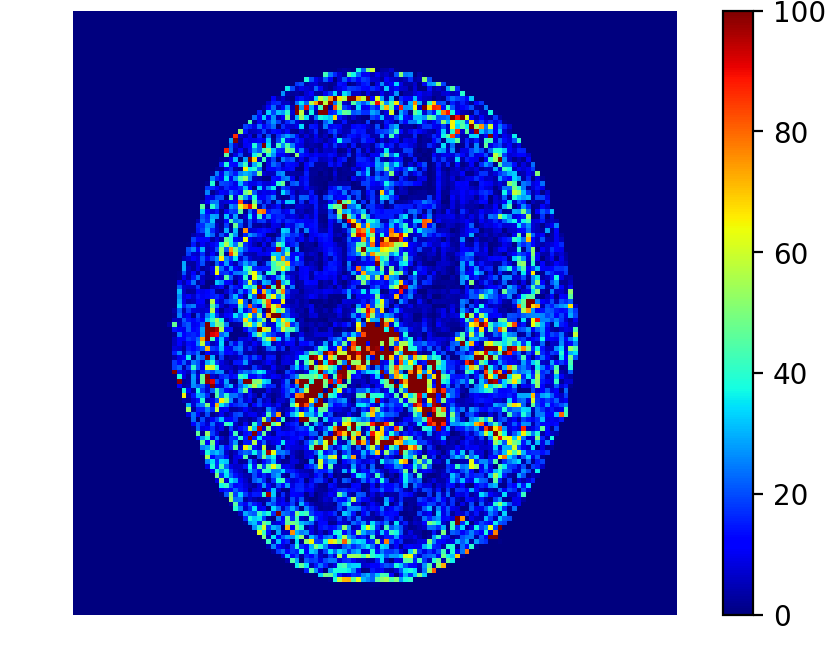}
	\end{minipage} 
	\begin{minipage}[b]{0.19\linewidth}
		\raggedright 
		\includegraphics[width=3.5cm, height=3cm,trim=0cm 0cm 0cm 0cm,clip]{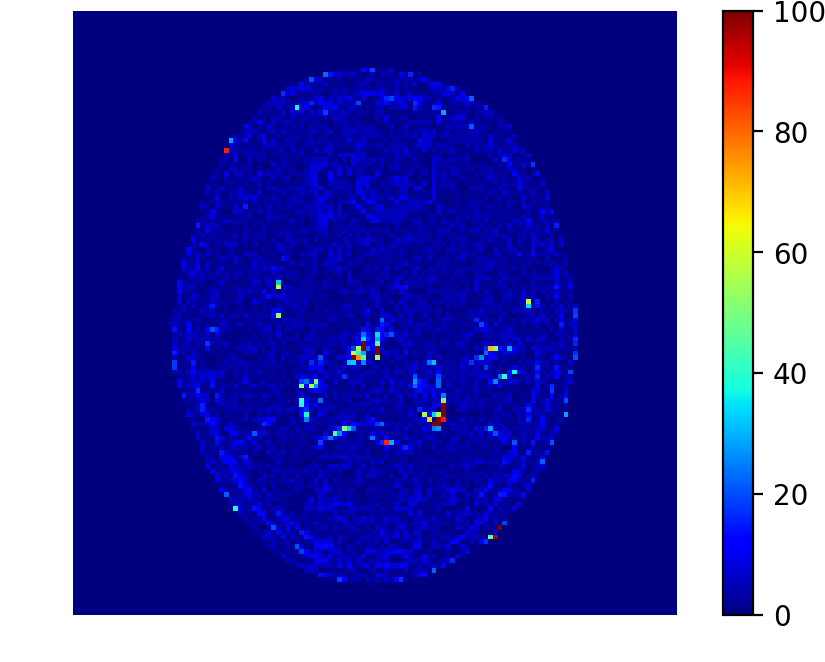}
	\end{minipage} 
	\begin{minipage}[b]{0.19\linewidth}
		\raggedright 
		\includegraphics[width=3.6cm, height=3cm,trim=0cm 0cm 0cm 0cm,clip]{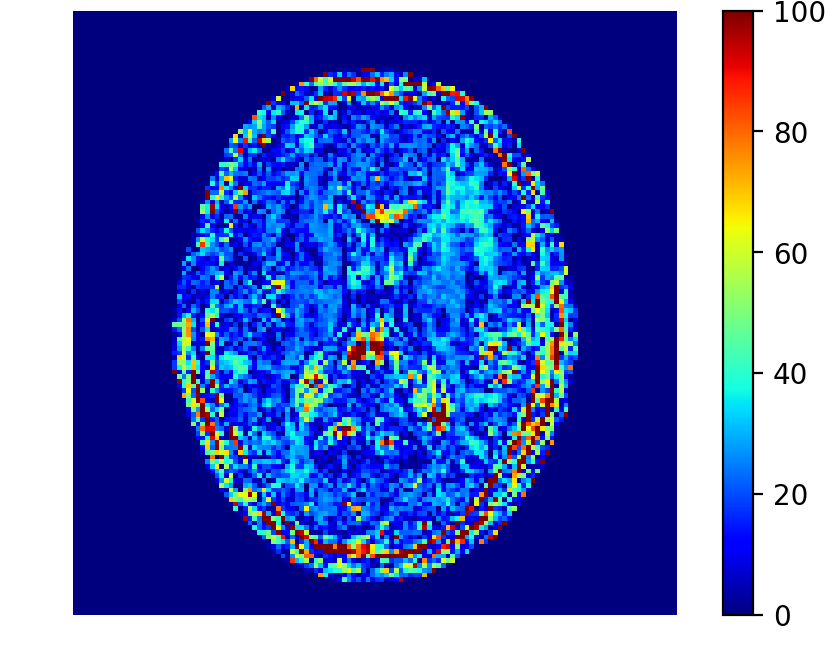}
	\end{minipage} 
	\begin{minipage}[b]{0.19\linewidth}
		\raggedright 
		\includegraphics[width=3.6cm, height=3cm,trim=0cm 0cm 0cm 0cm,clip]{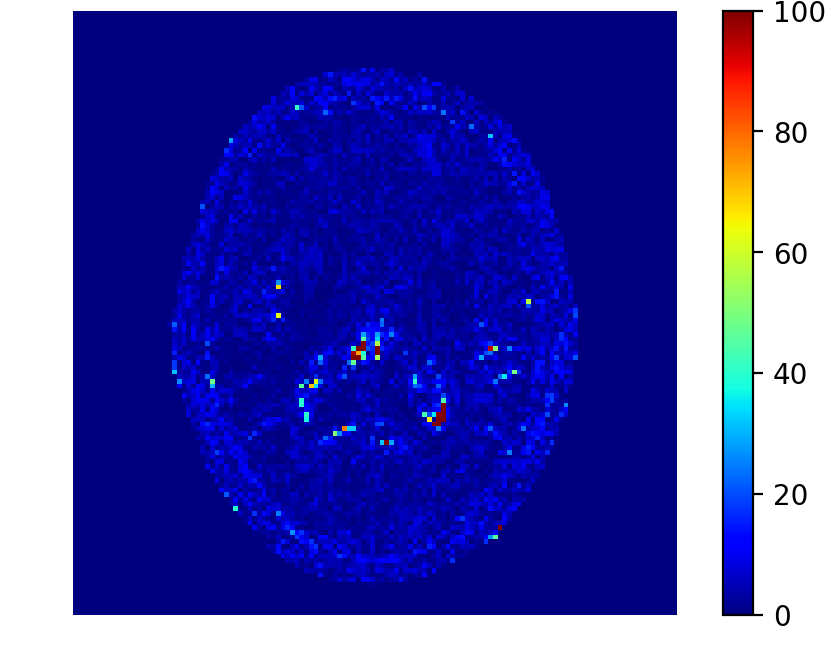}
	\end{minipage} 
	\begin{minipage}[b]{0.19\linewidth}
		\raggedright 
		\includegraphics[width=3.5cm, height=3cm,trim=0cm 0cm 0cm 0cm,clip]{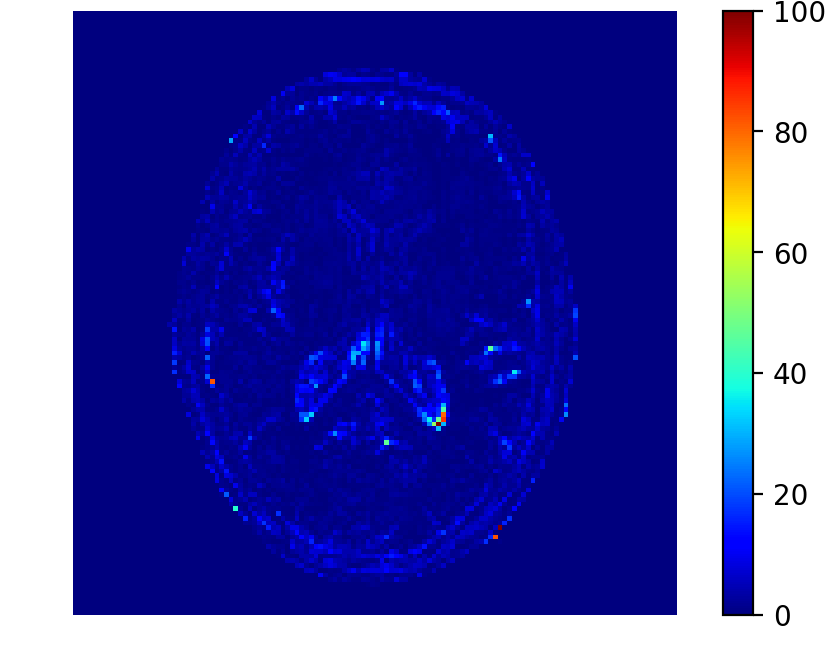}
	\end{minipage} 
	\\
	\begin{minipage}[b]{1\linewidth}
		\centering \footnotesize 
		(b) T2 estimation (top row) and residual errors (bottom row).
	\end{minipage}
	\\
	\vspace{+0.2cm}
	\begin{minipage}[b]{0.19\linewidth}
		\centering \scriptsize 
		BLIP~\cite{davies2014compressed}	
		\\ SNR = 21.06/14.87 dB
	\end{minipage} 
	\begin{minipage}[b]{0.19\linewidth}
		\centering \scriptsize 
		FLOR~\cite{mazor2018low}	
		\\ SNR = 36.44/25.65 dB
	\end{minipage} 
	\begin{minipage}[b]{0.19\linewidth}
		\centering \scriptsize  
		CNN~\cite{hoppe2017deep}				
		\\ SNR = 27.44/15.05 dB
	\end{minipage} 
	\begin{minipage}[b]{0.19\linewidth}
		\centering \scriptsize  
		FNN~\cite{cohen2018mr}
		\\ SNR = 38.17/25.43 dB
	\end{minipage}
	\begin{minipage}[b]{0.19\linewidth}
		\centering \scriptsize  
		HYDRA		
		\\ SNR = 37.86/28.35 dB
	\end{minipage}
	\\
	
	\vspace{-0.3cm}
	
	\caption{
		\footnotesize
		Visual results of testing on anatomical dataset with k-space subsampling factor 15\% using Gaussian pattern with $L=1000$. Comparison between BLIP~\cite{davies2014compressed}, FLOR~\cite{mazor2018low}, CNN by Hoppe et al.~\cite{hoppe2017deep}, FNN by Cohen et al.~\cite{cohen2018mr} and HYDRA. 
	}
	\label{Fig:GaussSubsample0_15_L1000}
\end{figure*}

We also investigated the performance with respect to the number of time frames. In particular, we increased $L$ from 200 to 1000 and kept other experiment settings the same as before. The quantitative results are shown in Table~\ref{Tab:GaussSubsamp_L1000}. It is noticed that given more time frames, all the methods show better performance. Moreover, the performance of learning-based methods, including~CNN~\cite{hoppe2017deep}, FNN~\cite{cohen2018mr} and HYDRA, improve more than model-based techniques~\cite{ma2013magnetic,davies2014compressed,mazor2018low}. In particular, our approach outperforms competing algorithms quantitatively in terms of PSRN, SNR, and RMSE, as well as demonstrates visual advantage, as shown in Figure~\ref{Fig:KSpaceGaussRatio_15} and Figure~\ref{Fig:GaussSubsample0_15_L1000}.

\subsubsection{k-space subsampling using spiral trajectories}

\begin{figure}[t] 
	\centering
	\begin{minipage}[b]{0.24\linewidth}
		\centering
		\includegraphics[width = 3.0cm, trim=0cm 0cm 0cm 0cm,clip ]{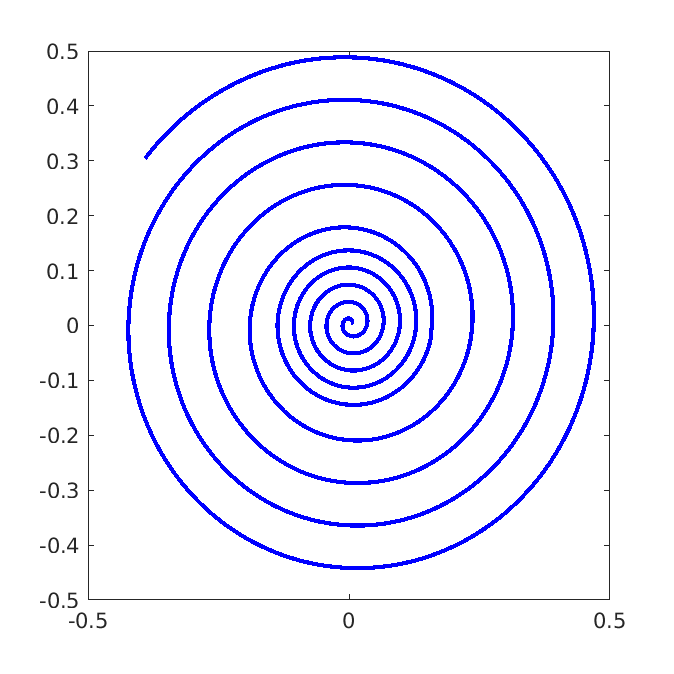}
	\end{minipage} 
	\begin{minipage}[b]{0.24\linewidth}
		\centering
		\includegraphics[width = 3.0cm, trim=0cm 0cm 0cm 0cm,clip ]{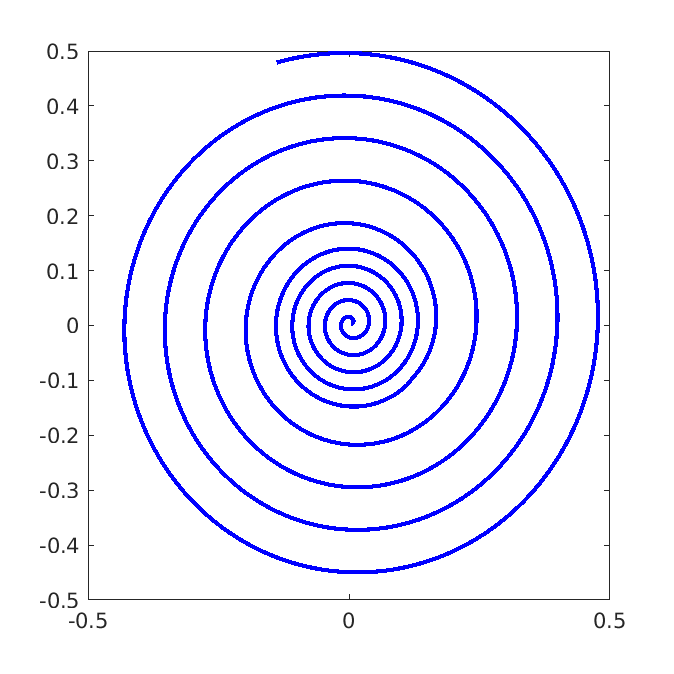}
	\end{minipage} 
	\begin{minipage}[b]{0.24\linewidth}
		\centering
		\includegraphics[width = 3.0cm, trim=0cm 0cm 0cm 0cm,clip ]{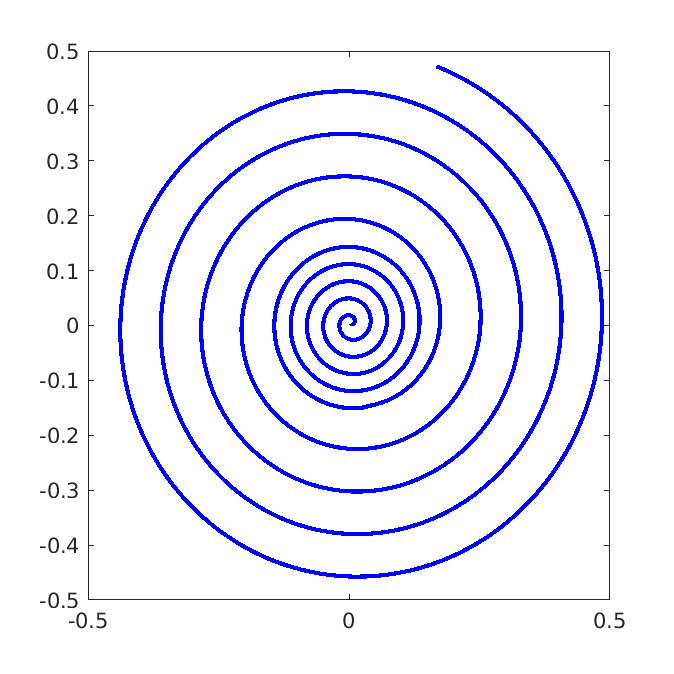}
	\end{minipage} 
	\begin{minipage}[b]{0.24\linewidth}
		\centering
		\includegraphics[width = 3.0cm, trim=0cm 0cm 0cm 0cm,clip ]{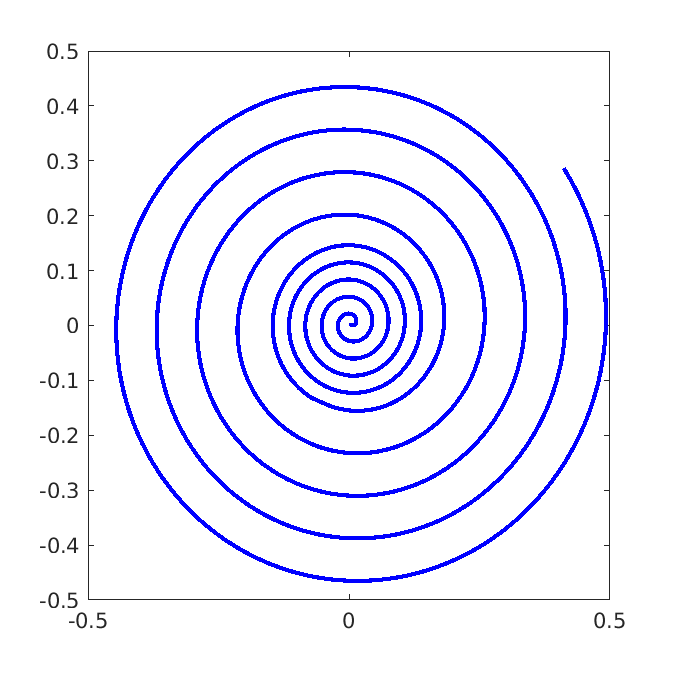}
	\end{minipage} 
	
	\caption{
		A series of spiral trajectories for k-space subsampling.
	}
	\label{Fig:SpiralSamplePatterns}
\end{figure}

We carried out additional experiments with widely used non-Cartesian sampling patterns -- variable density spiral trajectories~\cite{lee2003fast,mazor2018low}. A set of spiral trajectories used in the experiments are shown in Figure~\ref{Fig:SpiralSamplePatterns}. They have FOV of 24 and rotation angle difference of 7.5 degrees between any two adjacent spirals to spread out the alias artifacts. Given such spiral trajectories, data were subsampled to acquire 1488 k-space samples in each time frame, leading to a subsampling ratio of 9\% which is defined by the number of acquired samples in the k-space domain divided by the number of pixels in a frame. This setting closely matches the original MRF paper~\cite{ma2013magnetic} where each single spiral trajectory samples 1450 k-space points (leading to a subsampling ratio around 9\%) and any two adjacent spiral trajectories have a rotation angle of 7.5 degrees. 

\begin{figure*}[tb]
	\begin{multicols}{2}  
		\centering
		\begin{minipage}[b]{0.48\linewidth}
			\centering
			\includegraphics[width = 4cm, height=3.2cm]{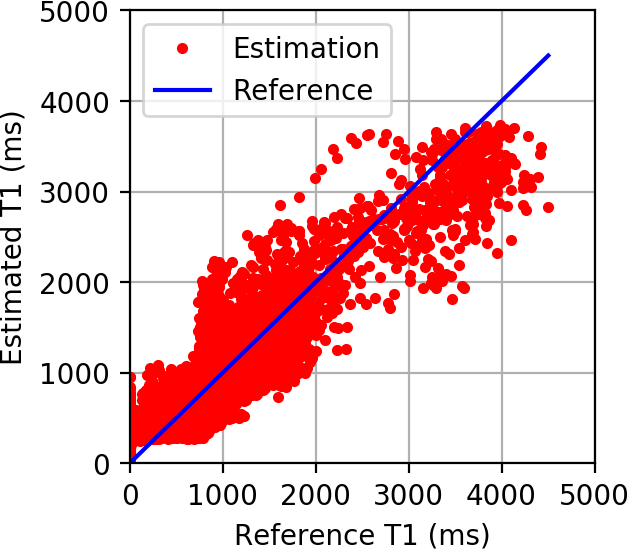}
		\end{minipage} 
		\begin{minipage}[b]{0.48\linewidth}
			\centering
			\includegraphics[width = 4cm, height=3.2cm]{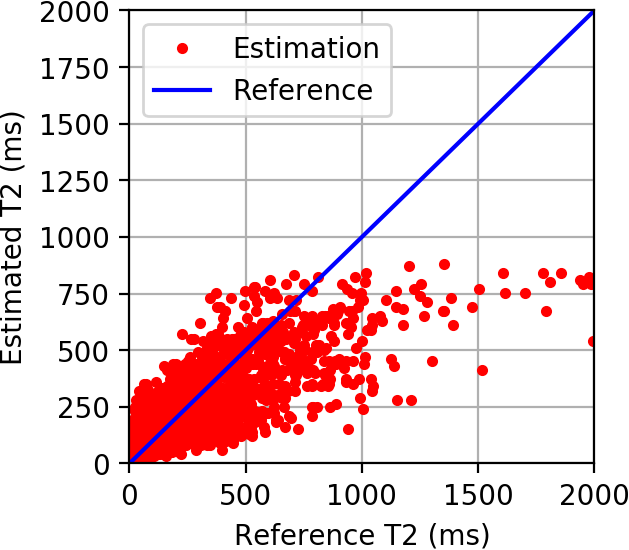}
		\end{minipage}
		\\
		\begin{minipage}[b]{1\linewidth}
			\centering
			\footnotesize (a) T1, T2 estimations using Ma et al.~\cite{ma2013magnetic}. 
		\end{minipage}
		\\
		\vspace{+0.3cm}
		\begin{minipage}[b]{0.48\linewidth}
			\centering
			\includegraphics[width = 4cm, height=3.2cm]{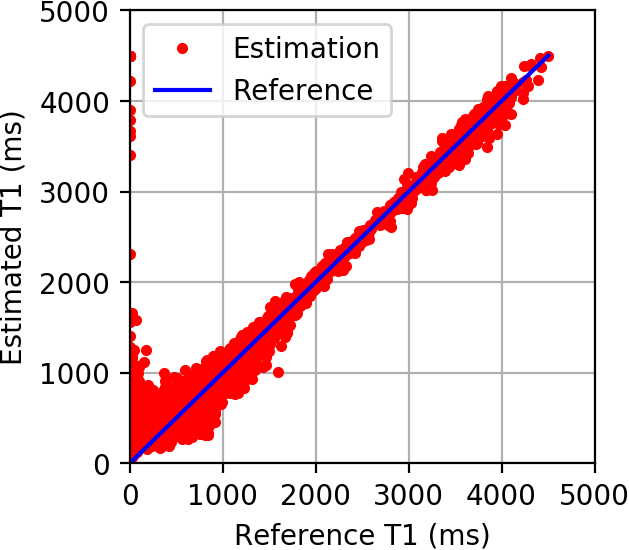}
		\end{minipage}
		\begin{minipage}[b]{0.48\linewidth}
			\centering
			\includegraphics[width = 4cm, height=3.2cm]{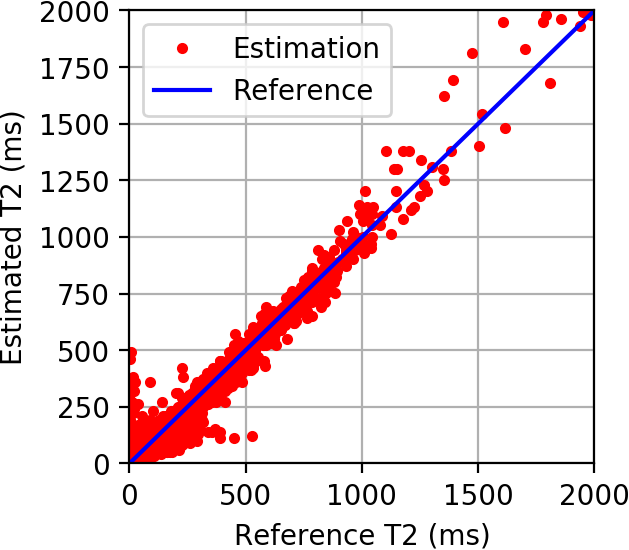}
		\end{minipage}
		\\
		\begin{minipage}[b]{1\linewidth}
			\centering
			\footnotesize (b) T1, T2 estimations using BLIP~\cite{davies2014compressed} 
		\end{minipage}
		\\
		\vspace{+0.3cm}
		\begin{minipage}[b]{0.48\linewidth}
			\centering
			\includegraphics[width = 4cm, height=3.2cm]{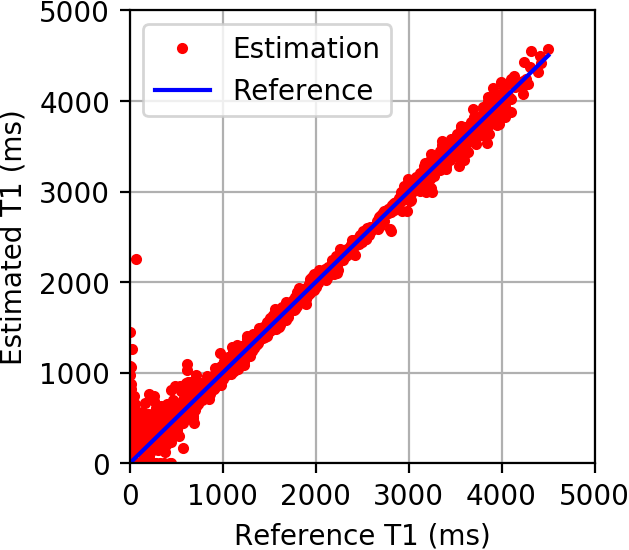}
		\end{minipage}
		\begin{minipage}[b]{0.48\linewidth}
			\centering
			\includegraphics[width = 4cm, height=3.2cm]{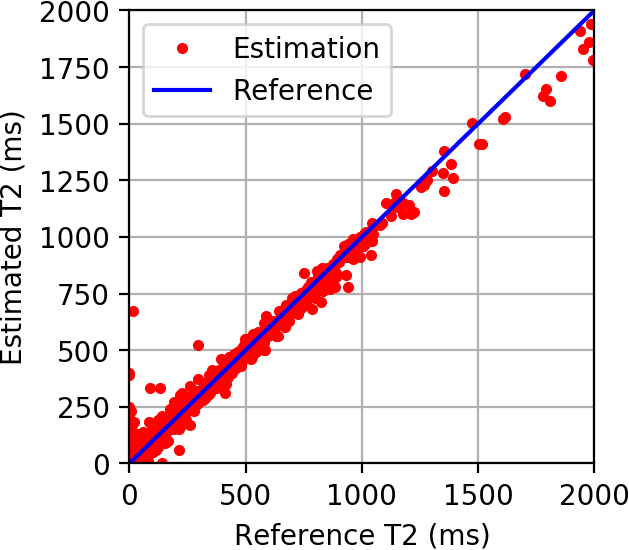}
		\end{minipage}
		\\
		\begin{minipage}[b]{1\linewidth}
			\centering
			\footnotesize (c) T1, T2 estimations using FLOR~\cite{mazor2018low} 
		\end{minipage}
		\\
		\vspace{+0.3cm}
		\begin{minipage}[b]{0.48\linewidth}
			\centering
			\includegraphics[width = 4cm, height=3.2cm]{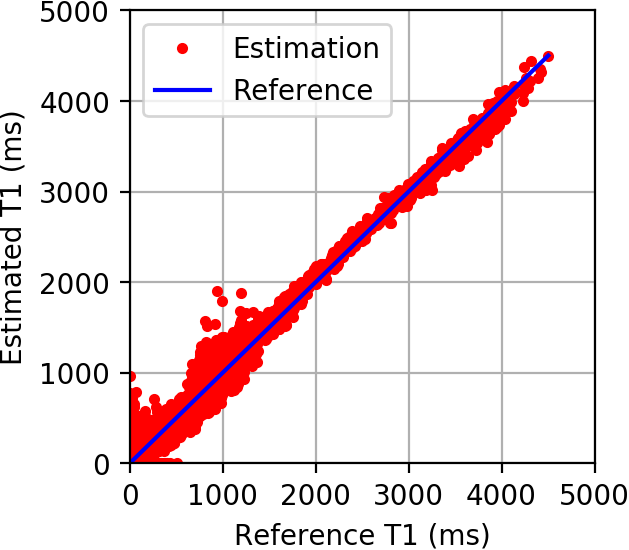}
		\end{minipage}
		\begin{minipage}[b]{0.48\linewidth}
			\centering
			\includegraphics[width = 4cm, height=3.2cm]{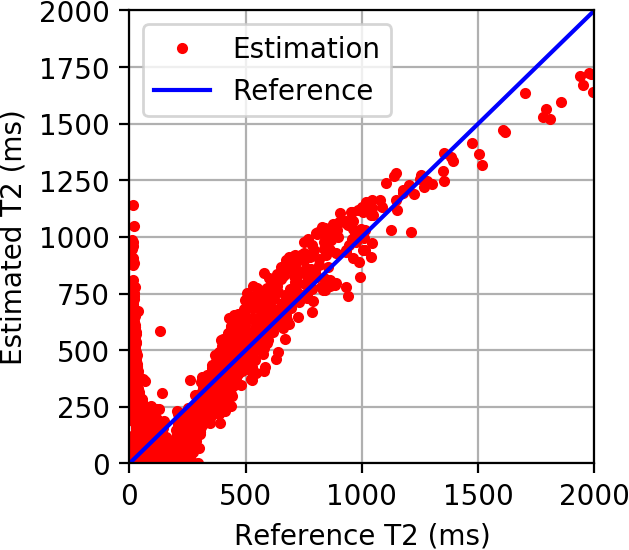}
		\end{minipage}
		\\
		\begin{minipage}[b]{1\linewidth}
			\centering
			\footnotesize (d) T1, T2 estimations using CNN~\cite{hoppe2017deep}.
		\end{minipage}
		\\
		\vspace{+0.3cm}
		\begin{minipage}[b]{0.48\linewidth}
			\centering
			\includegraphics[width = 4cm, height=3.2cm]{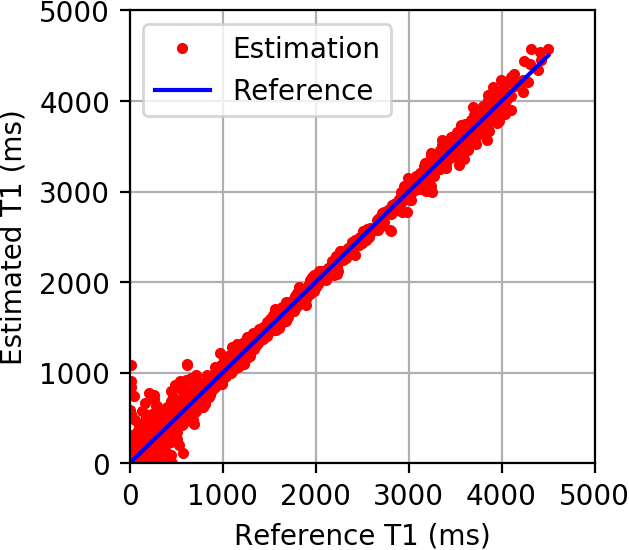}
		\end{minipage}
		\begin{minipage}[b]{0.48\linewidth}
			\centering
			\includegraphics[width = 4cm, height=3.2cm]{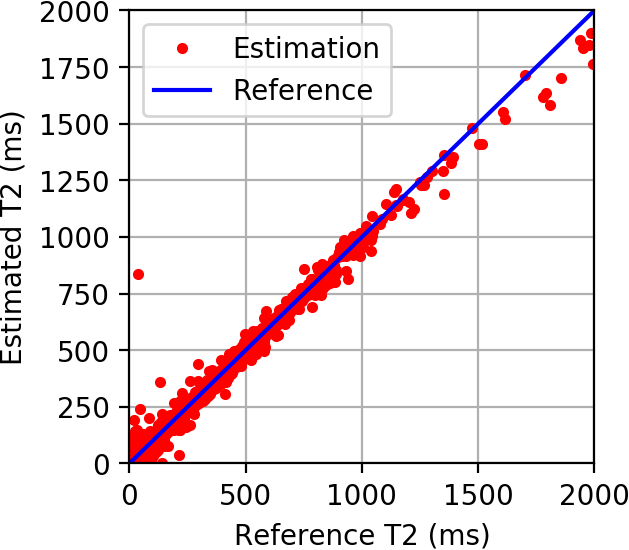}
		\end{minipage}
		\\
		\begin{minipage}[b]{1\linewidth}
			\centering
			\footnotesize (e) T1, T2 estimations using FNN~\cite{cohen2018mr}.
		\end{minipage}
		\\
		\vspace{+0.3cm}
		\begin{minipage}[b]{0.48\linewidth}
			\centering
			\includegraphics[width = 4cm, height=3.2cm]{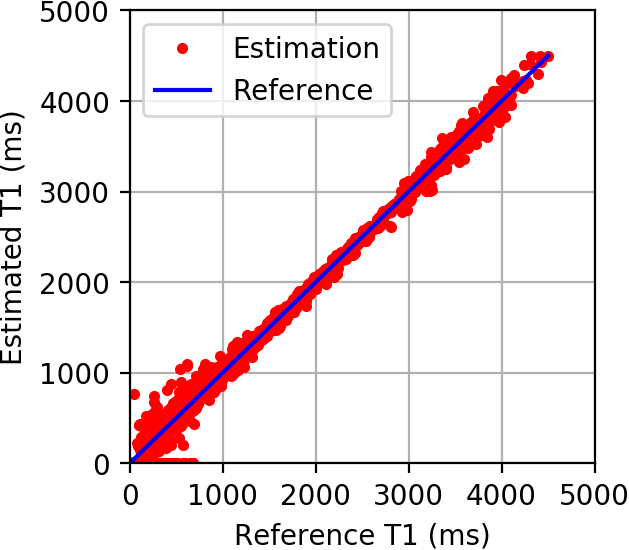}
		\end{minipage}
		\begin{minipage}[b]{0.48\linewidth}
			\centering
			\includegraphics[width = 4cm, height=3.2cm]{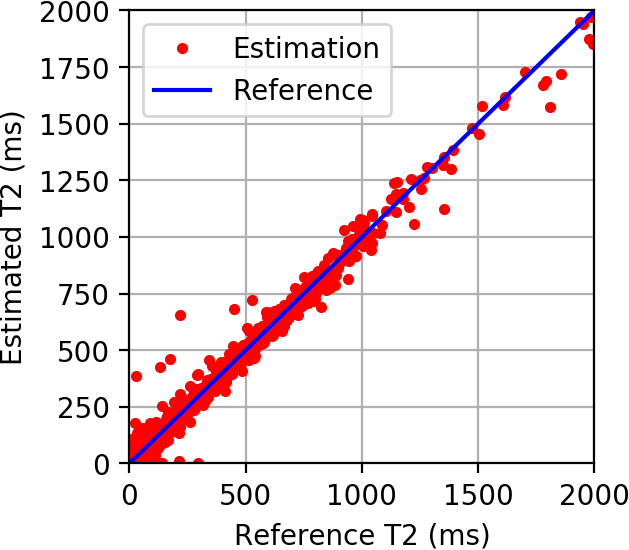}
		\end{minipage}
		\\
		\begin{minipage}[b]{1\linewidth}
			\centering
			\footnotesize (f) T1, T2 estimations using HYDRA.
		\end{minipage}
		\\
		\vspace{+0.3cm}
		
	\end{multicols}
	
	\vspace{-0.3cm}
	
	\caption{
		\footnotesize
		Testing on the anatomical dataset with k-space subsampling factor 9\% using spiral trajectories and 1000 time frames. Subfig. (a) - (f) show the results using Ma et al.~\cite{ma2013magnetic}, BLIP~\cite{davies2014compressed}, FLOR~\cite{mazor2018low}, CNN by Hoppe et al.~\cite{hoppe2017deep}, FNN by Cohen et al.~\cite{cohen2018mr} and HYDRA.
	}
	\label{Fig:KSpaceSpiralRatio_09}
\end{figure*}

\begin{figure*}[t]
	\centering 
	\begin{minipage}[b]{0.3\linewidth}
		\centering 
		\includegraphics[width=3.6cm, height=3cm,trim=0cm 0cm 0cm 0.4cm,clip]{./Figures/DeepMRF_ResNet_CNN/T1_true.png}
	\end{minipage}
	\begin{minipage}[b]{0.3\linewidth}
		\centering 
		\includegraphics[width=3.6cm, height=3cm,trim=0cm 0cm 0cm 0.4cm,clip]{./Figures/DeepMRF_ResNet_CNN/T2_true.png}
	\end{minipage}
	\\
	\begin{minipage}[b]{1\linewidth}
		\centering \footnotesize 
		(a) T1 Reference (left) and T2 Reference (right).
	\end{minipage}
	\\
	\vspace{+0.2cm}
	\begin{minipage}[b]{0.19\linewidth}
		\raggedright 
		\includegraphics[width=3.6cm, height=3cm,trim=0cm 0cm 0cm 0cm,clip]{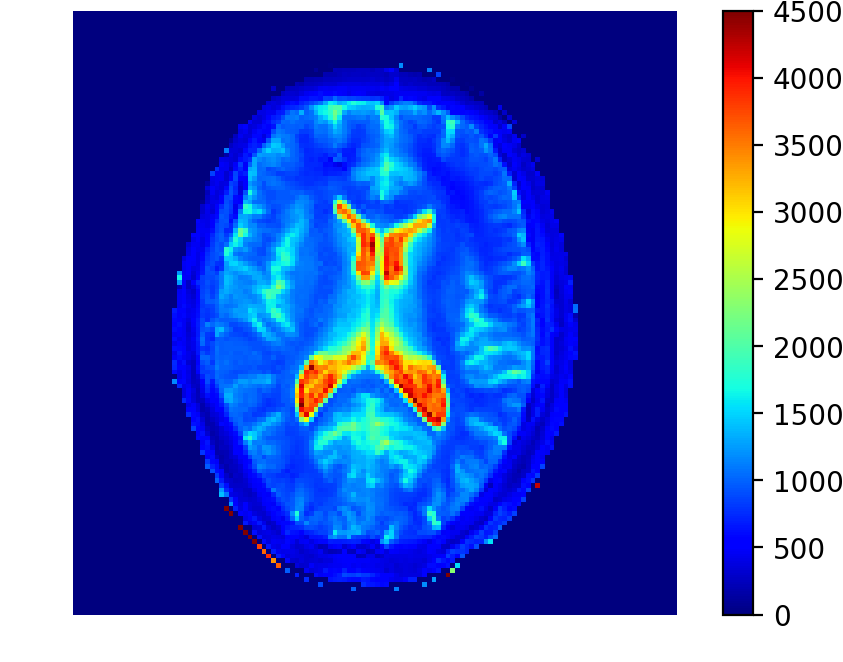}
	\end{minipage} 
	\begin{minipage}[b]{0.19\linewidth}
		\raggedright 
		\includegraphics[width=3.6cm, height=3cm,trim=0cm 0cm 0cm 0cm,clip]{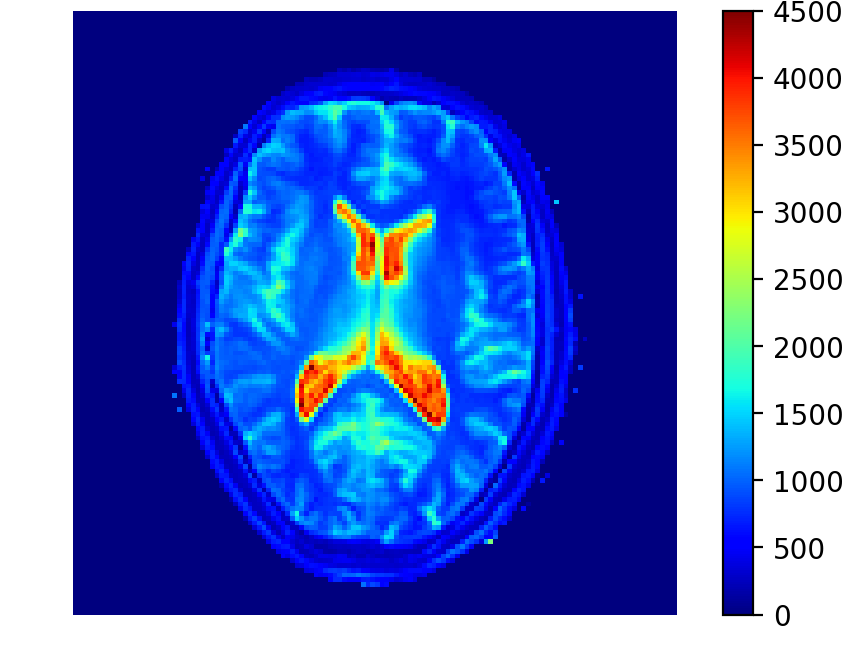}
	\end{minipage} 
	\begin{minipage}[b]{0.19\linewidth}
		\raggedright 
		\includegraphics[width=3.6cm, height=3cm,trim=0cm 0cm 0cm 0cm,clip]{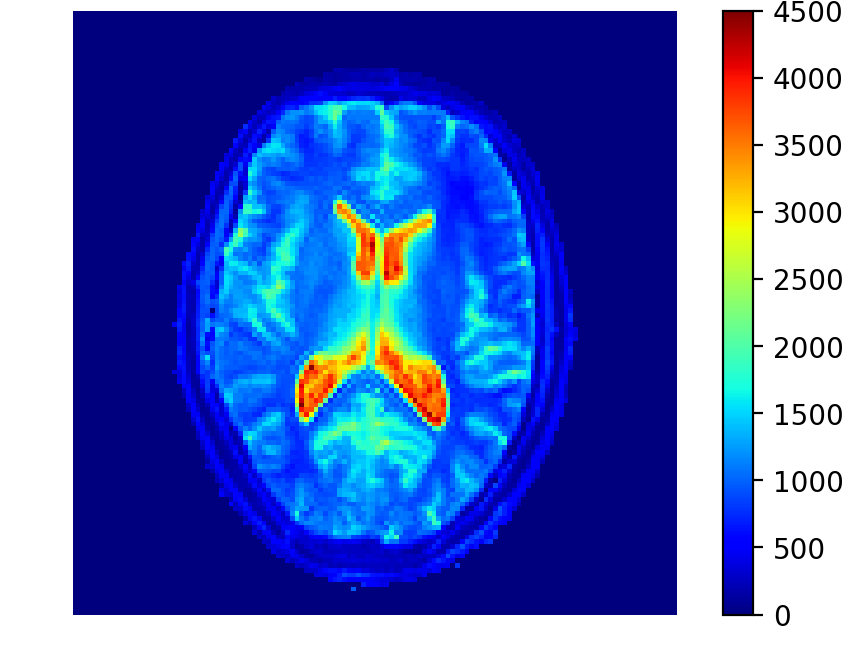}
	\end{minipage} 
	\begin{minipage}[b]{0.19\linewidth}
		\raggedright 
		\includegraphics[width=3.6cm, height=3cm,trim=0cm 0cm 0cm 0cm,clip]{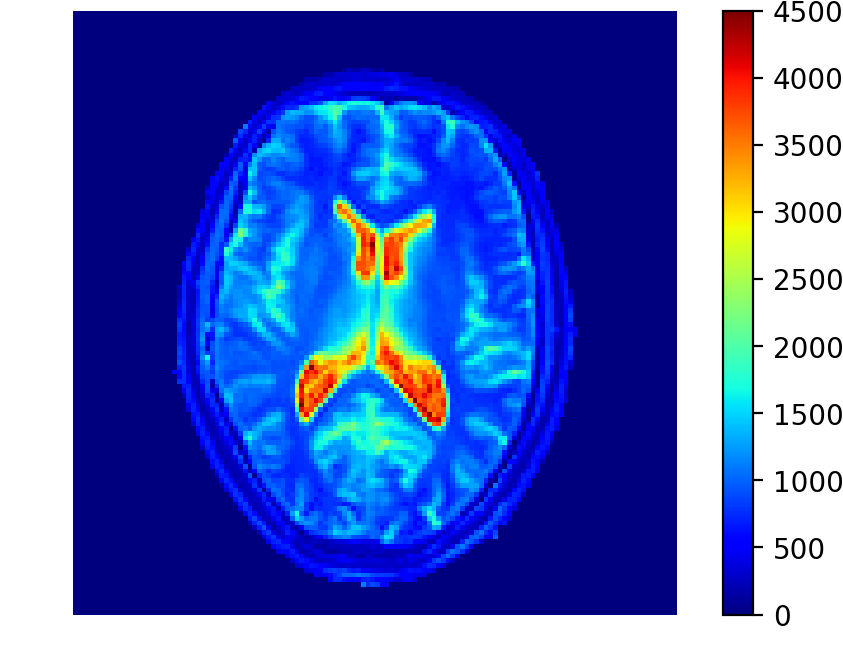}
	\end{minipage} 
	\begin{minipage}[b]{0.19\linewidth}
		\raggedright 
		\includegraphics[width=3.6cm, height=3cm,trim=0cm 0cm 0cm 0cm,clip]{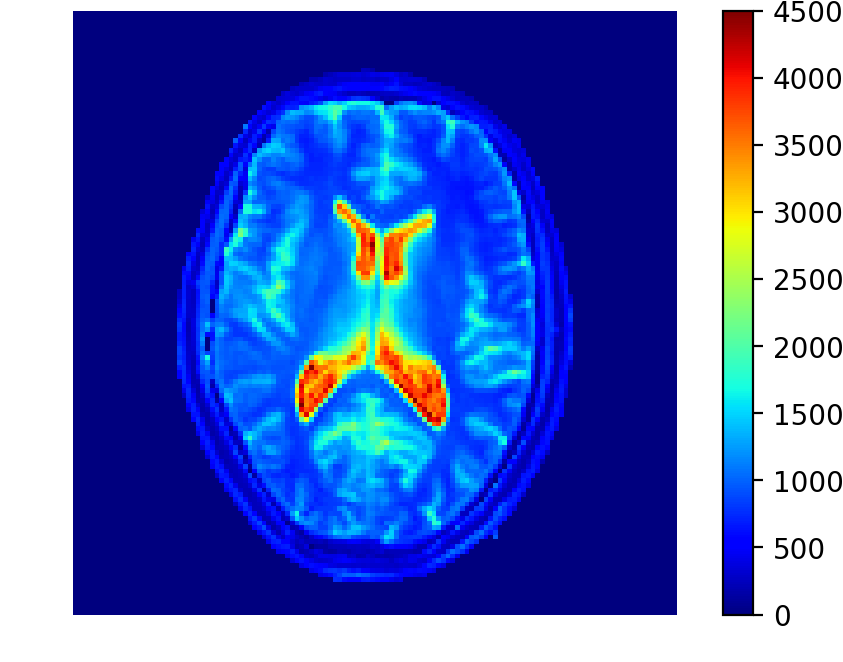}
	\end{minipage} 
	\\
	\begin{minipage}[b]{0.19\linewidth}
		\raggedright 
		\includegraphics[width=3.5cm, height=3cm,trim=0cm 0cm 0cm 0cm,clip]{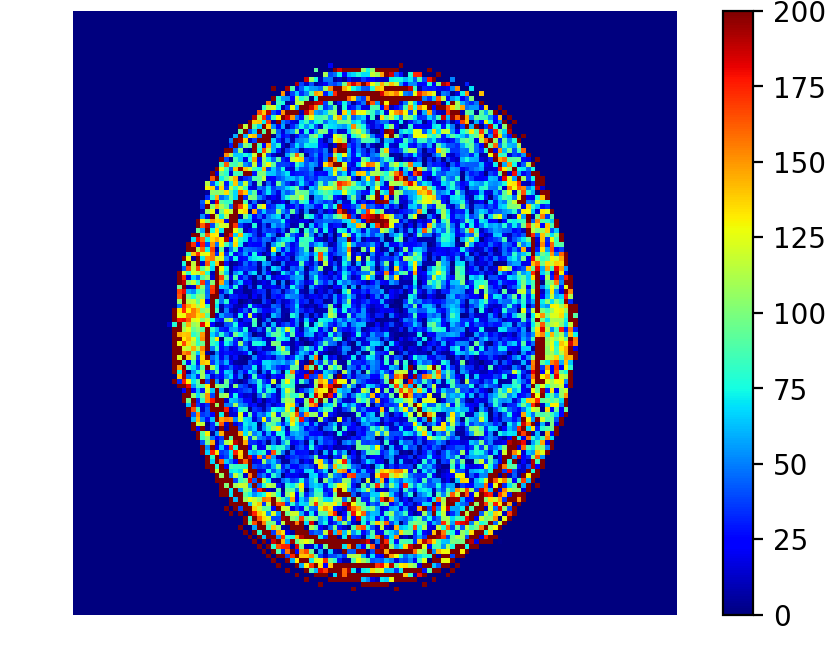}
	\end{minipage} 
	\begin{minipage}[b]{0.19\linewidth}
		\raggedright 
		\includegraphics[width=3.5cm, height=3cm,trim=0cm 0cm 0cm 0cm,clip]{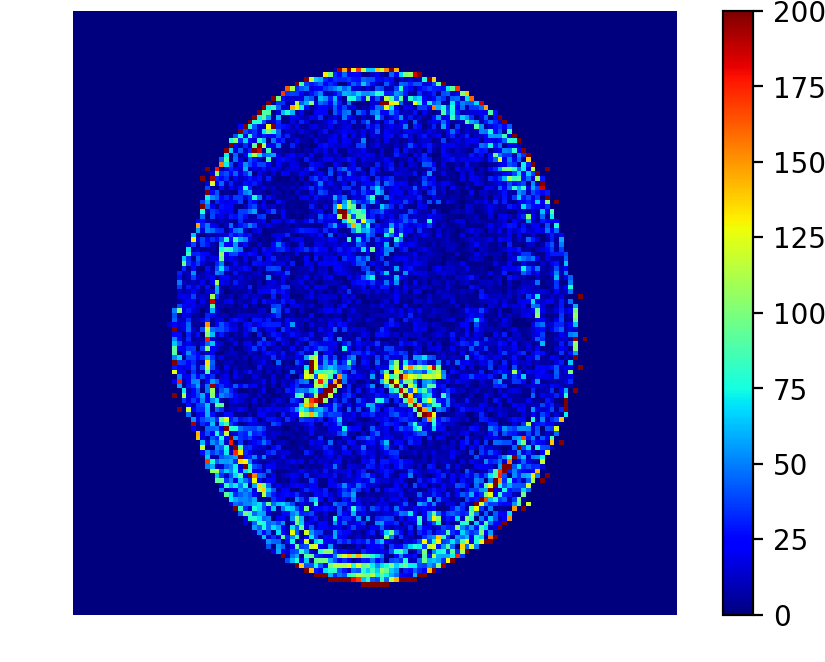}
	\end{minipage} 
	\begin{minipage}[b]{0.19\linewidth}
		\raggedright 
		\includegraphics[width=3.6cm, height=3cm,trim=0cm 0cm 0cm 0cm,clip]{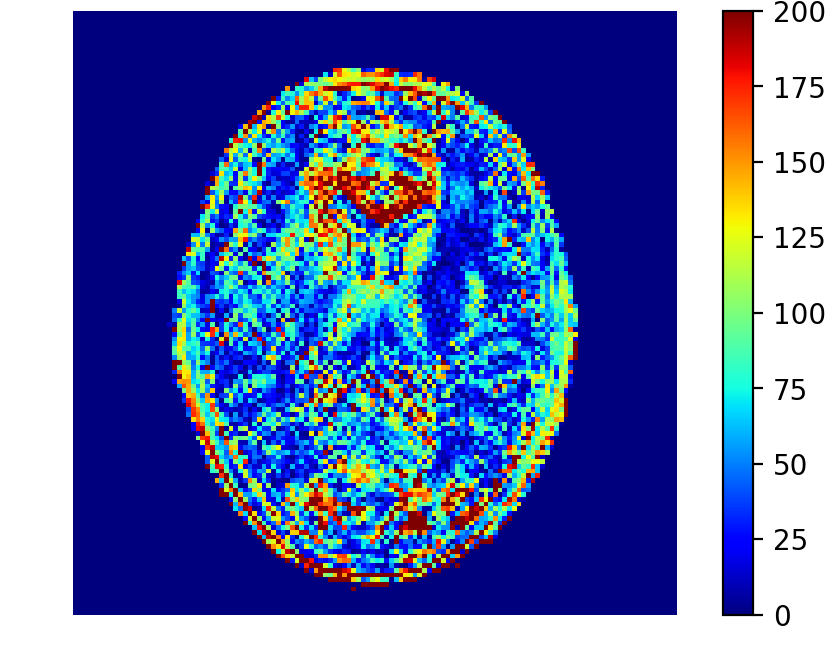}
	\end{minipage} 
	\begin{minipage}[b]{0.19\linewidth}
		\raggedright 
		\includegraphics[width=3.6cm, height=3cm,trim=0cm 0cm 0cm 0cm,clip]{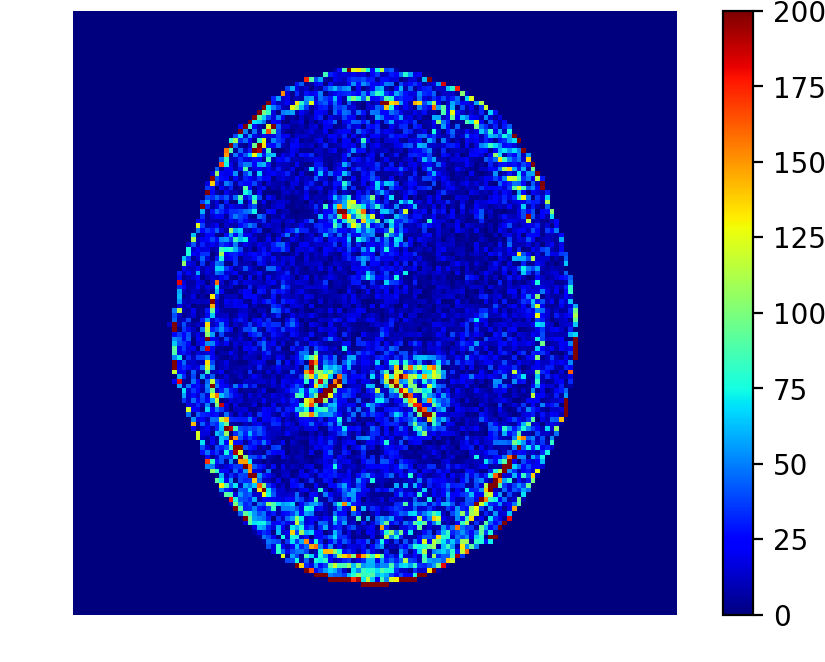}
	\end{minipage} 
	\begin{minipage}[b]{0.19\linewidth}
		\raggedright 
		\includegraphics[width=3.5cm, height=3cm,trim=0cm 0cm 0cm 0cm,clip]{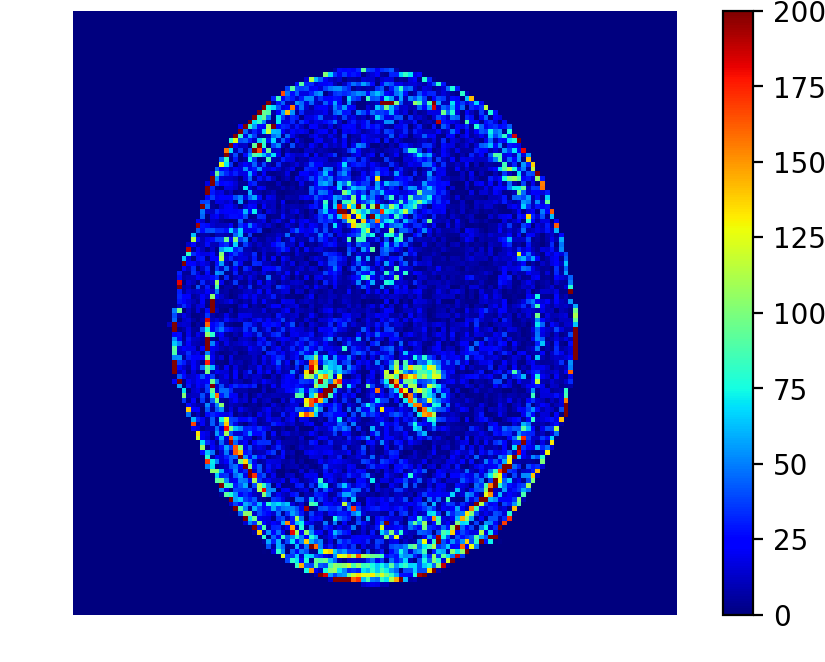}
	\end{minipage} 
	\\
	\begin{minipage}[b]{1\linewidth}
		\centering \footnotesize 
		(b) T1 estimation (top row) and residual errors (bottom row).
	\end{minipage}
	\\
	\vspace{+0.2cm}
	\begin{minipage}[b]{0.19\linewidth}
		\raggedright 
		\includegraphics[width=3.6cm, height=3cm,trim=0cm 0cm 0cm 0cm,clip]{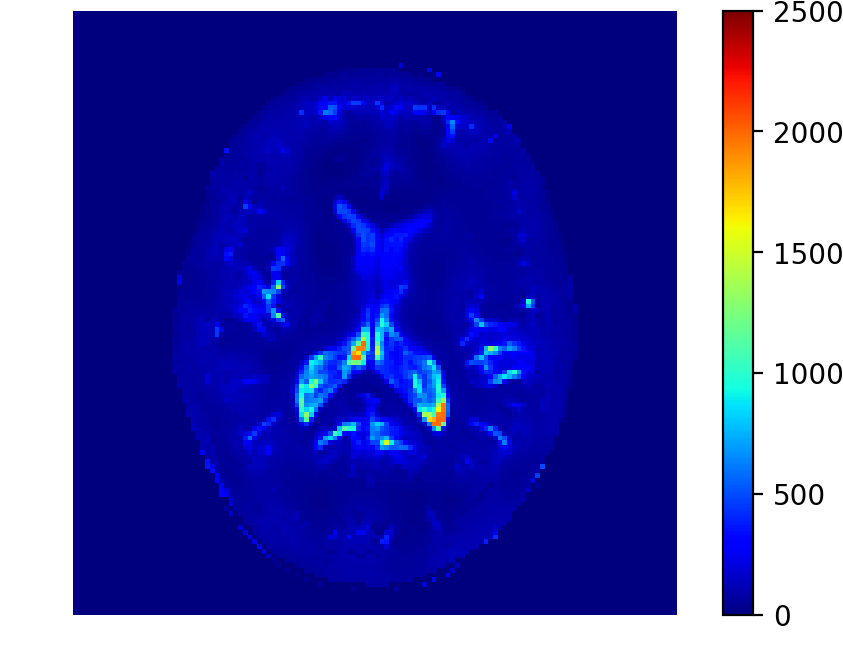}
	\end{minipage} 
	\begin{minipage}[b]{0.19\linewidth}
		\raggedright 
		\includegraphics[width=3.6cm, height=3cm,trim=0cm 0cm 0cm 0cm,clip]{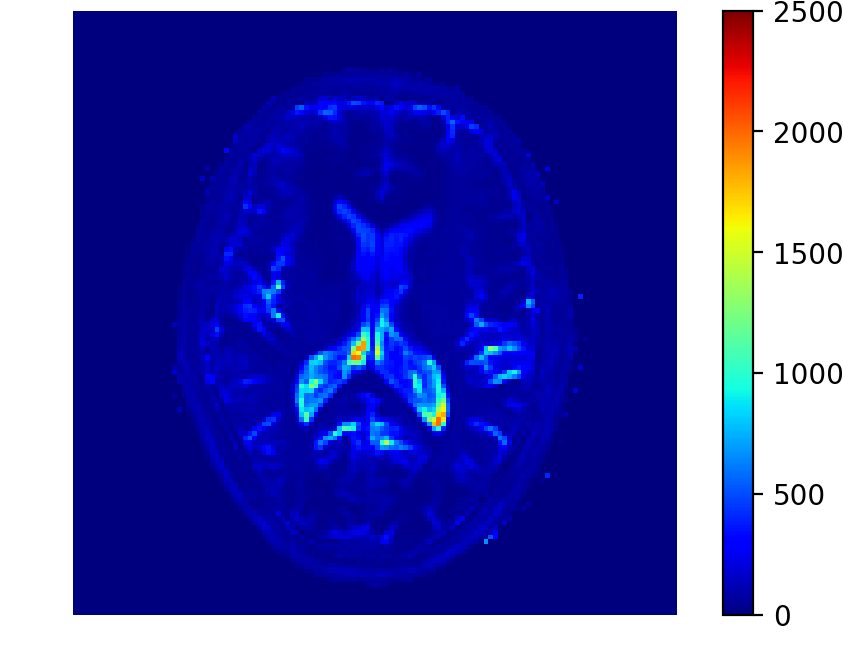}
	\end{minipage} 
	\begin{minipage}[b]{0.19\linewidth}
		\raggedright 
		\includegraphics[width=3.6cm, height=3cm,trim=0cm 0cm 0cm 0cm,clip]{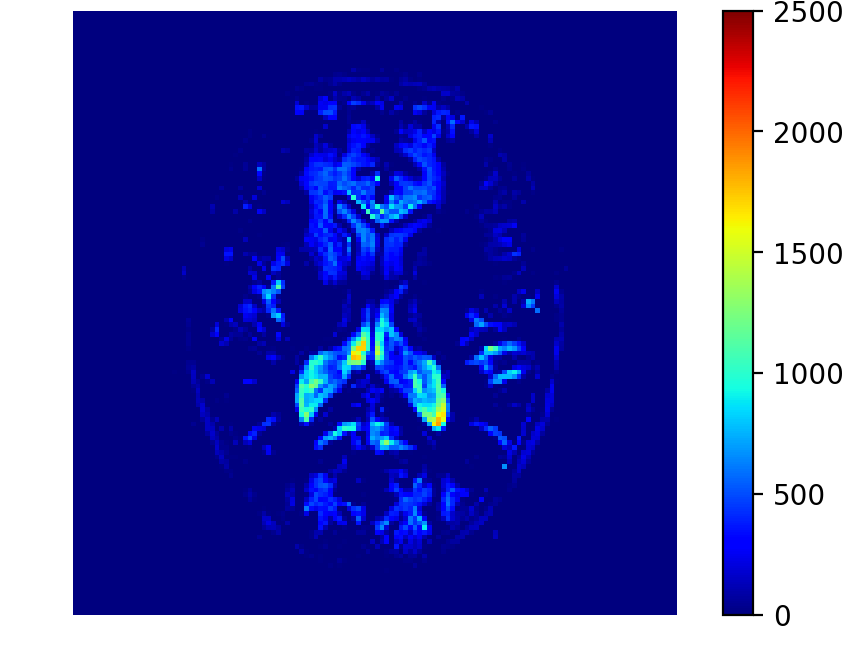}
	\end{minipage} 
	\begin{minipage}[b]{0.19\linewidth}
		\raggedright 
		\includegraphics[width=3.6cm, height=3cm,trim=0cm 0cm 0cm 0cm,clip]{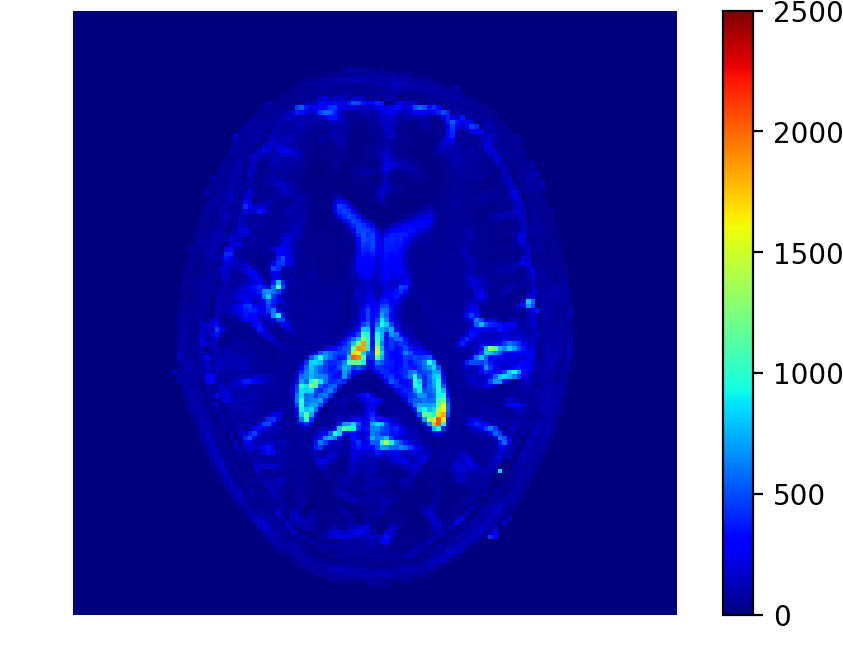}
	\end{minipage} 
	\begin{minipage}[b]{0.19\linewidth}
		\raggedright 
		\includegraphics[width=3.6cm, height=3cm,trim=0cm 0cm 0cm 0cm,clip]{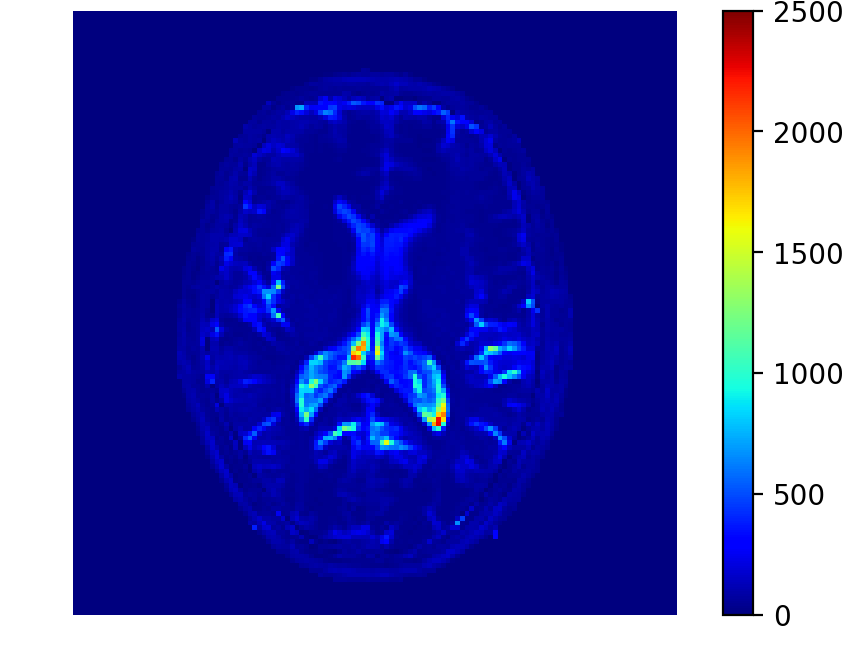}
	\end{minipage} 
	\\
	\begin{minipage}[b]{0.19\linewidth}
		\raggedright 
		\includegraphics[width=3.5cm, height=3cm,trim=0cm 0cm 0cm 0cm,clip]{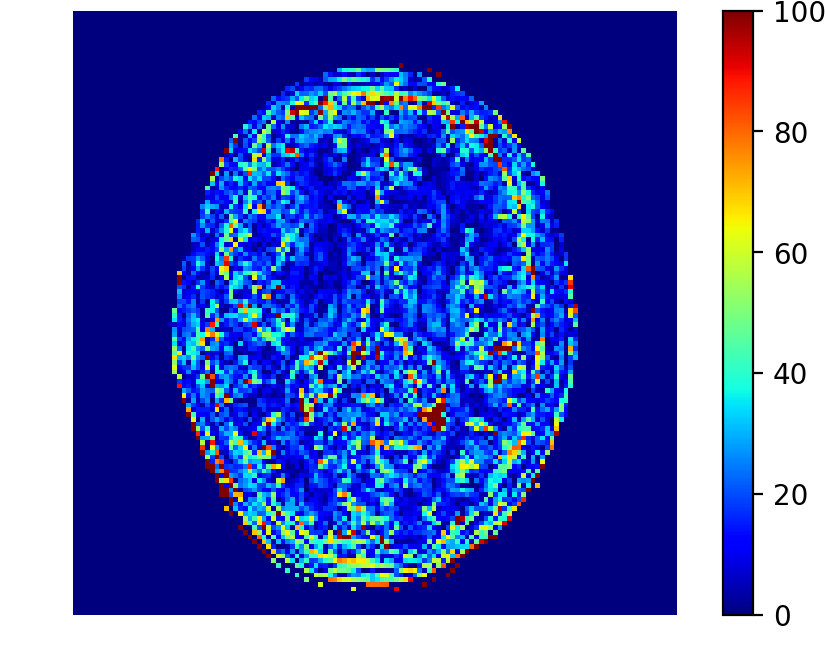}
	\end{minipage} 
	\begin{minipage}[b]{0.19\linewidth}
		\raggedright 
		\includegraphics[width=3.5cm, height=3cm,trim=0cm 0cm 0cm 0cm,clip]{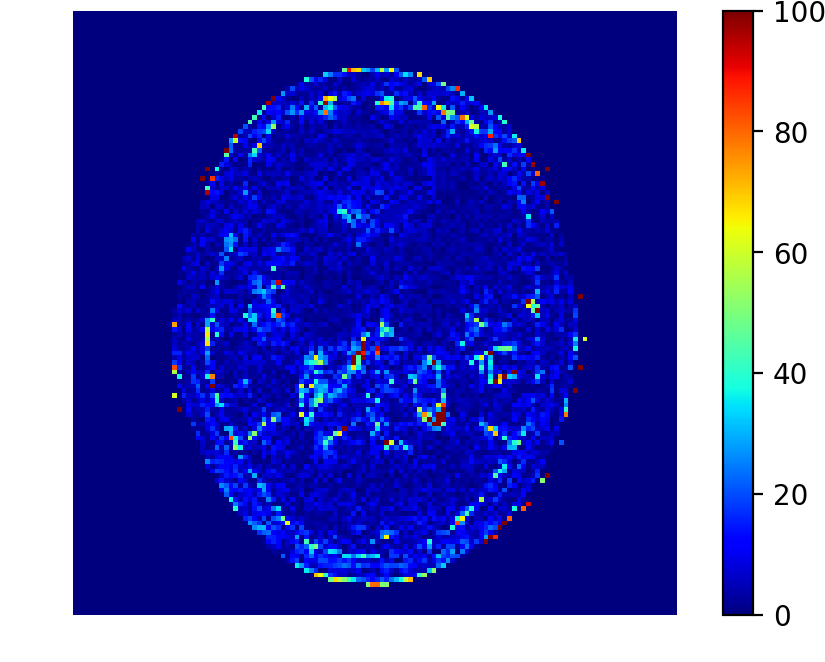}
	\end{minipage} 
	\begin{minipage}[b]{0.19\linewidth}
		\raggedright 
		\includegraphics[width=3.6cm, height=3cm,trim=0cm 0cm 0cm 0cm,clip]{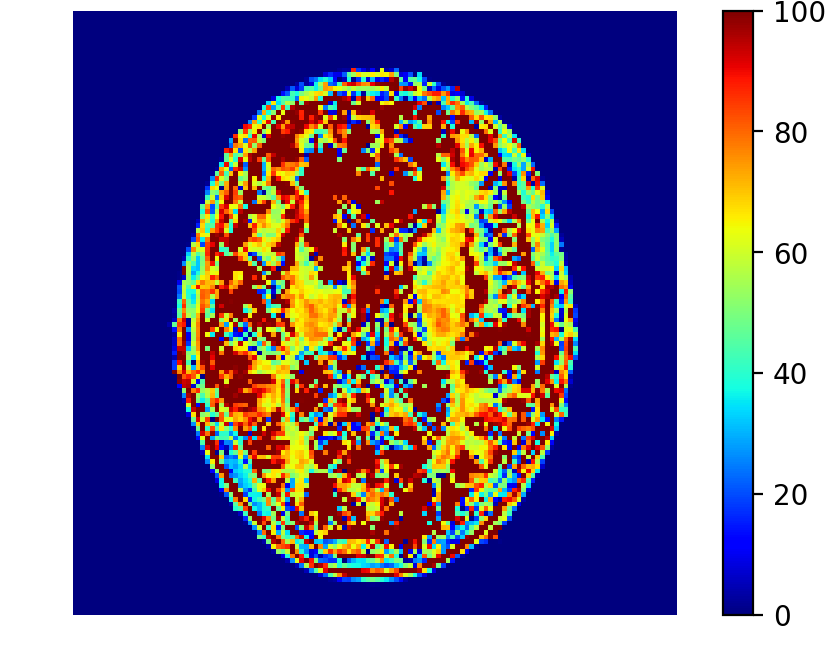}
	\end{minipage} 
	\begin{minipage}[b]{0.19\linewidth}
		\raggedright 
		\includegraphics[width=3.6cm, height=3cm,trim=0cm 0cm 0cm 0cm,clip]{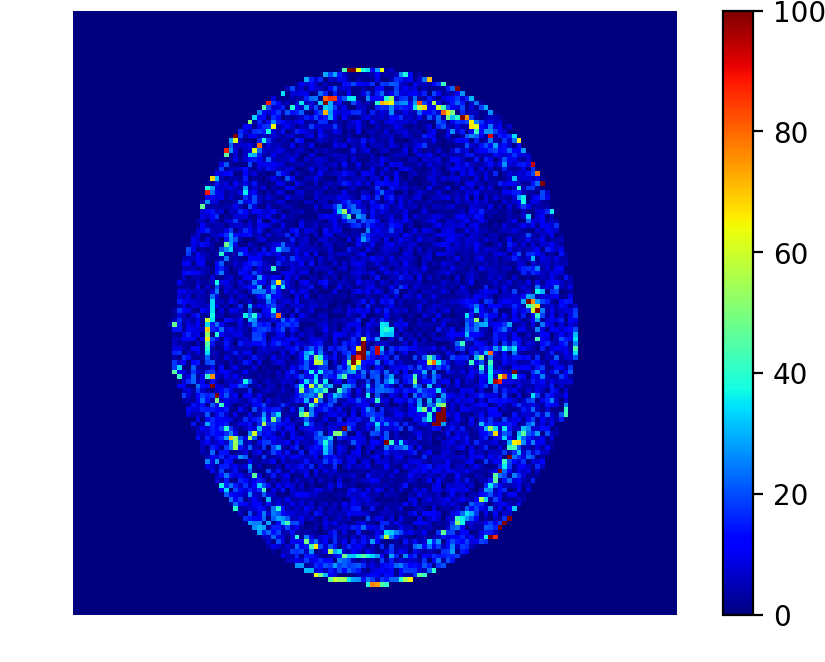}
	\end{minipage} 
	\begin{minipage}[b]{0.19\linewidth}
		\raggedright 
		\includegraphics[width=3.5cm, height=3cm,trim=0cm 0cm 0cm 0cm,clip]{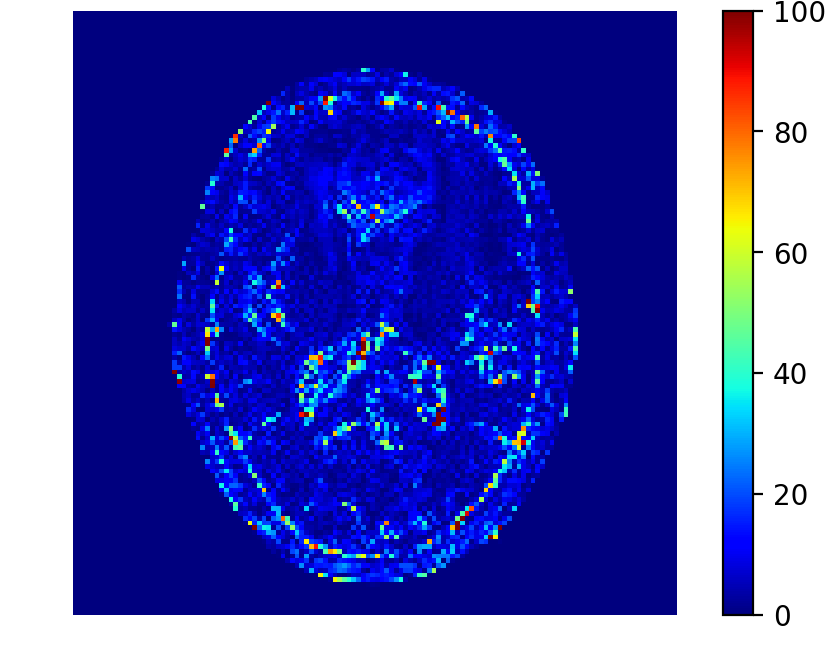}
	\end{minipage} 
	\\
	\begin{minipage}[b]{1\linewidth}
		\centering \footnotesize 
		(c) T2 estimation (top row) and residual errors (bottom row).
	\end{minipage}
	\\
	\vspace{+0.2cm}
	\begin{minipage}[b]{0.19\linewidth}
		\centering \scriptsize 
		BLIP~\cite{davies2014compressed}	
		\\ SNR = 15.03/15.22 dB		
	\end{minipage} 
	\begin{minipage}[b]{0.19\linewidth}
		\centering \scriptsize 
		FLOR~\cite{mazor2018low}	
		\\ SNR = 24.88/20.38 dB
	\end{minipage} 
	\begin{minipage}[b]{0.19\linewidth}
		\centering \scriptsize  
		CNN~\cite{hoppe2017deep}				
		\\ SNR = 21.45/4.48 dB
	\end{minipage} 
	\begin{minipage}[b]{0.19\linewidth}
		\centering \scriptsize  
		FNN~\cite{cohen2018mr}
		\\ SNR = 25.84/20.30 dB
	\end{minipage}
	\begin{minipage}[b]{0.19\linewidth}
		\centering \scriptsize  
		HYDRA		
		\\ SNR = 27.02/21.04 dB
	\end{minipage}
	\\
	
	\caption{
		Visual results of testing on anatomical dataset with k-space subsampling factor 9\% using spiral trajectories with $L=1000$. Comparison between BLIP~\cite{davies2014compressed}, FLOR~\cite{mazor2018low}, CNN by Hoppe et al.~\cite{hoppe2017deep}, FNN by Cohen et al.~\cite{cohen2018mr} and HYDRA. 
	}
	\label{Fig:SpiralSubsample0_09_L1000}
\end{figure*}

In the case of spiral subsampling, during the signature restoration, SParse Uniform ReSampling (SPURS) algorithm~\cite{kiperwas2017spurs} was exploited to implement nonuniform Fourier transform between k-space domain and image domain, as SPURS has proved to achieve smaller approximation errors while maintaining low computational cost comparing with other resampling methods, such as nonuniform-FFT algorithm~\cite{fessler2003nonuniform} and regularized Block Uniform ReSampling (rBURS)~\cite{rosenfeld2002new}. In the experiments, 1000 density variable spiral trajectories were used for k-space subsampling, leading to 1000 time frames. The quantitative and qualitative reconstruction results demonstrate that HYDRA outperforms competing methods with smaller estimation errors, as shown in Table~\ref{Tab:SpiralSubsamp0_09_L1000}, Figure~\ref{Fig:SpiralSubsample0_09_L1000} and Figure~\ref{Fig:KSpaceSpiralRatio_09}. 

\vspace{-0.3cm}

\section{Discussion}
\label{sec:Discussion}

\vspace{-0.3cm}

\subsection{Relation to previous works}

\vspace{-0.3cm}

Our low-rank based signature restoration method is adapted from FLOR~\cite{mazor2018low} by removing the operation of projecting the temporal signal onto a dictionary. Thus, the signature restoration does not require a simulated dictionary, and saves computational cost. Although recent works~\cite{cohen2018mr,hoppe2017deep} exploit neural networks to perform parameter mapping, replacing dictionaries and lookup-tables used in conventional MRF reconstruction approaches, our technique is different from these methods~\cite{cohen2018mr,hoppe2017deep}. We design a deep nonlocal residual CNN for capturing signature-to-parameter mapping which is organically combined with low-rank based de-aliasing techniques for signature restoration. In this way, our algorithm can bypass some of the issues associated with other techniques: 
(1) The input dimension issue. The proposed approach can ingest temporal signatures with different lengths without the need to change the structure of the network. This is due to the fact that we rely on convolutional neural networks (CNNs) rather than fully-connected neural networks (FNNs) such as the model used in \cite{cohen2018mr}. 
(2) The k-space subsampling issue. The proposed approach involves a hybrid of a neural network with a low-rank based de-aliasing approach. Thus it is able to deal with correlations both over time and space via exploiting low-rank regularization and convolution operation. This enables our work to handle k-space subsampling situations.
(3) The complex mappings issue. By exploiting a residual network structure, our method can be successfully extended to deeper levels and thus obtain a better capacity to learn complex signature-to-parameter mapping functions.
(4) Distortion and corruption issue. Due to the subsampling in k-space, the restored temporal signatures suffer from local distortion and corruption. Such deviation may lead to performance degeneration in the second stage. By incorporating non-local operations in the network design, our method is able to capture global features and find most relevant components for inference, thereby reducing interference of local distortion and corruption.

\vspace{-0.3cm}

\subsection{Computational complexity}

\vspace{-0.3cm}

HYDRA involves two main stages: the low-rank based signature restoration stage and the network based parameter restoration stage. Even though the time cost for parameter restoration is longer than previous methods~\cite{cohen2018mr,hoppe2017deep}, the time cost in the this stage is only a small fraction of the total time consumption, as the computational complexity is dominated by the signature restoration stage. In other words, the computational burden of HYDRA lies in the SVD calculation in the first stage. Hence, fast SVD methods can be employed to dramatically improve the efficiency of signature restoration.

\subsection{Model storage requirements}

Regarding the storage requirement (in double precision), HYDRA needs only 2.1 megabytes to store the network with 0.5 million parameters, while it requires 108 megabytes to store a simulated dictionary of size $80100 \times 200$ and 551 megabytes for size $80100 \times 1000$. Note that the dictionary volume will grow exponentially with the number of parameters, but the space required for storing a network is not strictly limited by the dictionary density once the topology of the network is fixed, thus significantly alleviating the storage burden inherent to the exponential growth of multi-dimensional dictionaries.

\subsection{Impact of providing continuous T1/T2 values}

Providing continuous T1/T2 values is an advantage of neural network based parameter mapping over dictionary matching. This property may find promising applications in some practical scenarios, for exampling, monitoring sensitive changing of pathology condition over time, such as multiple sclerosis~\cite{manfredonia2007normal,papadopoulos2010ti}, stroke~\cite{bernarding2000histogram}, and treatment responses~\cite{mcsheehy2010quantified,weidensteiner2014tumour}, where the differences in T1 and T2 values between healthy and diseased tissues or between disease stages could be very small~\cite{jiang2017repeatability}. On the other hand, to fulfil this potential of network based MRF techniques, prerequisites on the accuracy and precision of MRI measurements are needed. Taking T1/T2 quantification as an example, even for the inversion recovery spin echo (IR-SE) / multiple single-echo spin echo MRI sequences which are considered as the gold standard for T1/T2 quantification, there exist variations of 2\% - 9\% on the measured relaxation times~\cite{jiang2017repeatability}. Such anatomical measurement uncertainties and model imperfections may weaken the advantage and clinical impact of providing continuous T1/T2 values using network based MRF techniques to some extent. Therefore, improving the accuracy of gold standard approaches in the future would contribute to making the most of the potential of neural networks in the MRF domain.

\section{Conclusion}
\label{sec:Conclusion}

\vspace{-0.2cm}

We proposed a hybrid deep MRF approach which combines low-rank based signature restoration with learning-based parameter restoration. In our approach, a low-rank based de-aliasing method is used to restore clean signatures from subsampled k-space measurements. Then, a 1D deep nonlocal residual CNN is developed for efficient signature-to-parameter mapping, replacing the time-consuming dictionary matching operation in conventional MRF techniques. Our approach demonstrates advantages in terms of inference speed, accuracy and storage requirements over competing MRF methods as no dictionary is needed for recovery.


\clearpage

\begin{table}[t]
	\centering
	\small 
	\caption{
		Brief description of experiment types and settings.
	}
	\begin{tabular}{p{0.2\columnwidth}|p{0.7\columnwidth} }
		\hline \hline
		Experiment & Settings \\
		\hline
		Training & 
		\tabincell{l}{
			Input: $\mathbf{D}$, size $K \times L = 80100 \times 200$. \\
			Groundtruth: $\mathbf{LUT}$, size $80100 \times 2$. \\
			k-space subsampling factor $\beta$: not available.\\
		}
		\\
		\hline
		\tabincell{l}{Testing on \\ synthetic data} & 
		\tabincell{l}{
			Input: $\bX$, size $N \times L = 80000 \times 200$. \\
			Groundtruth: $\boldsymbol{\Theta}^{T12}$, size $80000 \times 2$. \\
			k-space subsampling factor $\beta$: not available.\\
		}
		\\
		\hline
		\tabincell{l}{Testing on \\ anatomical data} &
		\tabincell{l}{
			Input: $\bY$, size $Q \times L = 16384 \beta \times 200$ or $16384 \beta \times 1000$. \\
			Reference: $\boldsymbol{\Theta}^{T12}$, size $N \times 2 = 16384 \times 2$. \\
			k-space subsampling factor $\beta$: 70\%, 15\% using Gaussian patterns.\\
		}
		\\
		\hline
		\tabincell{l}{Testing on \\ anatomical data} &
		\tabincell{l}{
			Input: $\bY$, size $Q \times L = 16384 \beta \times 1000$. \\
			Reference: $\boldsymbol{\Theta}^{T12}$, size $N \times 2 = 16384 \times 2$. \\
			k-space subsampling factor $\beta$: 9\% using spiral trajectories. \\
		}
		\\
		\hline \hline
	\end{tabular}
	\label{Tab:ExperimentsTypes}
\end{table}

\begin{table}[t]
	\centering
	\footnotesize
	\caption{
		Testing on synthetic dataset. Comparing parameter restoration performance, in terms of PSNR, SNR, RMSE and correlation coefficient.%
	}
	\begin{tabular}{p{0.14\columnwidth}| c |c |c |c}
		\hline \hline
		& Dict. Match. & CNN~\cite{hoppe2017deep} & FNN~\cite{cohen2018mr} & Proposed  \\
		& T1  /  T2   & T1  /  T2  & T1  /  T2  & T1  /  T2  \\
		\hline
		PSNR (dB) & 59.15 / 52.31 & 62.96 / 49.64 & 58.97 / 54.96 & \textbf{79.30 / 72.99} \\
		SNR (dB) & 55.23 / 47.15 & 59.05 / 44.49 & 55.06 / 49.81 & \textbf{75.38 / 67.83} \\
		RMSE (ms) & 5.515 / 4.847 & 3.554 / 6.591 & 5.63 / 3.57 & \textbf{0.542 / 0.448} \\
		CorrCoef & 1.00 / 1.00 & 1.00 / 1.00 & 1.00 / 1.00 & {1.00} / {1.00} \\
		time cost (s) & 464.10 & 2.87 & \textbf{1.58} & 8.2 \\
		\hline \hline
	\end{tabular}
	\label{Tab:TestingSynthetic}
\end{table}

\begin{table}[t]
	\centering
	\footnotesize
	\caption{
		Testing on synthetic dataset involving detailed T1 / T2 examples that are not on the training grid and their intervals are much smaller than the training grid intervals. D.M. denotes dictionary matching. T1 and T2 errors are defined as the difference between estimated values and groundtruth values.
	}
	\begin{tabular}{c| c |c |c |c| c |c |c |c}
		\hline \hline
		& \multicolumn{4}{c|}{T1 Estimation} & \multicolumn{4}{c}{T1 Errors} \\
		\hline
		Truth & D.M. & \cite{hoppe2017deep} & \cite{cohen2018mr} & Ours & D.M. & \cite{hoppe2017deep} & \cite{cohen2018mr} & Ours  \\
		\hline
		1005.0 & 1001.0 & 1002.9 & 1009.3 & 1004.8 & -4.0 & -2.1 & 4.3 & \textbf{-0.2} \\
		1005.5 & 1001.0 & 1003.3 & 1010.0 & 1005.3 & -4.5 & -2.3 & 4.5 & \textbf{-0.2} \\
		1006.0 & 1011.0 & 1003.6 & 1010.6 & 1005.8 & 5.0 & -2.4 & 4.6 & \textbf{-0.2} \\
		1006.5 & 1011.0 & 1004.1 & 1011.2 & 1006.3 & 4.5 & -2.5 & 4.7 & \textbf{-0.2} \\
		1007.0 & 1011.0 & 1004.5 & 1011.8 & 1006.8 & 4.0 & -2.5 & 4.8 & \textbf{-0.3} \\
		\hline
		RMSE & - & - & - & - & 4.4 & 2.3 & 4.6 & \textbf{0.2} \\
		\toprule
		\hline
		& \multicolumn{4}{c|}{T2 Estimation} & \multicolumn{4}{c}{T2 Errors} \\
		\hline
		Truth & D.M. & \cite{hoppe2017deep} & \cite{cohen2018mr} & Ours & D.M. & \cite{hoppe2017deep} & \cite{cohen2018mr} & Ours  \\
		\hline
		505.0 & 501.0 & 513.6 & 504.3 & 505.2 & -4.0 & 8.6 & -0.7 & \textbf{0.2} \\
		505.5 & 511.0 & 514.1 & 504.8 & 505.7 & 5.5 & 8.6 & -0.7 & \textbf{0.2} \\
		506.0 & 501.0 & 514.7 & 505.3 & 506.2 & -5.0 & 8.6 & -0.7 & \textbf{0.2} \\
		506.5 & 511.0 & 515.2 & 505.9 & 506.8 & 4.5 & 8.7 & -0.6 & \textbf{0.3} \\
		507.0 & 511.0 & 515.7 & 506.4 & 507.3 & 4.0 & 8.7 & -0.6 & \textbf{0.3} \\
		\hline
		RMSE & - & - & - & - & 4.6 & 8.6 & 0.7 & \textbf{0.2} \\
		\hline \hline
	\end{tabular}
	\label{Tab:TestingSynthetic_Contin}
\end{table}

\begin{table}[t]
	\centering
	\footnotesize 
	\caption{
		Testing on anatomical dataset with full k-space sampling. Comparing parameter restoration performance, in terms of PSNR, SNR, RMSE and correlation coefficient.
	}
	\begin{tabular}{p{0.14\columnwidth}| c |c |c |c |c }
		\hline \hline
		& Dict. Match. & CNN~\cite{hoppe2017deep} & FNN~\cite{cohen2018mr} & Proposed basic & Proposed nonlocal \\
		& T1  /  T2   & T1  /  T2  & T1  /  T2  & T1  /  T2 & T1  /  T2  \\
		\hline
		PSNR (dB) & 56.64 / 52.04 & 54.06 / 49.88 & 54.53 / 54.36 & 56.59 / 60.01 & 56.47 / 62.56 \\
		SNR (dB) & 42.20 / 27.81 & 39.63 / 25.66 & 40.09 / 30.07 & 42.15 / 35.76 & 42.03 / 38.32 \\
		RMSE (ms) & 6.623 / 6.252 & 8.912 / 8.015 & 8.45 / 4.78 & 6.661 / 2.498 & 6.76 / 1.86 \\
		CorrCoef & 1.00 / 1.00 & 1.00 / 1.00 & 1.00 / 1.00 & 1.00 / 1.00 & {1.00} / {1.00} \\
		time cost (s) & 84.56 & 0.69 & 0.41 & 1.6 & 2.1 \\
		\hline \hline
	\end{tabular}
	\label{Tab:TestingPhantom}
\end{table}

\begin{table}[t]
	\centering
	\scriptsize
	\caption{
		Testing on anatomical dataset with k-space subsampling ratio 70\% and 15\% using Gaussian patterns and 200 time frames. 
	}
	\begin{tabular}{c| c |c |c| c |c | c | c } 
		\hline \hline
		& \multicolumn{7}{c}{k-space subsampling factor $\beta$ = 70\%} \\
		\hline
		& Ma et al.~\cite{ma2013magnetic} & BLIP~\cite{davies2014compressed} & FLOR~\cite{mazor2018low}  
		& CNN~\cite{hoppe2017deep} & FNN~\cite{cohen2018mr} & \multicolumn{2}{c}{Proposed} \\
		& & & & & & basic & nonlocal \\
		\hline
		PSNR (dB) & 23.69 / 38.17 & 45.67 / 47.84 & 50.11 / 50.85 & 49.71 / 45.48 & 50.15 / 51.08 & 50.79 / 51.59 & 49.87 / 57.57 \\
		SNR (dB) & 8.73 / 13.84 & 31.28 / 23.49 & 35.67 / 26.48 & 35.26 / 21.19 & 35.70 / 26.67 & 36.34 / 27.19 & 35.42 / 33.30 \\
		RMSE (ms) & 294.32 / 30.87 & 23.42 / 10.14 & 14.01 / 7.17 & 14.71 / 13.31 & 13.99 / 6.98 & 12.99 / 6.57 & 14.44 / 3.31 \\
		time cost (s) & 72.88 & 75.70 & 85.35 & 23.72 & 23.53 & 24.85 & 26.3 \\		%
		\hline \hline
		& \multicolumn{7}{c}{k-space subsampling factor $\beta$ = 15\%} \\
		\hline
		& Ma et al.~\cite{ma2013magnetic} & BLIP~\cite{davies2014compressed} & FLOR~\cite{mazor2018low}  
		& CNN~\cite{hoppe2017deep} & FNN~\cite{cohen2018mr} & \multicolumn{2}{c}{Proposed} \\
		& & & & & & basic & nonlocal \\
		\hline
		PSNR (dB) & 27.94 / 32.84 & 35.45 / 39.25 & 44.95 / 46.11 & 43.74 / 35.98 & 45.03 / 45.90 & 45.23 / 44.44 & 45.39 / 51.32 \\
		SNR (dB) & 13.50 / 8.61 & 20.99 / 14.58 & 30.51 / 21.89 & 29.23 / 12.26 & 30.58 / 21.32 & 30.76 / 19.78 & 30.91 / 26.99 \\
		RMSE (ms) & 180.3 / 57.03 & 76.01 / 27.25 & 25.46 / 12.37 & 29.27 / 39.73 & 25.21 / 12.68 & 24.65 / 15.00 & 24.20 / 6.79 \\
		time cost (s) & 106 & 112.8 & 121.7 & 24.54 & 24.36 & 25.67 & 27.31 \\
		\hline \hline
	\end{tabular}
	\label{Tab:TestingPhantomSubsamplingNonlocal}
\end{table}
%
\begin{table}[t]
	\centering
	\footnotesize 
	\caption{
		Testing on anatomical dataset with k-space subsampling ratio 15\% using Gaussian patterns and 1000 time frames. 
	}
	\begin{tabular}{c| c |c |c| c |c | c  } 
		\hline \hline
		& Ma et al.~\cite{ma2013magnetic} & BLIP~\cite{davies2014compressed} & FLOR~\cite{mazor2018low}  
		& CNN~\cite{hoppe2017deep} & FNN~\cite{cohen2018mr} & Proposed \\
		\hline
		PSNR (dB) & 27.53 / 33.28 & 35.50 / 39.10 & 50.90 / 50.04 & 41.96 / 39.21 & 52.62 / 49.86 & 52.32 / 52.79 \\
		SNR (dB) & 13.09 / 9.05 & 21.06 / 14.87 & 36.44 / 25.65 & 27.44 / 15.05 & 38.17 / 25.43 & 37.86 / 28.35 \\
		RMSE (ms) & 189.09 / 54.21 & 75.53 / 27.74 & 12.83 / 7.87 & 35.91 / 27.37 & 10.52 / 8.04 & 10.89 / 5.74 \\
		\hline \hline
	\end{tabular}
	\label{Tab:GaussSubsamp_L1000}
\end{table}

\begin{table}[t]
	\centering
	\scriptsize
	\caption{
		Testing on anatomical dataset with k-space subsampling ratio 9\% using spiral trajectories and 1000 time frames. 
	}
	\begin{tabular}{c| c |c |c| c |c | c  } 
		\hline \hline
		& Ma et al.~\cite{ma2013magnetic} & BLIP~\cite{davies2014compressed} & FLOR~\cite{mazor2018low}  
		& CNN~\cite{hoppe2017deep} & FNN~\cite{cohen2018mr} & Proposed \\
		\hline
		PSNR (dB) & 26.66 / 30.44 & 29.35 / 39.47 & 39.32 / 44.60 & 35.68 / 27.74 & 40.26 / 44.70 & 41.45 / 45.41 \\
		SNR (dB) & 12.22 / 6.21 & 15.03 / 15.22 & 24.88 / 20.38 & 21.45 / 4.48 & 25.84 / 20.30 & 27.02 / 21.04 \\
		RMSE (ms) & 209.01 / 75.18 & 153.37 / 26.57 & 48.67 / 14.72 & 73.96 / 102.57 & 43.65 / 14.55 & 38.08 / 13.41 \\
		\hline \hline
	\end{tabular}
	\label{Tab:SpiralSubsamp0_09_L1000}
\end{table}

\end{document}